\documentclass[twocolumn,superscriptaddress,showpacs,prd,aps,amsmath,amssymb,nofootinbib]{revtex4-1}
\usepackage{natbib}
\usepackage{graphicx,color}
\usepackage{amsmath,amssymb}
\usepackage{verbatim}
\usepackage{float}
\usepackage{wasysym}
\usepackage{amssymb,graphicx}
\usepackage{epsfig}
\usepackage{psfrag}
\usepackage{dsfont}
\usepackage{amsfonts}
\usepackage{mathrsfs}
\usepackage{multirow}
\usepackage{times}
\usepackage{bm}
\usepackage{hyperref}
\usepackage{xspace}
\hypersetup{
  colorlinks=true,        % false: boxed links; true: colored links
  linkcolor=blue,         % color of internal links
  citecolor=cyan,         % color of links to bibliography
}
\usepackage{pifont}% http://ctan.org/pkg/pifont
\usepackage{multirow}
\usepackage{verbatim}

\graphicspath{{./}}
\newcommand\NS{\text{\tiny{NS}}}
\newcommand\tmer{t_{\text{\tiny mer}}}
\newcommand\tbh{t_{\text{\tiny BH}}}
\hypersetup{
}

\begin{document}
\title{Magnetohydrodynamic Simulations of Binary Neutron Star Mergers in General Relativity:\\
  Effects of Magnetic Field Orientation on Jet Launching}
\author{Milton Ruiz}
\affiliation{Department of Physics, University of Illinois at
  Urbana-Champaign, Urbana, IL 61801}
\author{Antonios Tsokaros}
\affiliation{Department of Physics, University of Illinois at
  Urbana-Champaign, Urbana, IL 61801}
\author{Stuart L. Shapiro}
\affiliation{Department of Physics, University of Illinois at
  Urbana-Champaign, Urbana, IL 61801}
\affiliation{Department of Astronomy \& NCSA, University of
  Illinois at Urbana-Champaign, Urbana, IL 61801}
%
%%%%%%%%%%%%%%%%%%%%
%%%   ABSTRACT   %%%
%%%%%%%%%%%%%%%%%%%%
%
\begin{abstract}
  Binary neutron star mergers can be sources of gravitational waves coincident with
  electromagnetic counterpart emission across the spectrum. To solidify their role as
  multimessenger sources, we present fully 3D, general relativistic, magnetohydrodynamic
  simulations of highly spinning binary neutrons stars initially on quasicircular orbits
  that merge and undergo delayed collapse to a black hole. The binaries consist of two
  identical stars modeled as $\Gamma=2$ polytropes with spin $\chi_{\text{\tiny{NS}}}=
  0.36$ aligned along the direction of the total orbital angular momentum $L$. Each star
  is initially threaded by a dynamical unimportant interior dipole magnetic field. The
  field is extended into the exterior where a nearly force-free magnetosphere  resembles that
  of a pulsar. The magnetic dipole moment $\mu$ is either aligned or perpendicular to
  $L$ and has the same initial  magnitude for each orientation. For comparison, we also
  impose symmetry across the orbital plane in one case where $\mu$~in both stars is aligned
  along $L$. We find that the lifetime of the transient hypermassive neutron star remnant,
  the jet launching time, and the ejecta (which can give rise to a detectable kilonova) are very
  sensitive to the magnetic field orientation. By contrast, the physical properties of
  the black hole + disk remnant, such as the mass and spin of the black hole, the accretion
  rate, and the electromagnetic (Poynting) luminosity, are roughly independent of the initial
  magnetic field orientation. In addition, we find imposing symmetry across the orbital
  plane does not play a significant role in the final outcome of the mergers.
  Our results suggest that, as in the black hole-neutron star merger scenario, an incipient
  jet emerges only when the seed
  magnetic field has a sufficiently large-scale poloidal component aligned to the initial
  orbital angular momentum. The lifetime [$\Delta t\gtrsim 140(M_{\NS}/1.625M_\odot)\rm
    ms$] and Poynting luminosities~[$L_{\text{\tiny EM}}\simeq 10^{52}$erg/s] of the
  jet, when it forms,  are consistent with typical short gamma ray bursts, as well as with the
  Blandford--Znajek mechanism for launching jets.
\end{abstract}

\pacs{04.25.D-, 04.25.dk, 04.30.-w, 47.75.+f}
\maketitle

%%%%%%%%%%%%%%%%%%%%%%
%%%  Introduction  %%%
%%%%%%%%%%%%%%%%%%%%%%
\section{Introduction}
The  exciting prospect of simultaneous observations of both gravitational waves
(GWs) and  electromagnetic (EM) signals originating from  the coalescence and
merger of binary neutron stars~(NSNS) makes these systems, along with black
hole-neutron star~(BHNS) binaries, prime targets for the LIGO/Virgo scientific
collaboration in the era of multimessenger astronomy (MA). These systems had
long been hypothesized as progenitors of the same central engines that power
short-hard gamma-ray bursts~(sGRBs), see e.g.~\cite{Pac86ApJ,EiLiPiSc,NaPaPi},
which was strongly supported by the first detection of a kilonova associated
with the sGRB ``GRB130603B''~\cite{Tanvir:2013pia,Berger:2013wna}.

The strongest theoretical support for this hypothesis came from self-consistent,
fully general relativistic magnetohydrodynamic (GRMHD)~simulations of BHNS and
NSNS mergers~\cite{prs15,Ruiz:2018wah,Ruiz:2016rai,Ruiz:2017inq} that showed
that an incipient jet may be launched if the NS is suitably magnetized. Nevertheless,
the detection of GW170817~\cite{TheLIGOScientific:2017qsa} coincident with a sGRB
(event~GRB170817A~\cite{Monitor:2017mdv}), as well as its  association with kilonova
AT 2017gfo/DLT17ck~\cite{Valenti:2017ngx}, provides the best direct observational
evidence so far
that some sGRBs are indeed powered by NSNS mergers, or at least  by the merger
of a compact binary where at least one of the companions is a NS. Note that the
progenitor of GW170817 has been identified as an NSNS based on the masses of the
companions; depending on the spin priors of the binary companions, their inferred
masses are in the broad range of~$0.86-2.26M_\odot$, though the total mass of the
system is constrained to be~$2.73-3.29 M_\odot$~with~$90\%$~credibility
\cite{TheLIGOScientific:2017qsa}. These masses are consistent with astrophysical
observations of NSs~(see~e.g.~\cite{Ozel:2016oaf,Lattimer:2015nhk,Bogdanov:2019qjb,
  Bogdanov:2019ixe}), but it cannot rule out the presence of a stellar-mass
BH~\cite{Foucart:2018rjc}. Recently, X-ray observations have strongly suggested that
the rapidly rotating, giant star 2MASS J05215658+4359220 is the binary companion of
a noninteracting~$\sim 3M_\odot$~BH~\cite{Thompson637}. So, there may be a population
of stellar-mass BHs missed by X-ray observations that eventually may form 
GW170817-like binary systems. Mechanisms and routes by which stellar-mass BH formation
may arise in binaries with NS companions were recently discussed in~\cite{Yang:2017gfb}.

The GRMHD simulations of  BHNSs mergers reported in~\cite{prs15,Ruiz:2018wah}, in
which the  NSs are modeled as irrotational $\Gamma=2$~polytropes, have shown
that these systems, evolved from the late inspiral through tidal disruption, merger,
and settling, can launch a magnetically-supported incipient jet. The lifetime of the
jet [$\Delta
  t\sim 0.5(M_{\NS}/1.4M_\odot)$s] and its outgoing Poynting luminosity [
  $L_{\text{\tiny EM}}\sim 10^{51}\rm erg/s$] turn out to be consistent with
typical sGRBs~\cite{Bhat:2016odd,Lien:2016zny,Svinkin:2016fho,Ajello:2019zki}, as
well as with the Blandford-Znajek (BZ) mechanism for launching jets and their
associated Poynting luminosities~\cite{BZeffect77}. Here $M_{\NS}$ is the rest-mass
of the NS. The key requirement for jet
launching is the existence of a large-scale  poloidal magnetic field component with
a consistent sign in the vertical direction threading the BH + disk remnant
\cite{Beckwith08,Beckwith09}. These magnetic components can be obtained by endowing
the NS with a dipolar magnetic field resembling that of pulsars, with the dipole
moment along the direction of the total angular momentum of the system. The presence
of the dipole field ensures that the BH poles will be threaded with poloidal magnetic
lines before tidal disruption, whereby a significant poloidal component of the field will
remain after the disruption. Differential rotation in the accretion disk winds up the
field lines, converting poloidal flux into toroidal flux.  The magnetic field then
is amplified to~$\gtrsim~10^{15}$G above the BH poles and wound into a helical
funnel, inside which fluid elements from the accretion disk flow outward with Lorentz
factors at launch of $\gtrsim\Gamma_{\text{\tiny L}}=1.2$. We say that, at this
point, an incipient jet has emerged. By contrast, if the initial magnetic field is
confined to the NS interior, the frozen-in magnetic field following the NS disruption
is wound into a nearly toroidal configuration~(see e.g.~\cite{Etienne:2011ea}), and
hence jet formation is suppressed.

The GRMHD simulations of NSNS mergers, in which the NS is modeled as
$\Gamma=2$~polytrope, show that an incipient jet can be launched whether or not
the seed poloidal magnetic field is confined to the interior of the NS as long
as the binary forms a transient hypermassive remnant before undergoing delayed
collapse to a BH~\cite{Ruiz:2016rai,Ruiz:2017inq,Ruiz:2019ezy}. In this case,
in contrast to the prompt collapse case (see e.g.~\cite{Totani:2013lia,Paschalidis:2018tsa}
for possible EM counterparts in this case), { {the formation of a hypermassive neutron
star (HMNS) allows magnetic instabilities to amplify the magnetic energy to reach
equipartition with the plasma kinetic energy
before BH formation}}~\cite{Kiuchi:2014hja}.
Following the HMNS collapse, a magnetically--supported jet is then launched once the
regions above the BH poles approach force-free values ($B^2/8\,\pi \rho_0\gg 1$).
Here $B$ and $\rho_0$ are the strength of the magnetic field and the rest-mass
density, respectively. As in the BHNS case, the lifetime of the jet and its
associated Poynting luminosity are consistent with typical sGRBs
\cite{Bhat:2016odd,Lien:2016zny,Svinkin:2016fho,Ajello:2019zki}, as well as with the BZ mechanism.
Note that in the GRMHD simulations reported in~\cite{Kawamura:2016nmk,Ciolfi:2017uak,
Kiuchi:2014hja}, where the magnetic field is confined to the NS interior, neither an
outgoing outflow nor a jet were observed. The lack of a jet in~\cite{Kawamura:2016nmk,Ciolfi:2017uak},
where the effects of realistic equations of state (EOSs), mass ratios, and
orientations of the seed  poloidal magnetic field were probed, is likely due to the
incomplete development of the magneto-rotational-instability (MRI), which is required
to boost the magnetic field strength, though the formation of an  organized helical
structure above the BH was evident (see e.g.~Fig.~9 in~\cite{Kawamura:2016nmk}). The
absence of a jet in the very high-resolution studies in~\cite{Kiuchi:2014hja}, in
which an H4 EOS is used to model the NS, can be attributed to the persistent fall-back
environment that increases the downward ram pressure above the BH poles. These studies
may require longer simulations ($>39$ ms after merger) for a jet to emerge as long as
the matter fall-back timescale is shorter than the accretion disk
lifetime~\cite{Paschalidis:2016agf}.  In all  the above numerical studies reflection
symmetry across the orbital plane was imposed.

As the key requirement for jet launching in the NSNS scenario seems to be the amplification
of the magnetic field during the  HMNS epoch, one might tentatively conclude that NSNS mergers
undergoing delayed collapse to a BH can lead to jets under a wide variety of initial magnetic
field configurations. But, {\it is this enough?} It has been suggested that pure poloidal magnetic
field configurations may be unstable over an Alfv\'en timescale~(see e.g.
\cite{1973MNRAS.163...77M,1973MNRAS.162..339W}). Full 3D Newtonian simulations, and
recent full 3D general relativity simulations using the Cowling approximation, showed that
nonrotating and isolated stars endowed with pure poloidal components may relax into
a new configuration with both poloidal and toroidal magnetic components of similar
strengths~(see e.g.~\cite{Braithwaite:2005ps,Ciolfi:2012en,Ciolfi:2011xa}). But, {\it is
this new magnetic field configuration suitable for jet launching?} Moreover, if GW radiation
and magnetic turbulent viscosity drive the bulk of the HMNS into a purely axisymmetric
configuration, the sustained amplification of the magnetic field is, according to the
anti-dynamo theory~\cite{1978mfge.bookM}, no longer possible. If so, {\it is the BH
  + disk remnant an EM counterpart orphan?}

To address the above questions, we perform full 3D GRMHD simulations of NSNS configurations
in quasicircular orbits that merge and undergo delayed collapse to a BH.  The binaries consist
of two identical, uniformly rotating NSs modeled with a $\Gamma=2$ polytropic EOS with spin
$\chi_{\text{\tiny{NS}}}\equiv J_{\text{\tiny{ql}}}/(M/2)^2=0.36$, where $J_{\text{
    \tiny{ql}}}$ is the quasilocal angular momentum of the NS, and $M$ is the Arnowitt-Deser-Misner
(ADM) mass of the system \cite{Tsokaros:2018dqs}. We choose highly spinning NSNS configurations
to reduce computational costs, because, as we recently showed in~\cite{Ruiz:2019ezy}, the higher
the initial spin of the binary companions the shorter the jet launching time. { {We adopt
a $\Gamma=2$ polytropic EOS for a direct comparison with our previous results~(see e.g.
\cite{Ruiz:2016rai,Ruiz:2017inq,Ruiz:2019ezy}}}).
Each star is initially endowed  with a dipolar magnetic field of the same magnitude extending
from the stellar interior into its exterior and whose dipole moment is either aligned or
perpendicular to the direction of the total orbital angular momentum of the system $L$. We
consider the following configurations:
\begin{enumerate}
\item {\it Ali-Ali case}: The magnetic dipole moment in both stars is aligned to $L$.
\item {\it Ali-Per case}: The magnetic dipole moment in one of the stars is aligned to $L$,
  while in the other is perpendicular to it.
\item {\it Per-Per case}: The magnetic dipole moment in both stars is perpendicular to $L$.
\end{enumerate}
Note that these three cases can be used to infer the outcome of general cases in which the
dipole moment of the seed magnetic field is misaligned by an angle $\theta\leq 90^\circ$ to
the spin of the NS. For comparison
purposes, we also consider a second Ali-Ali case in which symmetry across the orbital plane
(equatorial symmetry) is imposed. This case has been treated previously in~\cite{Ruiz:2019ezy},
and it will be denoted here as {\it Ali-Ali (Eq)}.

As in~\cite{Ruiz:2019ezy}, in all our cases we find that, following the NSNS merger, magnetic
braking due to turbulent magnetic fields in the bulk of the transient HMNS induces
the formation of a uniformly rotating central core immersed in a low-density
Keplerian cloud of matter. Depending on the initial orientation of the magnetic
dipole moment, the HMNS collapses to a BH in a timescale of $\Delta t
\sim 24-74(M_{\NS}/1.625M_\odot)\rm ms$ 
following the NSNS merger, the shortest one being for the Ali-Ali and  Ali-Ali (Eq)
cases  and the longest one for the Per-Per case. The mass and spin of the BH, as well
as the rest-mass of the accretion disk, are roughly independent of the initial magnetic field
configuration. However, we find that, as in the BHNS cases reported in
\cite{Ruiz:2018wah}, an incipient jet is launched only when the system has initially
a large-scale poloidal magnetic field component aligned to the initial angular orbital momentum.
The lifetime of the incipient jet [$\Delta t\gtrsim 140(M_{\NS}/1.625M_\odot)\rm ms$] and
its outgoing Poynting luminosities [$L_{\text{\tiny EM}}\simeq 10^{52}\rm erg/s$], when it forms,
are consistent with typical sGRBs~\cite{Bhat:2016odd,Lien:2016zny,Svinkin:2016fho,Ajello:2019zki},
as well as with the BZ mechanism for launching jets~\cite{BZeffect}. We also observe that only
in Ali-Ali cases does a significant fraction of the rest-mass ($\gtrsim 10^{-3}M_\odot$)
become unbound and hence may lead to a kilonova signal  observable by current
telescopes, such as the Large Synoptic Survey Telescope (LSST)
\cite{Shibata:2017xdx,Rosswog}. { {Although our preliminary GRMHD simulations do not account for all
the physical processes involved in NSNS mergers, they indicate that, as the ejecta is highly
affected by the configuration of the magnetic field prior to the merger, NSNS merger models without
magnetic fields that are  used to explain the early part of the radioactive powered kilonova signal
(blue luminosity) linked to  GW170817 may overestimate the amount of escaping matter.}} 

We also probe whether different seed magnetic field orientations could be distinguishable by
current GW detectors. Assuming a source distance of $50$Mpc, we compute the match function
$\mathcal{M}_{\text{\tiny{GW}}}$ and find that the GWs of Ali-Ali and Per-Per are
distinguishable for a signal-to-noise ratio~$>25$, while in the other cases, they can be
distinguished with a signal-to-noise ratio~$>15$.
As GW150914 (first GW detection of BHBH) and GW170817 (first GW detection of NSNS) events were
observed with a signal-to-noise ratio of $24$ and $32.5$~\cite{LIGO_first_direct_GW,
  TheLIGOScientific:2017qsa}, respectively, current GW detectors may, in principle, be able to
distinguish effects induced by different magnetic field configurations.

The paper is organized as follows. A short summary of the numerical methods and their
implementation, initial data, grid setup, and global diagnostic checks are given in
Sec.~\ref{sec:IDMethods}.  For further details, readers are referred to~\cite{Ruiz:2019ezy}.
Sec.~\ref{sec:Alig_cases} contains a detailed comparison of the evolution of Ali-Ali in both
equatorial symmetry and full 3D. Secs.~\ref{sec:Ali-Per_cases} and~\ref{sec:Per-Per_cases}
describe the evolution of the Ali-Per and Per-Per cases, along with a comparison with the
previous cases. In Sec.~\ref{sec:GW_ftt} we assess the distinguishability of the GWs for the
different cases. We summarize our findings and conclude in Sec.\ref{sec:conclusion}. Throughout
the paper, we adopt geometrized units ($G=c=1$) except where stated otherwise. Greek
indices denote all four spacetime dimensions, while Latin indices imply
spatial parts only.
%
%%%%%%%%%%%%%%%%%%%%%
%%%  Method  & ID %%%
%%%%%%%%%%%%%%%%%%%%%
\section{Numerical Setup}
\label{sec:IDMethods}
The following section summarizes the key aspect of our numerical
approach.

%%%%%%%%%%%%%%%%%%%%%%%%%%
%%%  Numerical Method  %%%
%%%%%%%%%%%%%%%%%%%%%%%%%%
\paragraph*{{\bf Numerical Methods}:}

We use the GRMHD code developed by the Illinois Numerical Relativity Group
\cite{Etienne:2010ui}, which is embedded in the~\texttt{Cactus} infrastructure
\cite{cactusweb} and uses~\texttt{Carpet}~\cite{carpetweb} for moving boxes
refinement. It employs  the BSSN evolution equations~\cite{shibnak95,
  BS}, with fourth-order centered
spatial differencing, except on shift advection terms, where a fourth-order
upwind differencing is used, coupled to the puncture gauge conditions (see~Eq.
(2)-(4) in~\cite{Etienne:2007jg}). In all our evolution, we set the damping
coefficient $\eta$ appearing in the shift condition~to~$3.75/M$, with $M$
the ADM mass of the system. Time integration is performed using the Method
of Lines with a fourth-order Runge-Kutta integration scheme and a
Courant-Friedrichs-Lewy factor equal to~$0.5$. For numerical stability,
we add fifth-order Kreiss-Oliger dissipation~\cite{goddard06} in the BSSN
evolution equations. Also, a dissipation term in the evolution equation for
the conformal factor is added to damp the Hamiltonian constraint violations~(see Eq.~19
in~\cite{DMSB}).

%%%%%%%%%%%%%%%%%%%%%%
%%%  Initial data  %%%
%%%%%%%%%%%%%%%%%%%%%%
\paragraph*{\bf{Initial Data}:}

We use the Compact Object CALculator~({\tt COCAL}) code to generate the initial NSNS
configurations on a quasicircular orbit (see e.g.~\cite{Tsokaros:2015fea,Tsokaros:2016eik,
  Tsokaros:2018dqs} for numerical details). Specifically, we use the
$\Gamma=2$, spinning NSNS configuration listed in Table 1 of~\cite{Ruiz:2019ezy},
for which the ADM mass of the system is $M=4.43(M_{\NS}/1.625M_\odot)\rm km=3.00
(M_{\NS}/1.625M_\odot)M_\odot$,
and has an initial coordinate separation of $45\,(M_{\NS}/1.625 M_\odot)\rm km$.
Each binary companion has a quasilocal dimensionless spin parameter $\chi_{\NS}
\equiv J_{\text{\tiny ql}}/(M/2)^2\simeq 0.36$  [or a rotational period $T\simeq
  2.3 (M_{\NS}/1.625M_\odot)\rm ms$] aligned with orbital 
angular momentum of the system~\cite{Tsokaros:2018dqs}, a rest mass of $M_{\NS}=
1.625M_\odot(k/k_\text{\tiny L})^{1/2}$ and compactness ${\cal C}=0.138$. Here
$k_{\text{\tiny L}}=269.6\rm km^2$ is the polytropic constant used to generate
the initial data where $k\equiv P/\rho_0^\Gamma$. Note that for an $\Gamma=2$
polytrope, the maximum mass configuration has $\mathcal{C}=0.21$, and
$M_{\NS}^{\text{\tiny max}}=1.23M_{\NS}$.

\begin{figure*}
  \centerline{{\bf Equatorial symmetry}\hspace{5cm} {\bf Full 3D}}
  \centering
  \includegraphics[width=0.46\textwidth]{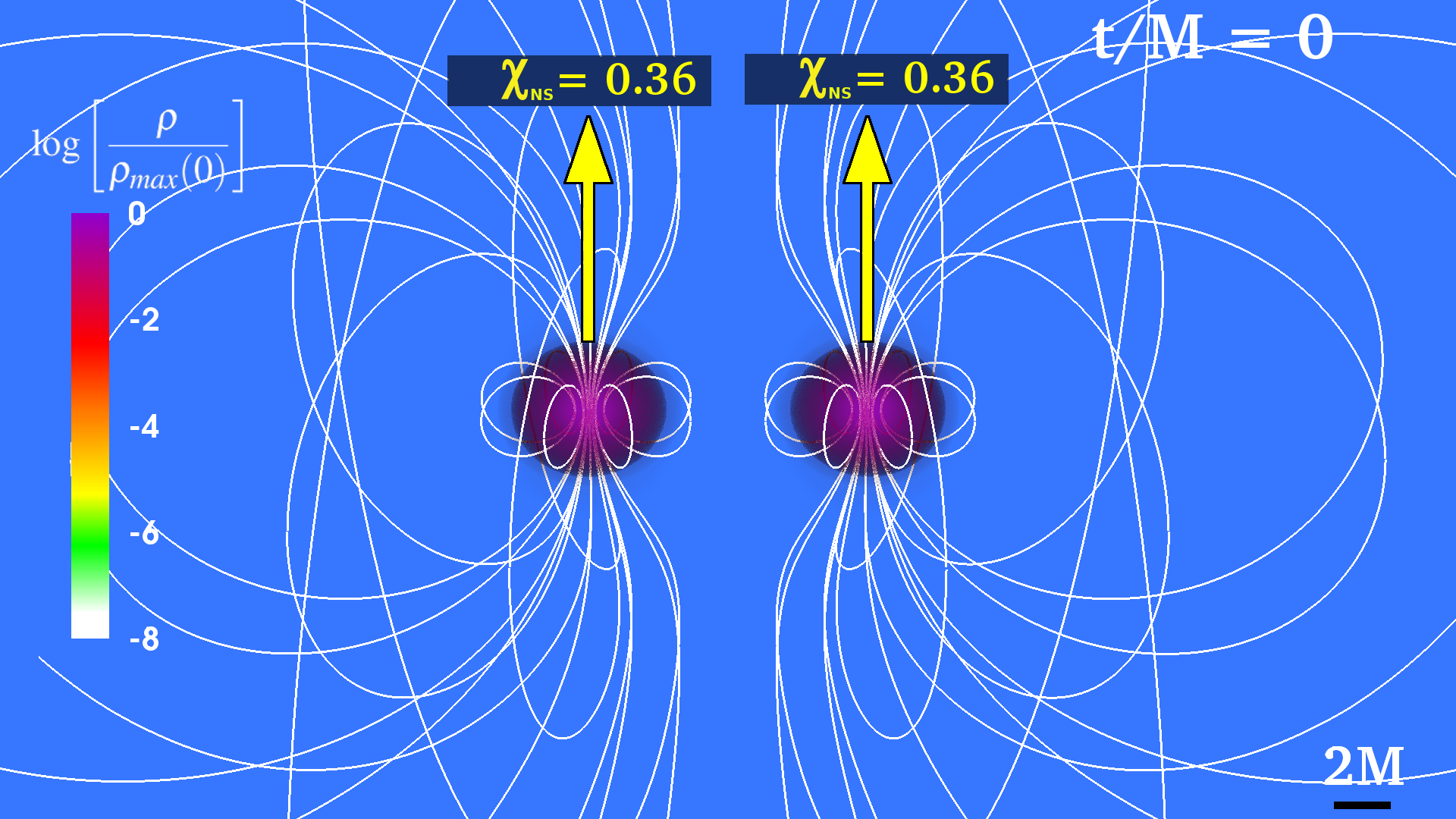}
  \includegraphics[width=0.46\textwidth]{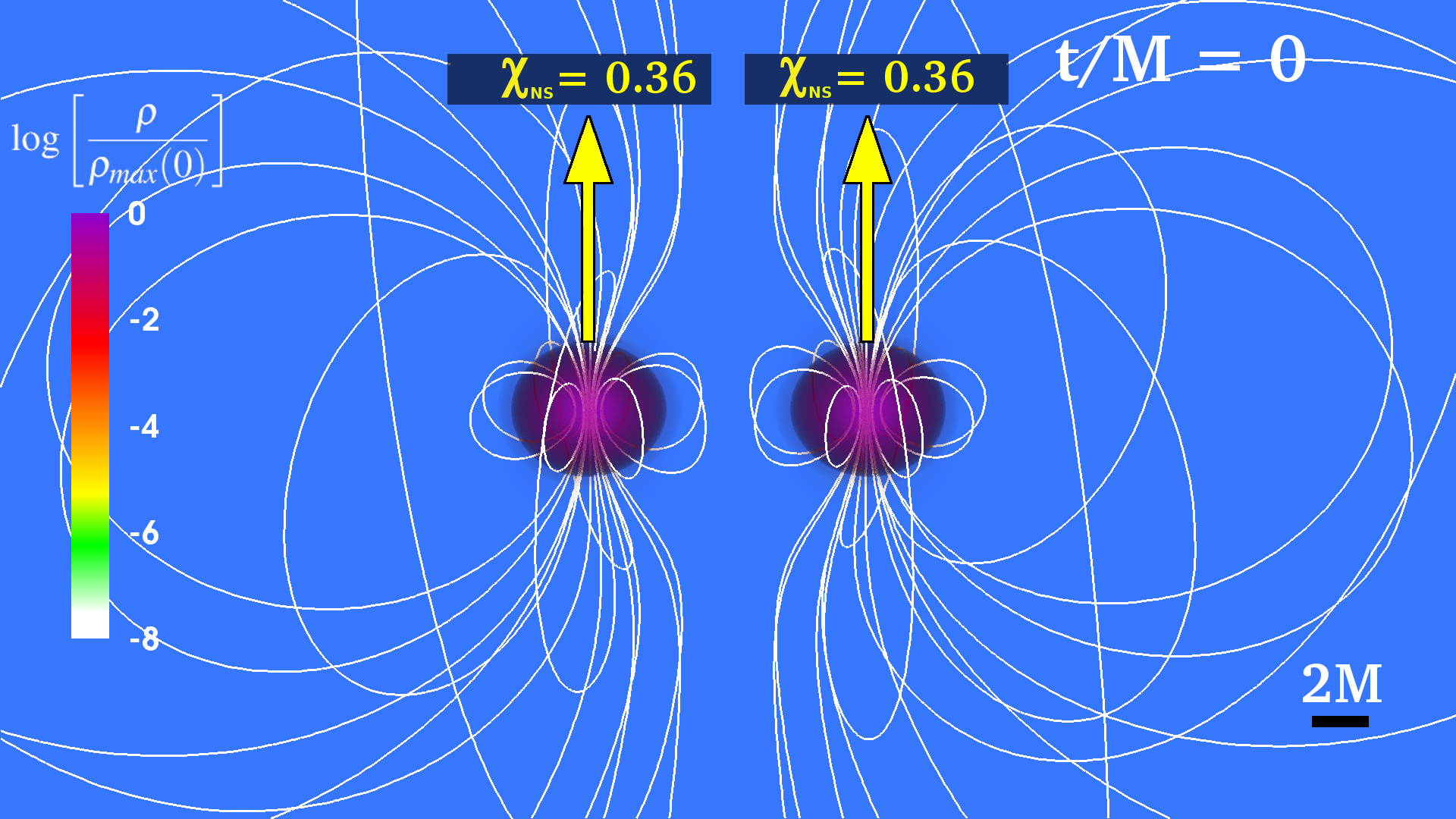}
  \includegraphics[width=0.46\textwidth]{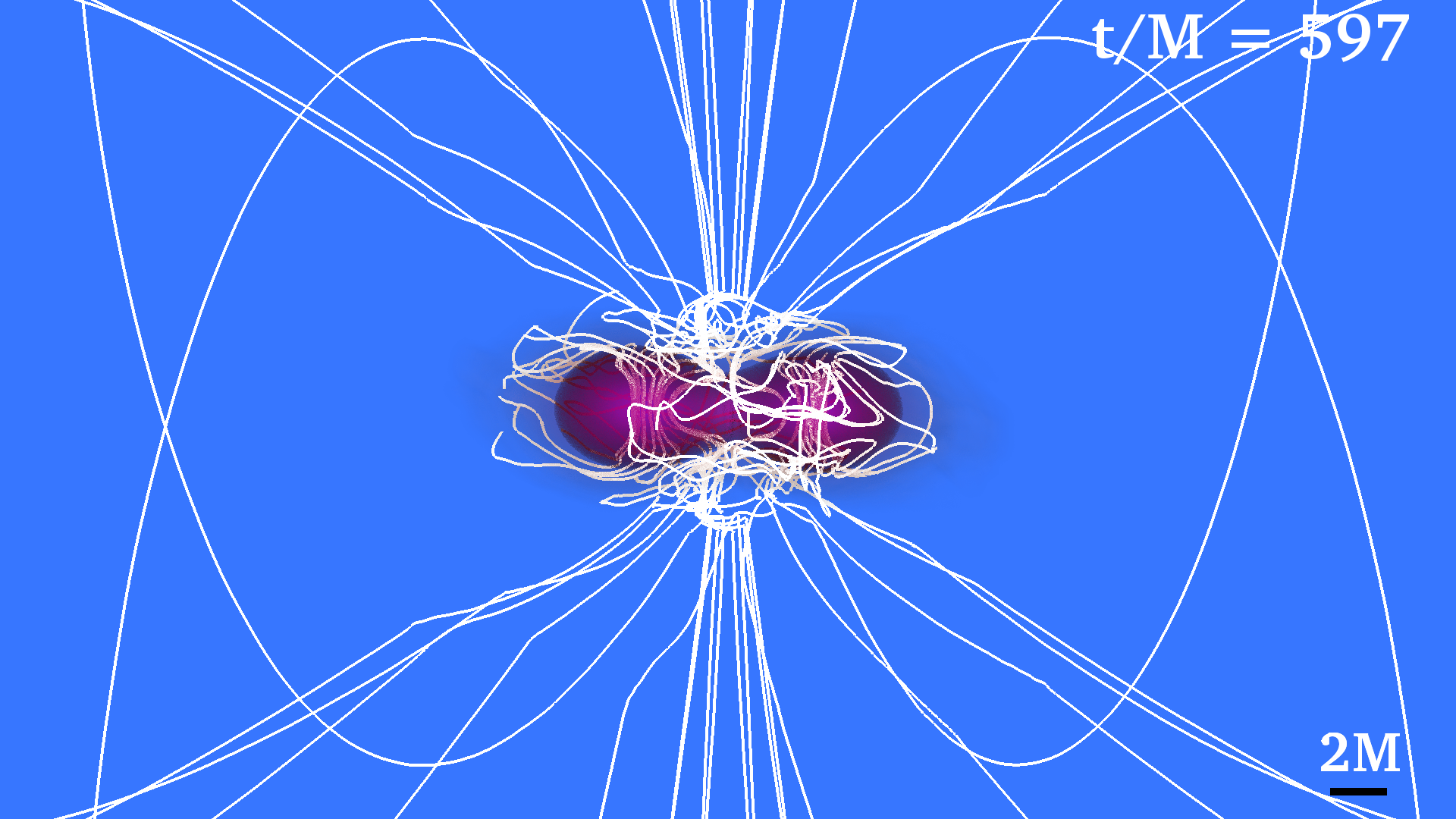}
  \includegraphics[width=0.46\textwidth]{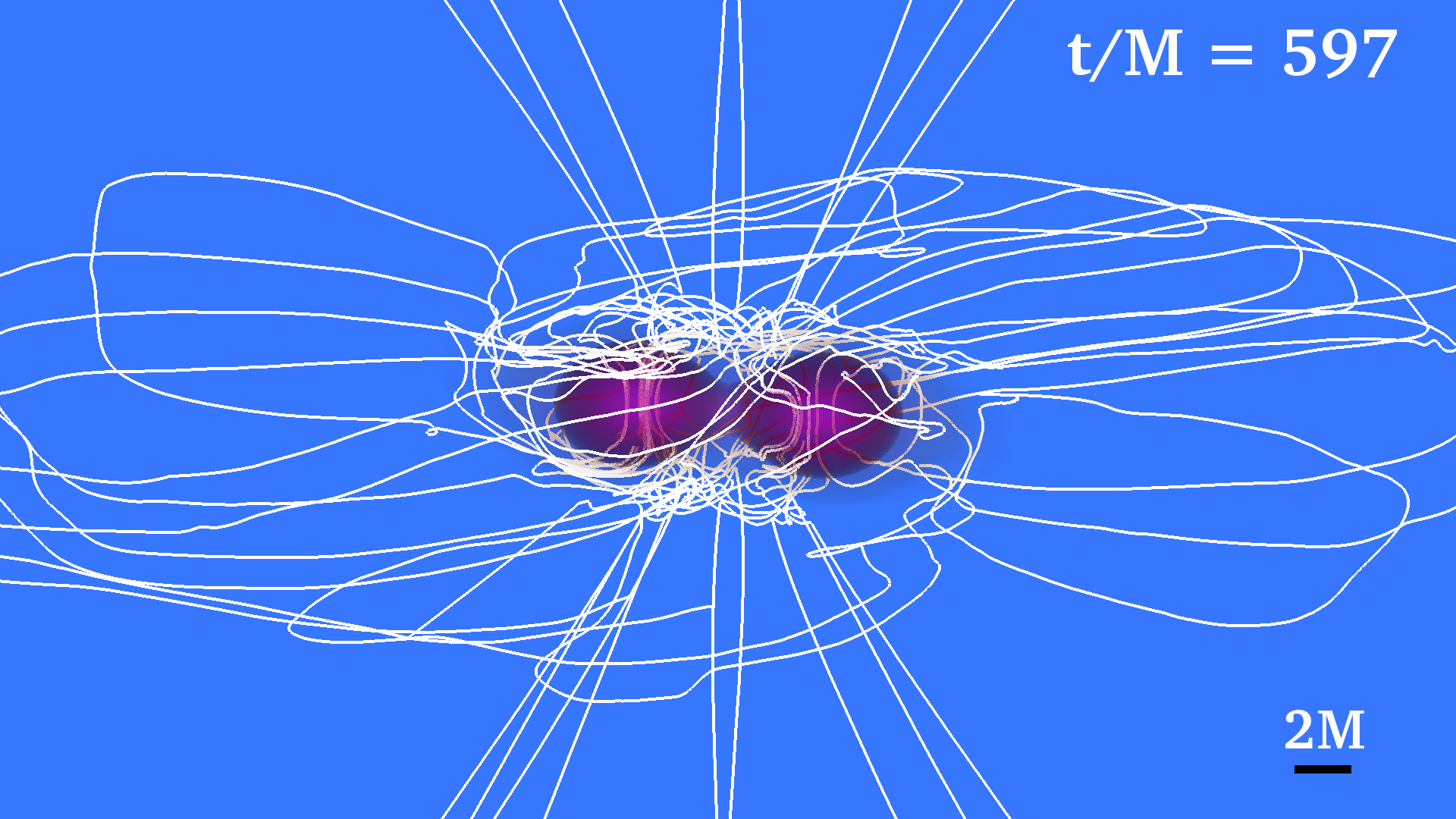}
  \includegraphics[width=0.46\textwidth]{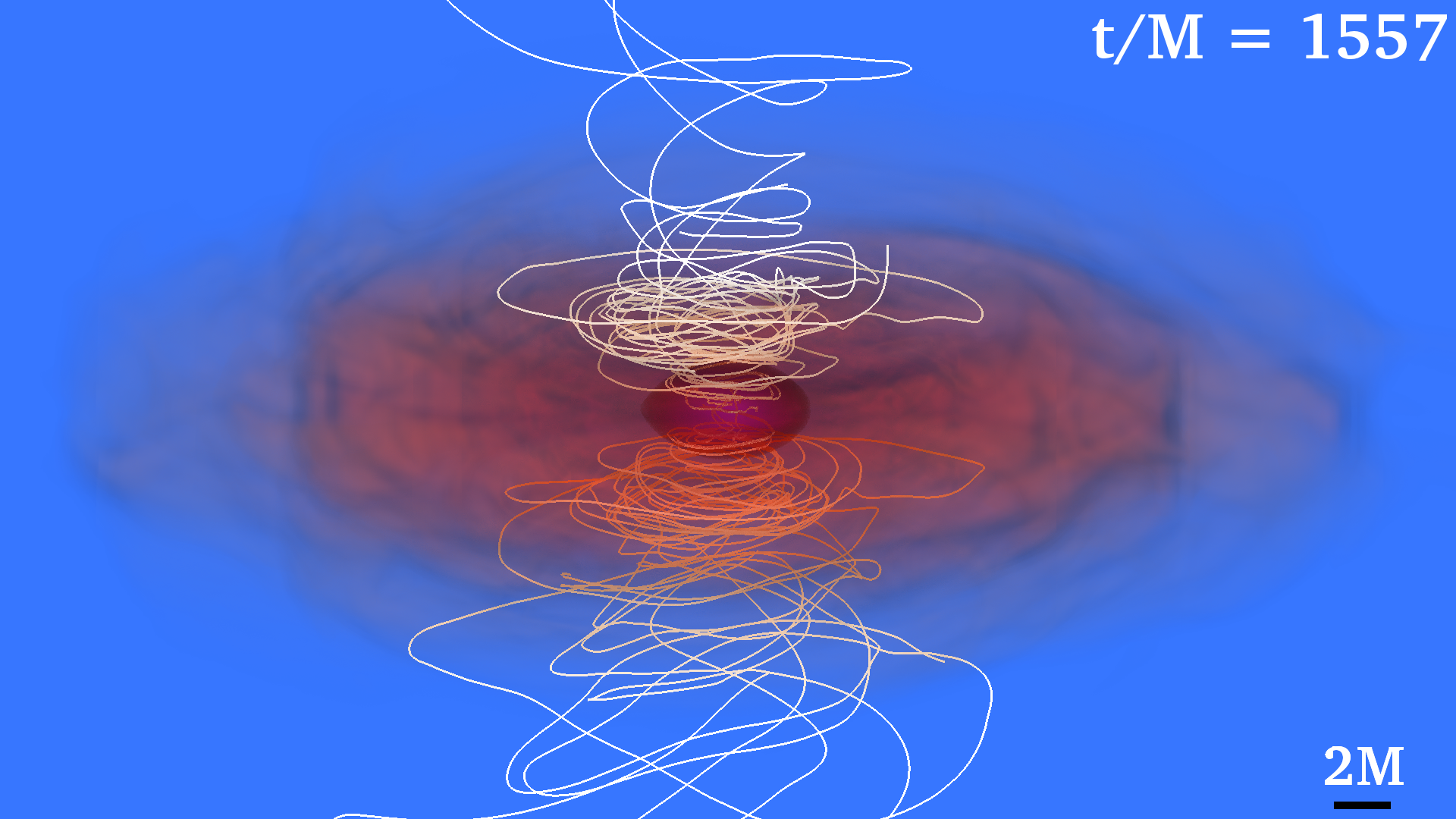}
  \includegraphics[width=0.46\textwidth]{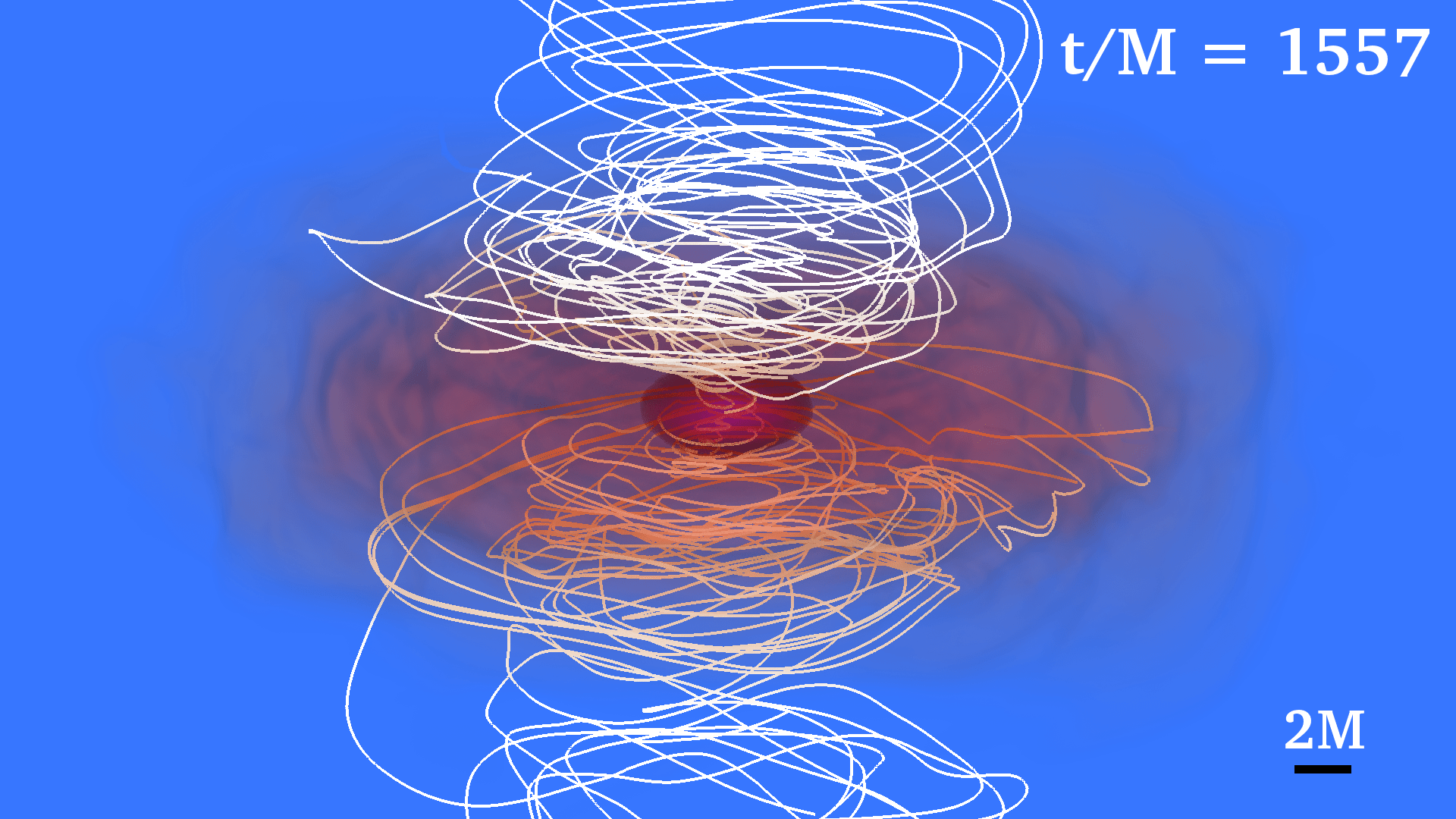}
  \includegraphics[width=0.46\textwidth]{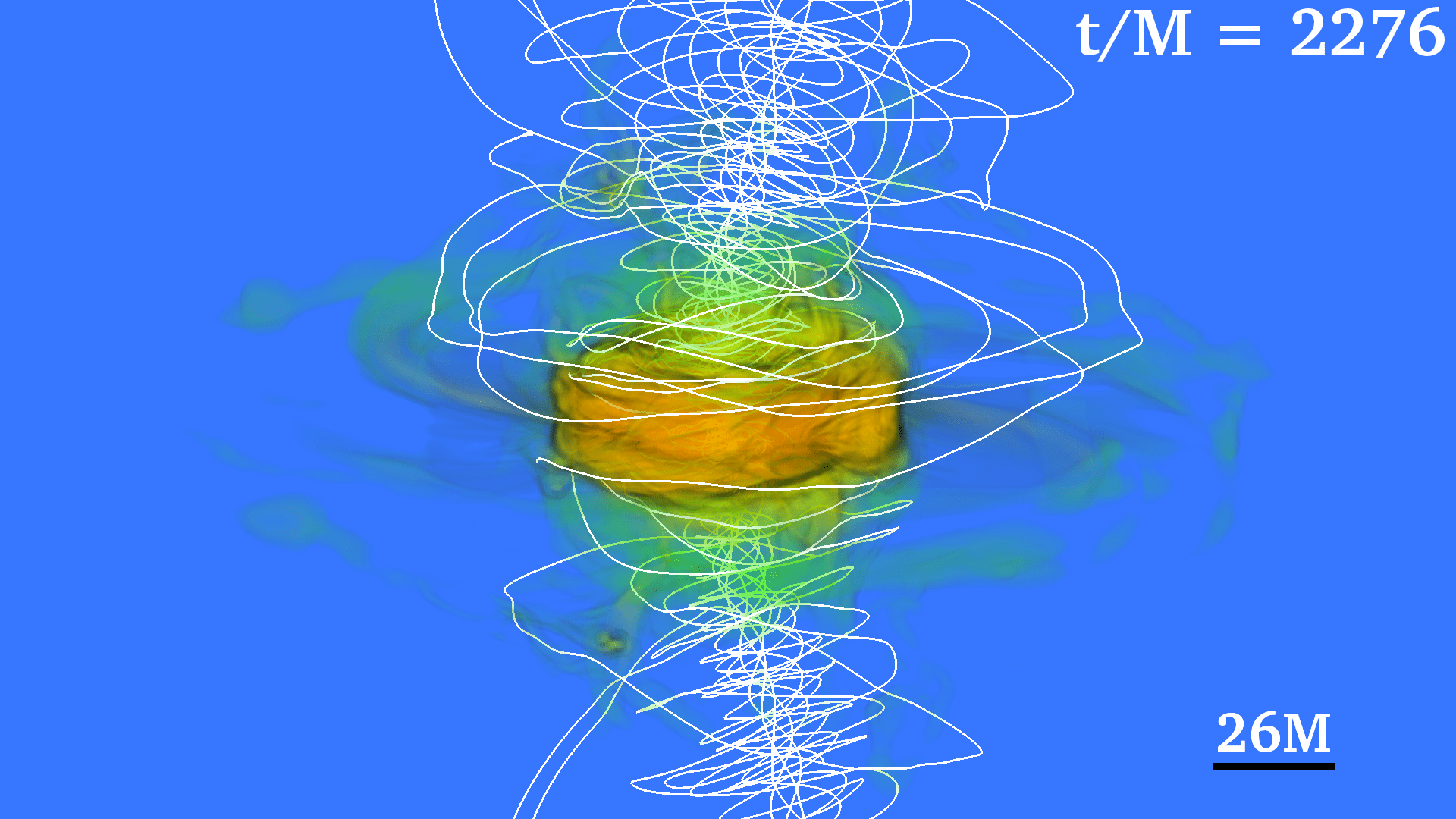}
  \includegraphics[width=0.46\textwidth]{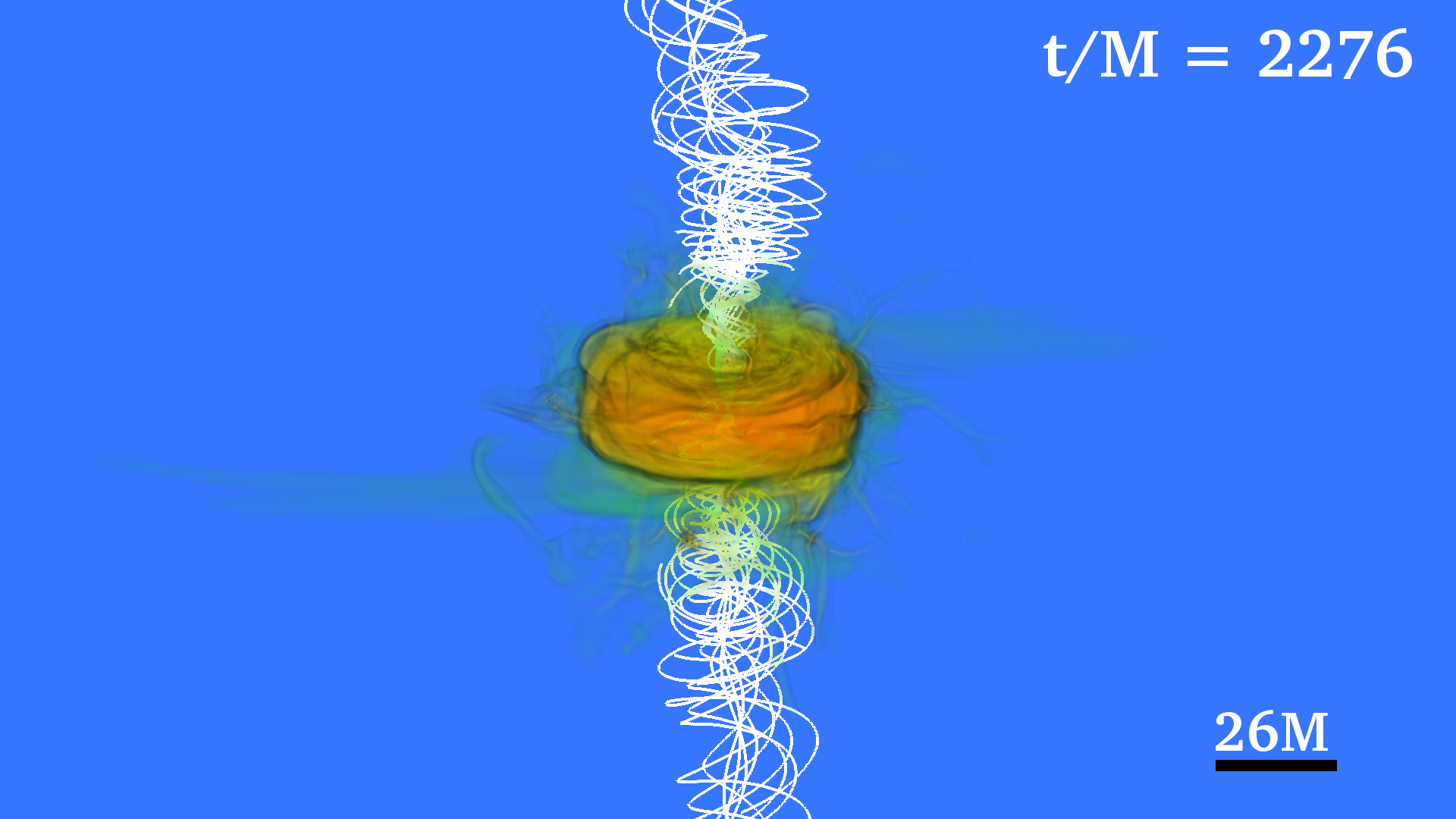}
  \includegraphics[width=0.46\textwidth]{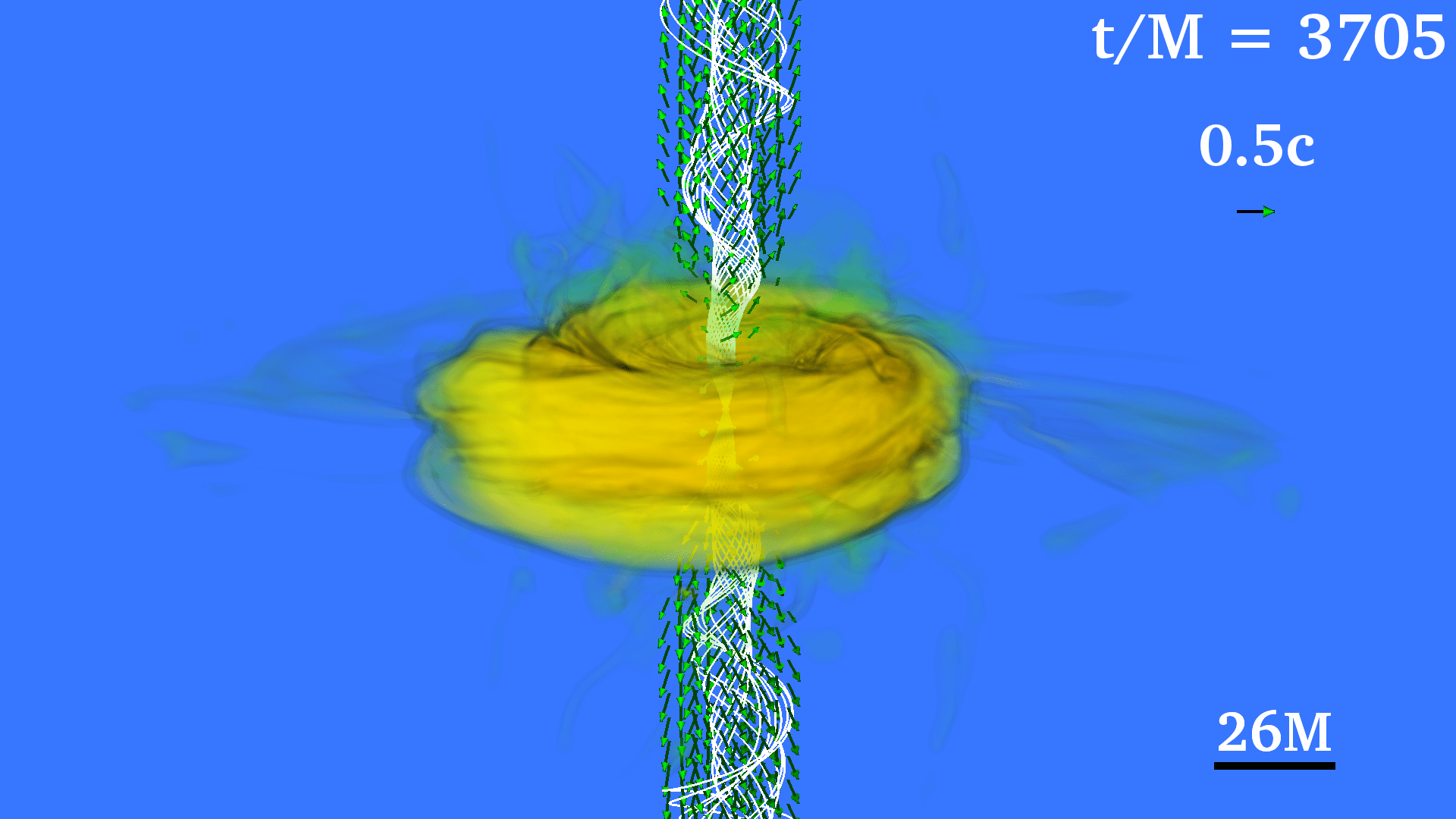}
  \includegraphics[width=0.46\textwidth]{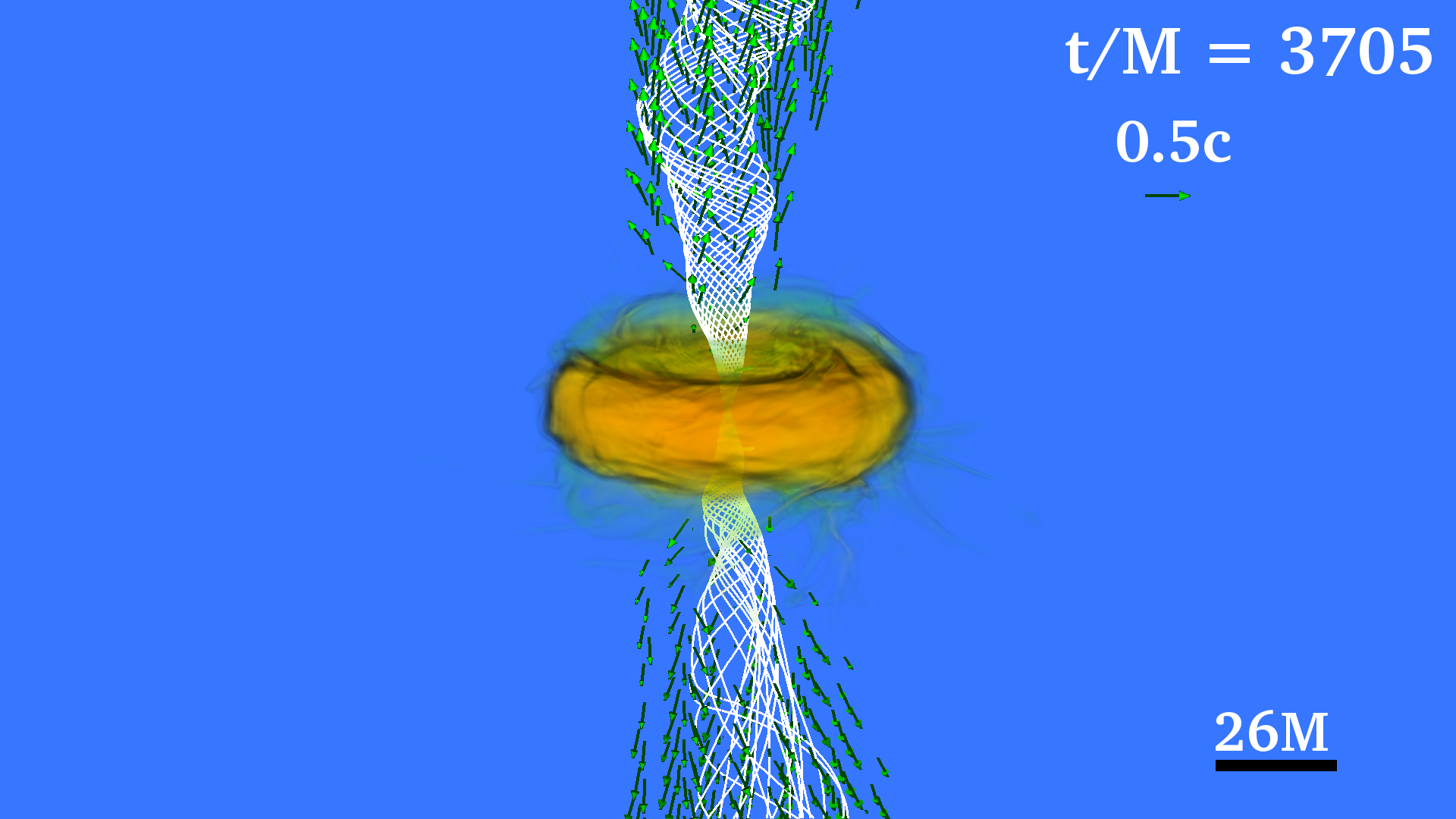}
  \caption{Volume rendering of rest-mass density $\rho_0$, normalized to the
    initial maximum value $\rho^{\rm max}_{0}=10^{14.78}(1.625\,M_\odot/
    M_{\NS})^2 \rm {g/cm}^3$ (log scale), at selected times for the Ali-Ali (Eq) case
    (left column) and the Ali-Ali case (right column). White lines represent the magnetic
    field lines, while arrows indicate plasma velocities. Bottom panels highlight
    the final configuration of the BH + disk remnant after an incipient jet has
    been launched. Here $M=1.47\times 10^{-2}(M_{\NS}/1.625M_\odot)\rm ms$= $4.4288
    (M_{\NS}/1.625M_\odot)\rm km$.
    \label{fig:Eq_vs_full}}
\end{figure*}

We initially endow the star with a dipole-like magnetic field whose dipole moment
is either aligned or perpendicular to the the total angular momentum of the system
$L$. Following~\cite{Ruiz:2019ezy}, the seed magnetic field in the aligned case is
generated by the vector potential (see~top panels in~Fig.~\ref{fig:Eq_vs_full})
%
%%%%%%%%%%%%%%%%%%%%%%%%%%
%%%  Vector potential  %%%
%%%%%%%%%%%%%%%%%%%%%%%%%%
\begin{eqnarray}
 A_\phi&=& \frac{\pi\,\varpi^2\,I_0\,r_0^2}{(r_0^2+r^2)^{3/2}}
 \left[1+\frac{15\,r_0^2\,(r_0^2+\varpi^2)}{8\,(r_0^2+r^2)^2}\right]\,,
 \label{eq:Aphi_pot}
\end{eqnarray}
induced by a current loop $I_0$  inside the star with radius $r_0$,
where $r^2=\varpi^2+z^2$, $\varpi^2= (x-x_{\text{\tiny{CM}}})^2 +(y-y_{
  \text{\tiny{CM}}})^2$, and $(x_{\text{\tiny{CM}}}, y_{\text{\tiny{CM}}})$
is the center of mass of the NS, defined here as the position of the
maximum value of the rest-mass density of each NS. We choose $I_0$ and $r_0$
such that the maximum value of the magnetic-to-gas pressure ratio is
$P_{\text{\tiny mag}}/ P_{\text{\tiny gas}}=0.003125$ at the center of
each star. With this choice, the resulting magnetic field strength at
the NS pole is initially ${B}_{\rm pole}\sim 10^{15.2}(1.625M_\odot/
M_{\NS})\rm G$. As pointed out in~\cite{Ruiz:2016rai}, this magnetic field
strength is used to mimic the result of exponential growth of the magnetic field
due to the Kelvin-Helmholtz instability (KHI), along with the 
MRI, triggered during the NSNS merger and HMNS formation. However,
this growth is captured only in very high-resolution [$\Delta x\lesssim 70\rm m$]
NSNS simulations~\citep{Kiuchi:2014hja,Kiuchi:2015sga}; during merger the rms
value of the magnetic field
strength is boosted from $B_{\text{\tiny rms}}\sim 10^{13}\rm G$ to
$B_{\text{\tiny rms}}\sim 10^{15.5}\rm G$, with local values up to
$B\sim 10^{17}\rm G$.

%%%%%%%%%%%%%%%%%%%%%%%%%%%
%%%  Ali-Ali comparison %%%
%%%%%%%%%%%%%%%%%%%%%%%%%%%

\begin{figure}
  \centering
  \includegraphics[width=0.44\textwidth]{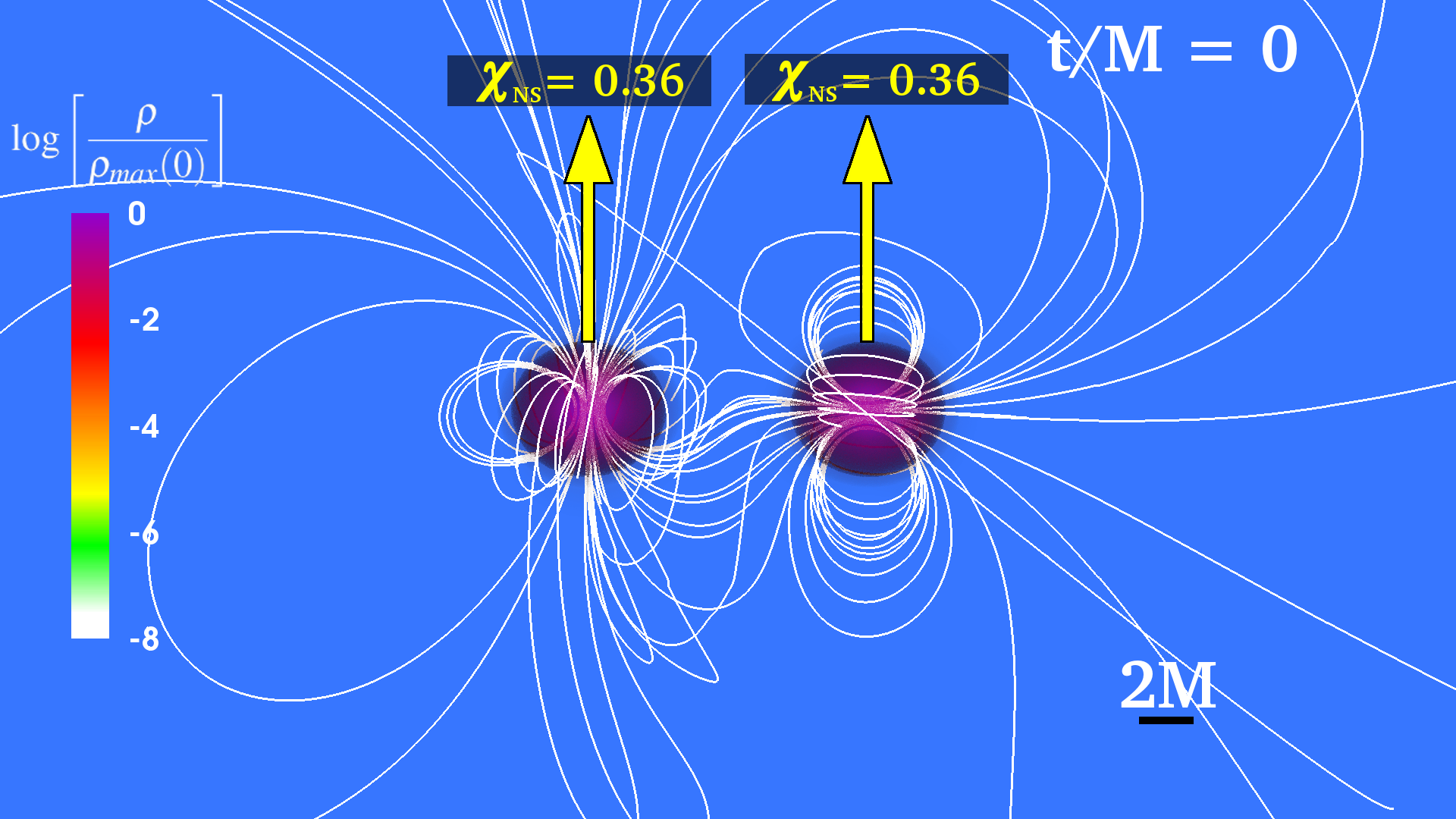}
  \includegraphics[width=0.44\textwidth]{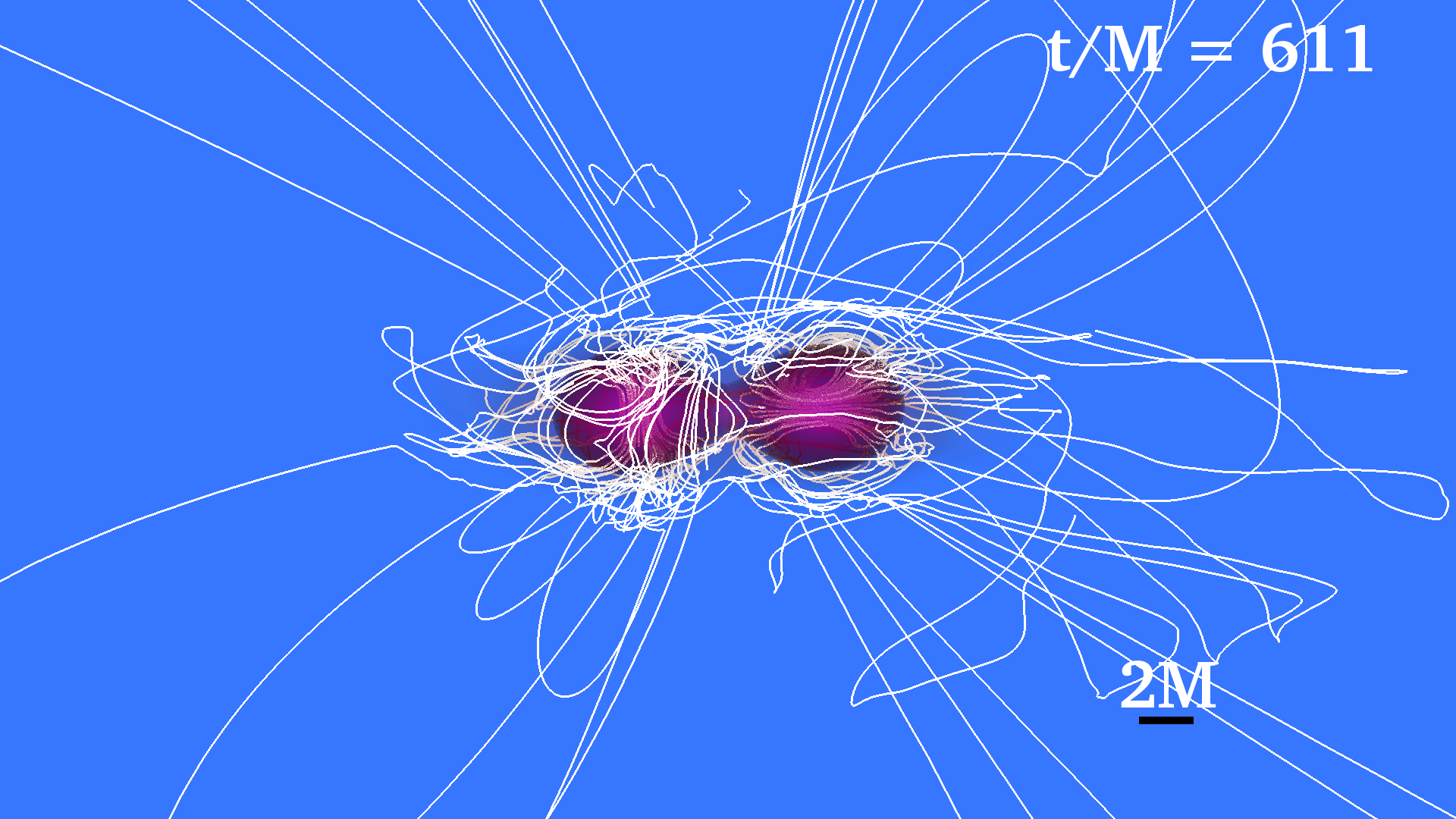}
  \includegraphics[width=0.44\textwidth]{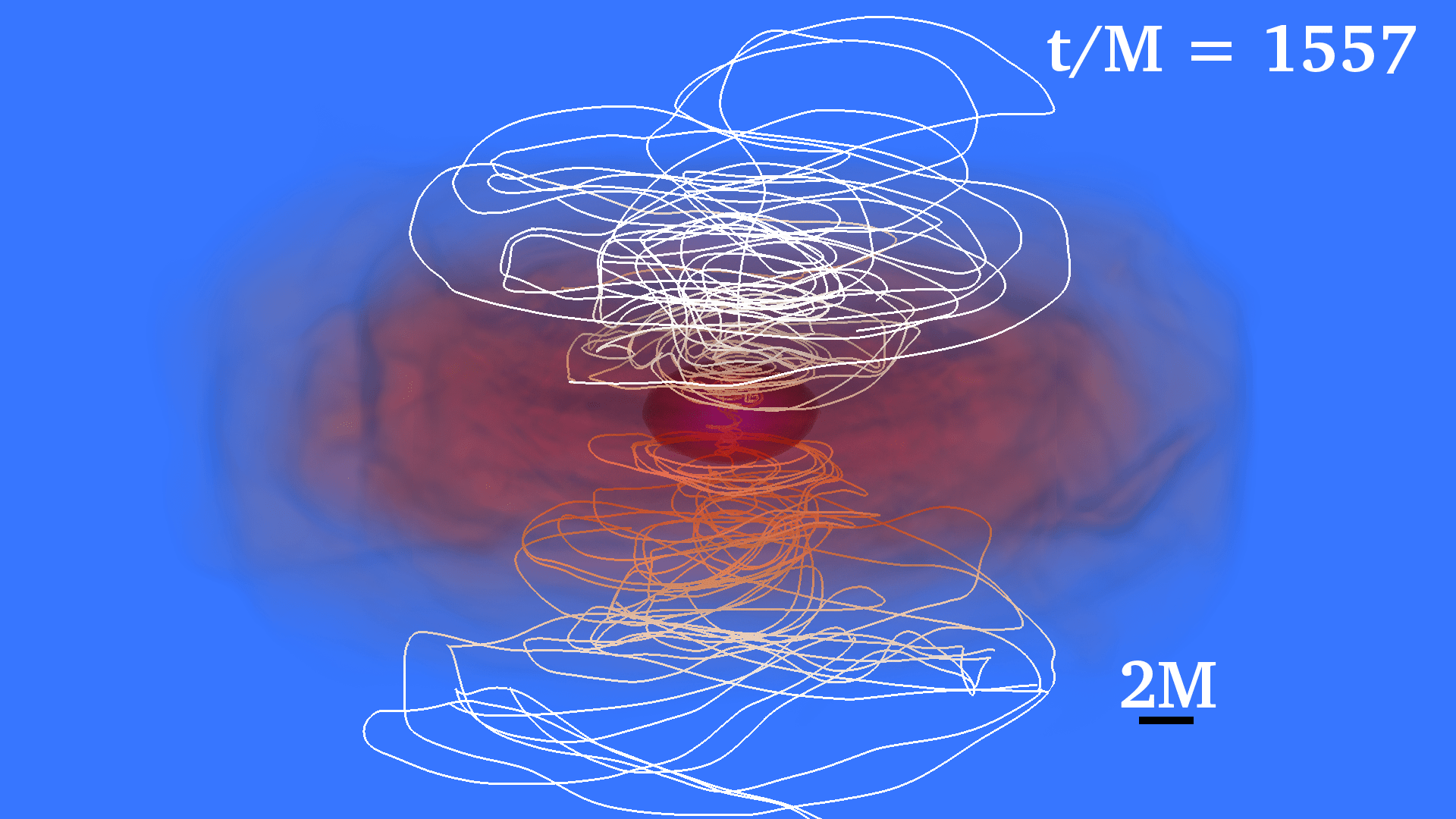}
  \includegraphics[width=0.44\textwidth]{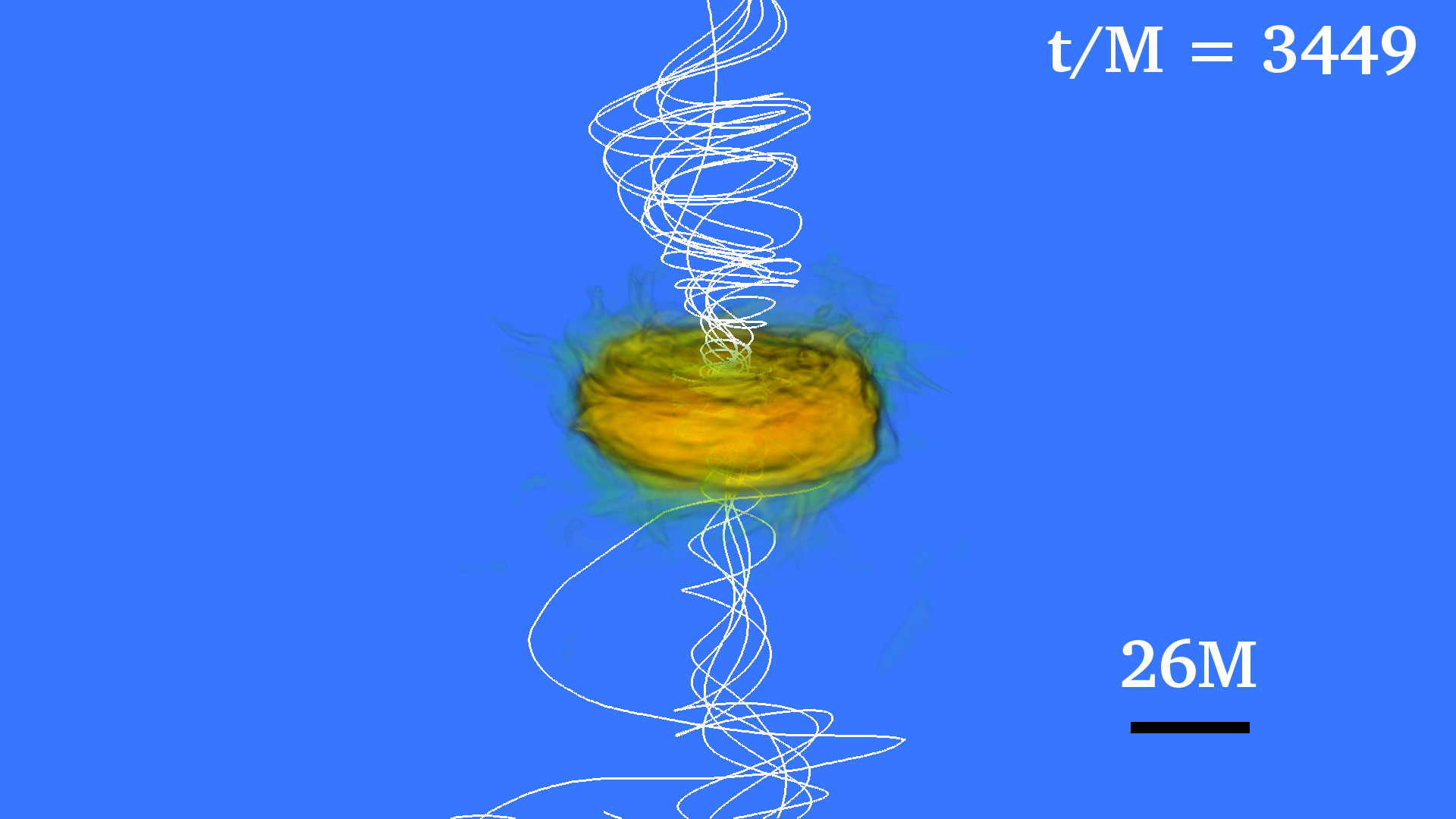}
  \includegraphics[width=0.44\textwidth]{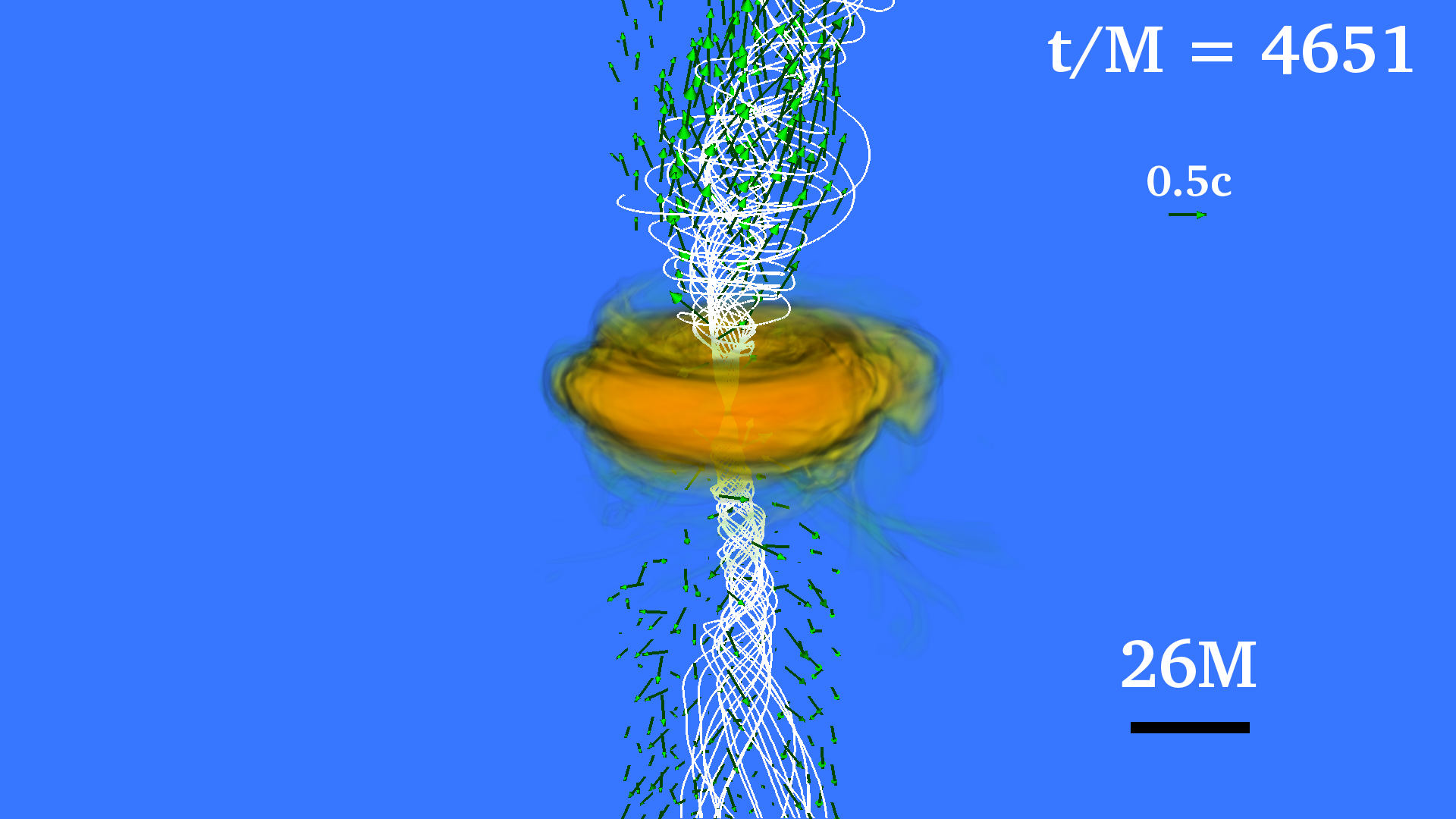}
  \caption{Same as Fig.~\ref{fig:Eq_vs_full} but for the Ali-Per case.
    \label{fig:per_ali}}
\end{figure}

In the perpendicular case, on the other hand, we keep the same
magnetic-to-gas-pressure ratio at the center of mass of the NS but
rotate counterclockwise the Cartesian components of the above vector
potential by $90^\circ$ (see top panel in Fig.~\ref{fig:per_ali} and
top left panel in Fig.~\ref{fig:per-per}).

Following~\cite{Ruiz:2018wah}, to mimic the ``force-free'' magnetosphere
surrounded the NS, and to reliably evolve the magnetic field outside
the star, we set a variable exterior, low-density magnetosphere such as
that $P_{\text{\tiny mag}}/ P_{\text{\tiny gas}}$ is $100$ everywhere. This density
increases the total rest-mass of the system by $\lesssim 0.5\%$~\cite{Ruiz:2016rai}.
For the subsequent evolution, as is typically done in standard hydrodynamics schemes,
we integrate the ideal GRMHD equations everywhere, imposing on top of the
magnetosphere a density floor in regions where $\rho_0^{\text{\tiny atm}}\leq
10^{-10} \rho_0^{\text{\tiny max}}$. Here $\rho_0^{\text{\tiny max}}$ is the initial
maximum rest-mass density of the system.

%%%%%%%%%%%%%%%%%%%%%%%%
%%%  Grid Structure  %%%
%%%%%%%%%%%%%%%%%%%%%%%%
\paragraph*{\bf{Grid structure}:}
In all simulations, we use seven refinement levels with two sets of nested
refinement boxes (one for each NS and centered in its center of mass),
differing in size and resolution by factors of two. The innermost refinement
level around each star has a side length of $\simeq 1.3\,R_{\NS}$, where
$R_{\NS}$ is the initial NS equatorial radius, and a grid spacing of $\sim 0.05
M = 0.227(M_{\NS}/1.625M_\odot)\rm km$. With this choice the initial NS
equatorial radius is  resolved by $\sim 66$ grid points, which matches the
resolution used in~\cite{Ruiz:2019ezy}. We also rerun the Per-Per case
at a resolution of $\sim 0.04 M = 0.177(M_{\NS}/1.625M_\odot)\rm km$.
Finally, the outer boundary is located at $267M\sim 1183(M_{\NS}/1.625M_{\odot})\rm km$.

%%%%%%%%%%%%%%%%%%%
%%% Diagnostics %%%
%%%%%%%%%%%%%%%%%%%
\paragraph*{{\bf Diagnostics}:}
\label{subsec:diagnostics}
%
%%%%%%%%%%%%%%%%%%%%%%%%%%
%%%  Ali-Perpendicular %%%
%%%%%%%%%%%%%%%%%%%%%%%%%%
%
%%%%%%%%%%%%%%%%%%%%%%
%%%  sanity check  %%%
%%%%%%%%%%%%%%%%%%%%%%
To analyze and check the reliability of the evolution of our binary systems,
we use  the following tools:

\begin{itemize}
\item {\it Global diagnostic checks}: To validate the numerical integration, {  {we monitor the
  $L_2$ norm of the normalized Hamiltonian
  and momentum constraints computed using~Eqs.~(40)-(41) in~\cite{Etienne:2007jg}. In all our
  cases (see Table~\ref{table:summary_key}) we find that during the inspiral and HMNS evolution
  phase, the constraints  remain below $\sim 0.02$}}. They peak at~$\lesssim 0.08$ during  BH
  formation and settle back to~$\lesssim 0.01$ after the BH + disk remnant reaches
  quasi-equilibrium. We also monitor the conservation of both the ADM mass~$M$~and the ADM
  angular $J$~computed using~Eqs.~(19)-(22) in~\cite{Etienne:2011ea}. By the end of the
  simulations we find that, in all configurations, the violation of the~$M_{\text{\tiny ADM}}$
  conservation is $\lesssim 1\%$, while the violation of the conservation of~$J_{\text{\tiny ADM}}$
  is $\sim 4\%$. Similar values were reported in our long-term, pure hydrodynamic simulations of
  spinning NSNS modeled by piecewise EOSs~\cite{Tsokaros:2019anx}.
  The above calculations take into account the GW radiation losses
  and the ejected material following merger; to measure the energy and angular momentum carried
  off by GWs, we use a modified version of the {\tt Psikadelia} thorn that computes $\Psi_4$
  \cite{Ruiz:2007yx} at different radii between $r_{\text{\tiny min}}\approx 30M\sim 133(M_{\NS}/
  1.625M_\odot)\rm km$ and $r_{\text{\tiny max}}\approx 170M\sim 752(M_{\NS}/1.625M_\odot)\rm km$.
  Around $\sim 0.8\%$ of the total energy, and $\sim 12\%$  of the angular momentum, is radiated
  away~(see~Table~\ref{table:summary_key}). The escaping mass, i.e. unbound fluid elements satisfying
  $-1 -u_t > 0$ with positive radial velocity, is computed as $M_{\text{\tiny esc}} = -\int d^3x\,
  \sqrt{\gamma}\,\alpha\,u^t\,\rho_0$ outside a coordinate radius $r > 30M\sim 133(M_{\NS}/1.625M_\odot)
  \rm km$. Here $\alpha$ is the lapse, $\gamma$ is the determinant of the 3-metric, $u^t$ is the
  time-component of the 4-velocity, and $\rho_0$ is the rest-mass density. Depending on the initial
  configuration of the magnetic field, between $10^{-4}\%$ and $0.16\%$ of the total rest-mass of
  the system is ejected.
  %
%%%%%%%%%%%%%%%%%%%%%
%%%  Per-Per case %%%
%%%%%%%%%%%%%%%%%%%%%
\begin{figure*}
  \centering
  \includegraphics[width=0.33\textwidth]{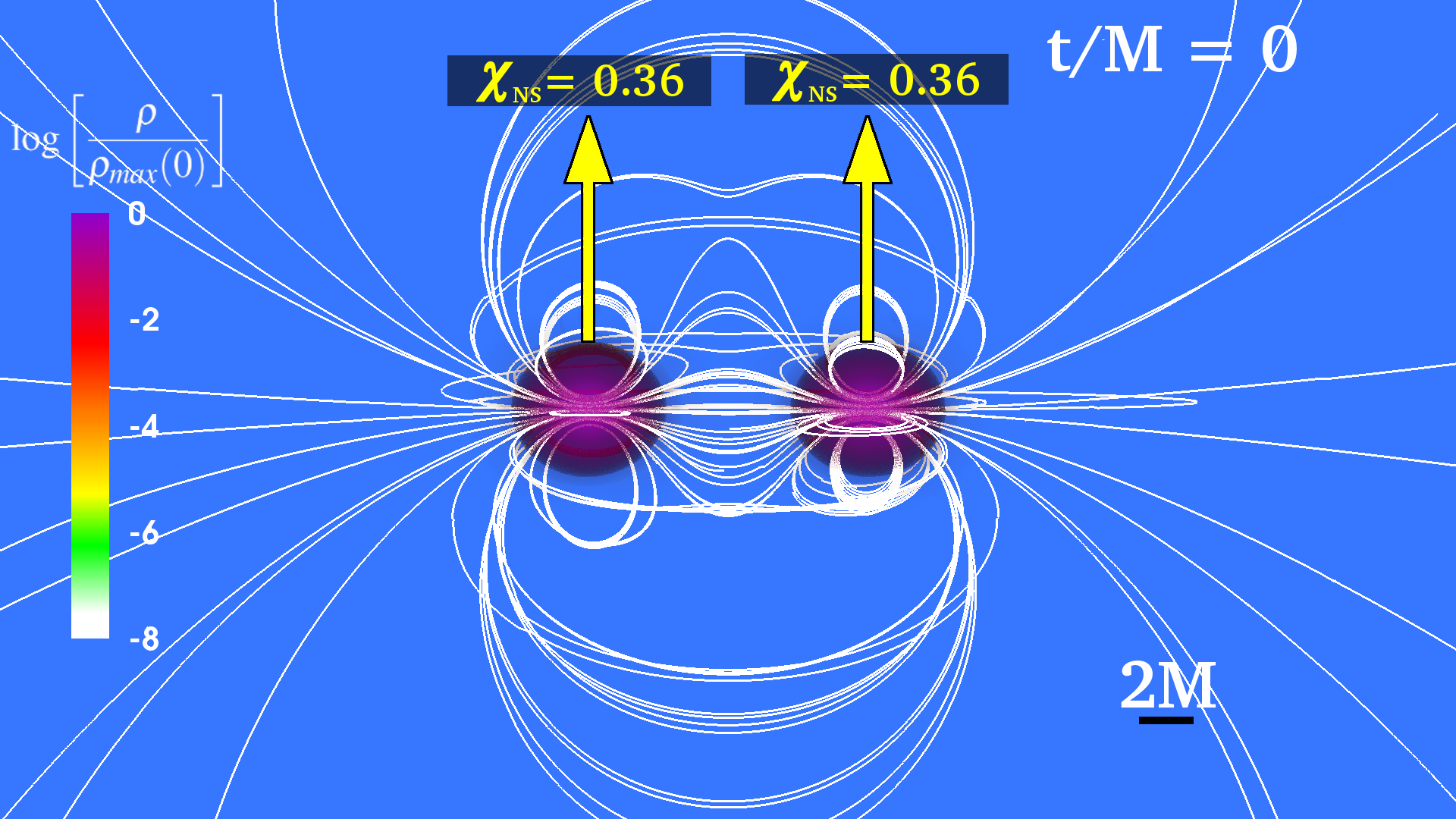}
  \includegraphics[width=0.33\textwidth]{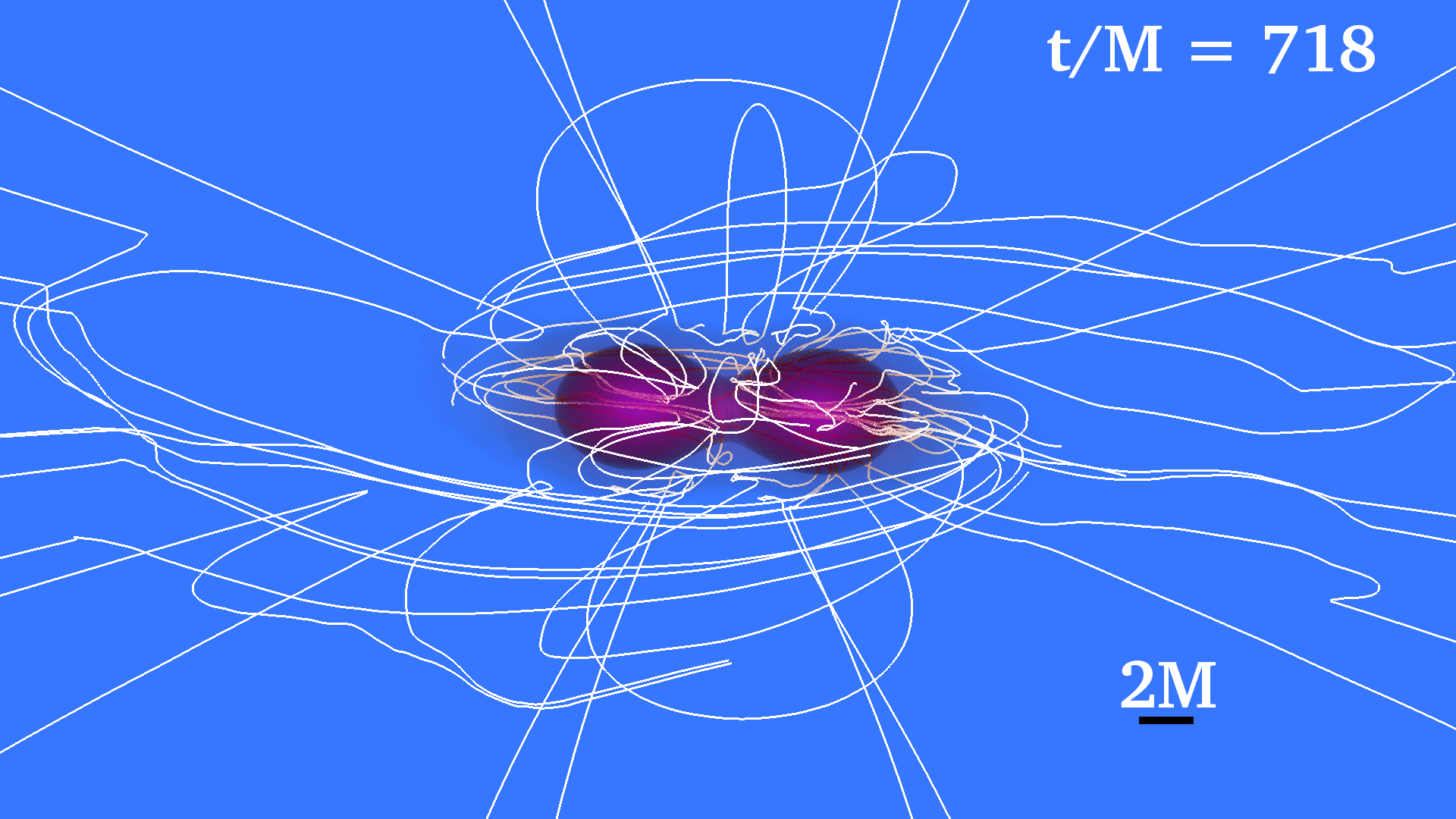}
  \includegraphics[width=0.33\textwidth]{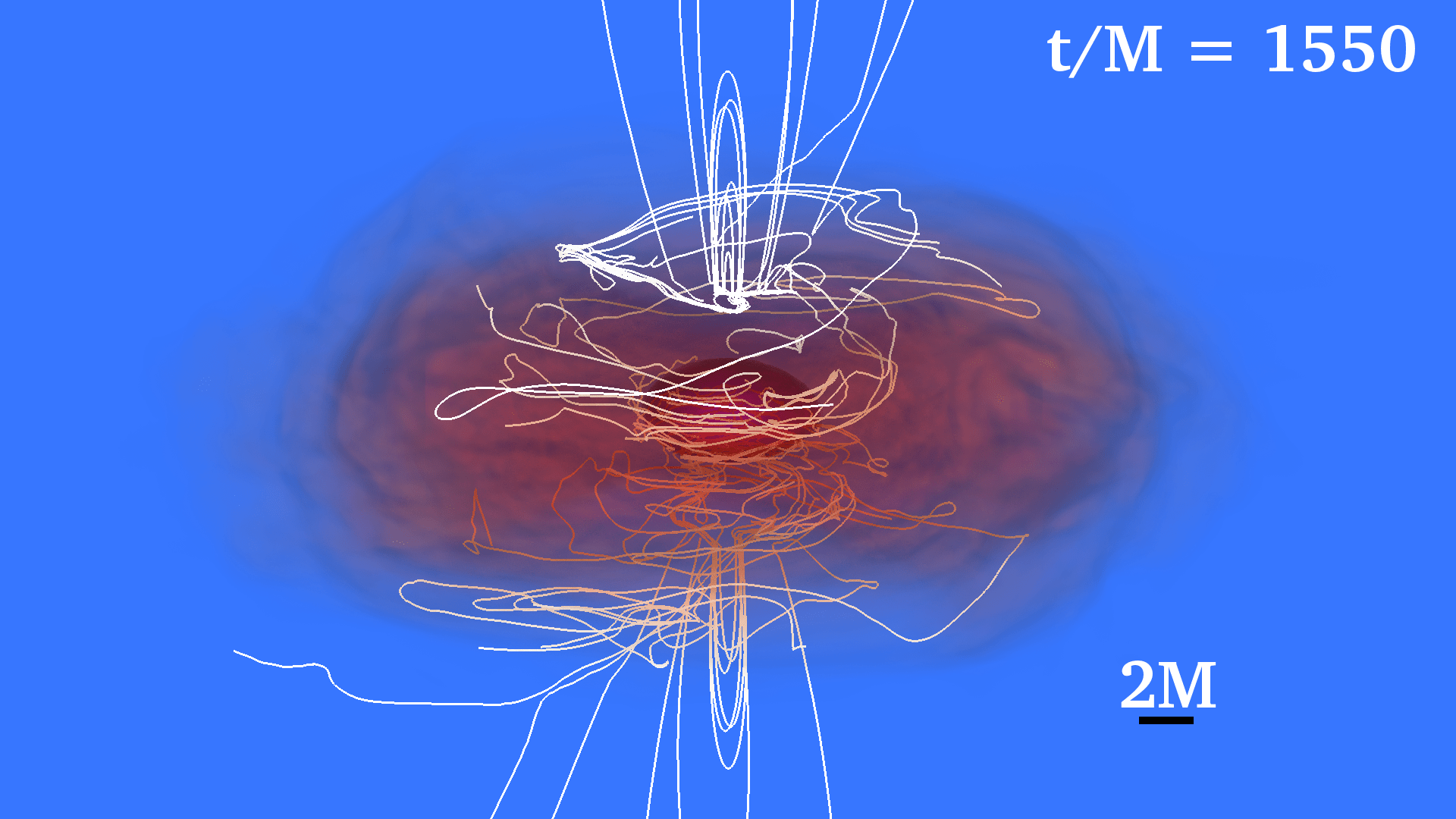}
  \includegraphics[width=0.33\textwidth]{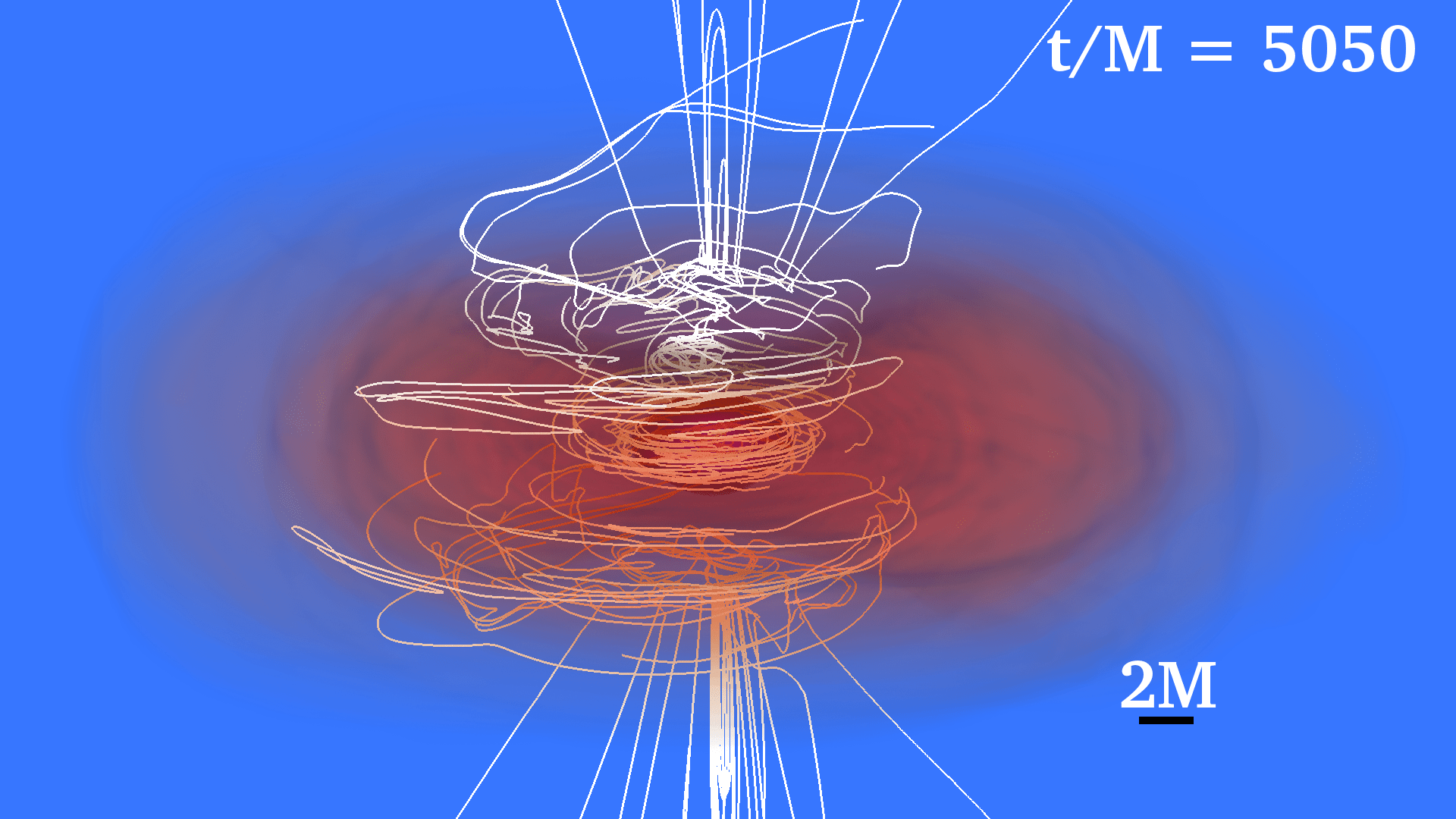}
  \includegraphics[width=0.33\textwidth]{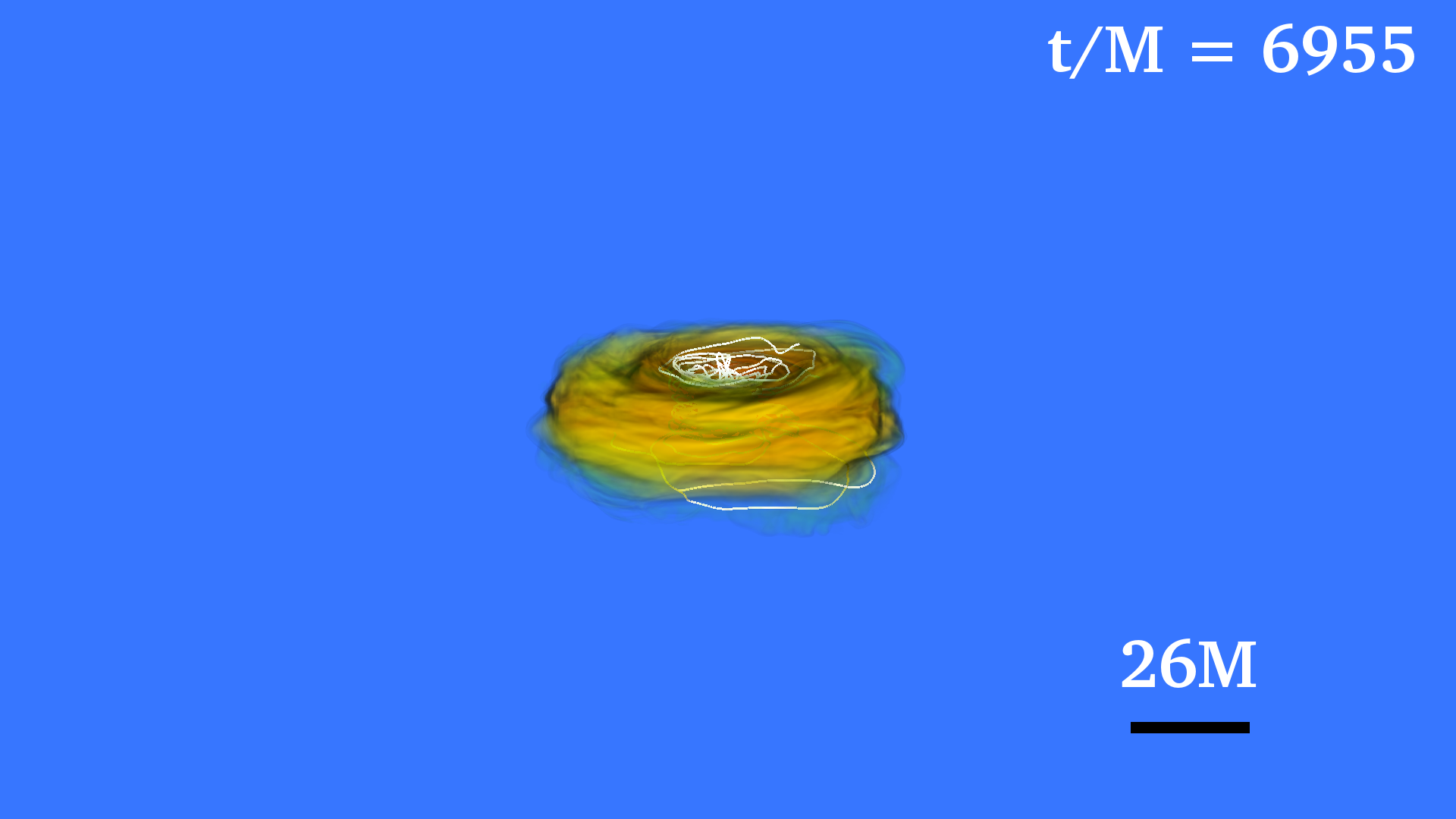}
  \includegraphics[width=0.33\textwidth]{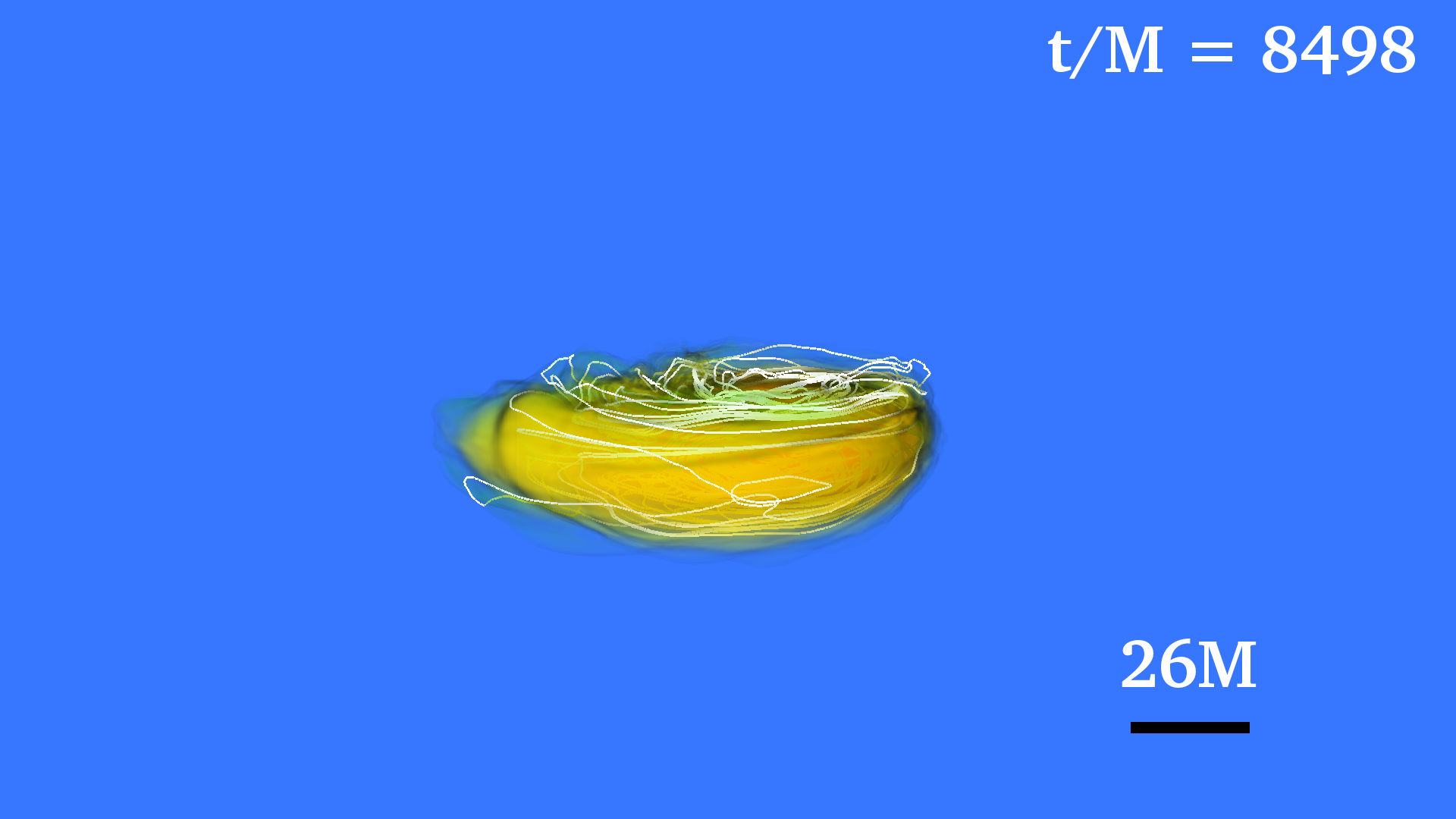}
  \caption{Volume rendering of rest-mass density $\rho_0$, normalized to the initial
    maximum value $\rho^{\rm max}_{0}=10^{14.78}(1.625\,M_\odot/
    M_{\NS})^2 \rm {g/cm}^3$ (log scale), at selected times for the Per-Per case.
    White lines represent the magnetic field lines. Bottom panels highlight the
    final configuration of the BH + disk remnant.
    \label{fig:per-per}}
\end{figure*}

%%%%%%%%%%%%%%%%%%%
%%% checks HMNS %%%
%%%%%%%%%%%%%%%%%%%

\item{\it Post-merger diagnostics}:
  To probe if magnetic instabilities are triggered during the formation and evolution of
  the transient HMNS, we monitor the growth of the magnetic energy~$\mathcal{M} =\int
  u^\mu u^\nu T^{(\text{\tiny{EM}})}_{\mu\nu}\,dV$ measured by a comoving observer. Once
  the HMNS has settled down, we compute the quality factor $Q_{\text{\tiny{MRI}}}\equiv
  \lambda_{\text{\tiny{MRI}}}/dx$ that measures the number of grid points per fastest-growing
  MRI mode. Here $\lambda_{\text{\tiny{MRI}}}$ is the fastest-growing MRI wavelength. A
  $Q_{\text{\tiny{MRI}}}\gtrsim 10$ is required to properly capture the MRI
  \cite{Shiokawa:2011ih,Sano:2003bf}. MHD turbulence is also diagnosed via the effective
  Shakura--Sunyaev parameter $\alpha_{\text{\tiny SS}}={T^{EM}_{\hat{r}\hat{\phi}}}/{P}$,
  where $T^{EM}_{\hat{r}\hat{\phi}}$ is the  $\hat{r}-\hat{\phi}$ component of the EM
  stress energy tensor computed using the orthornormal contravariant tetrad
  system $e^{\hat{i}}_l$ in the local fluid-frame~\cite{FASTEST_GROWING_MRI_WAVELENGTH}.
  Following~\cite{Gold:2013zma}, we report an azimuthally- and $z$-  averaged
  $\alpha_{\text{\tiny SS}}=\alpha_{\text{\tiny SS}}(r)$
  profile.  
  Finally, to measure the degree of differential rotation of the HMNS, we  monitor its
  azimuthally-averaged angular velocity $\Omega(t,r)$ using Eq.~2~in~\cite{Ruiz:2019ezy}.
 %  
%%%%%%%%%%%%%%%%%%%%%%%%%
%%% checks BH + disk  %%%
%%%%%%%%%%%%%%%%%%%%%%%%%
\item{\it Post-collapse diagnostics}:
  We~adopt~the {\tt AH\-Find\-er\-Di\-rect} thorn~\cite{ahfinderdirect} to locate
  and monitor the apparent horizon (AH), and the isolated horizon formalism~\cite{dkss03}
  to measure the mass of the BH $M_{\text{\tiny BH}}$ and its dimensionless spin parameter
  $a/M_{\text{\tiny BH}}$. Following BH formation, the outgoing EM Poynting
  luminosity is computed via $L=-\int T^{r(\text{\tiny{EM}})}_t\,\sqrt{-g}\,d\mathcal{S}$
  across different spherical surfaces of coordinate radii between $r_{\text{\tiny{ext}}}=46
  M\simeq 204(M_{\NS}/1.625M_\odot)\rm km$ and $r_{\text{\tiny{ext}}}=190 M \simeq 842(M_{\NS}/
  1.625M_\odot)\rm km$. To assess whether the magnetic field above the BH remnant poles is
  sufficiently strong to launch a jet we compute the force-free parameter $b^2/(2\rho_0)$,
  where $b^2=b^\mu b_\mu$, with $b^\mu=B^\mu_{(u)}/\sqrt{4\pi}$ the magnetic field
  measured by an observer co-moving with the fluid. When it exceeds $\sim 10-100$
  a jet is typically launched via the BZ mechanism~\cite{prs15}. 
  Finally, we compute the  rest-mass accretion rate $\dot{M}$ via Eq. A11
  in~\cite{Farris:2009mt}. 
\end{itemize}
%
%%%%%%%%%%%%%%%%
%%% Results  %%%
%%%%%%%%%%%%%%%%
%
\section{Results}
\label{sec:results}
\begin{figure}
  \centering
    \includegraphics[width=0.49\textwidth]{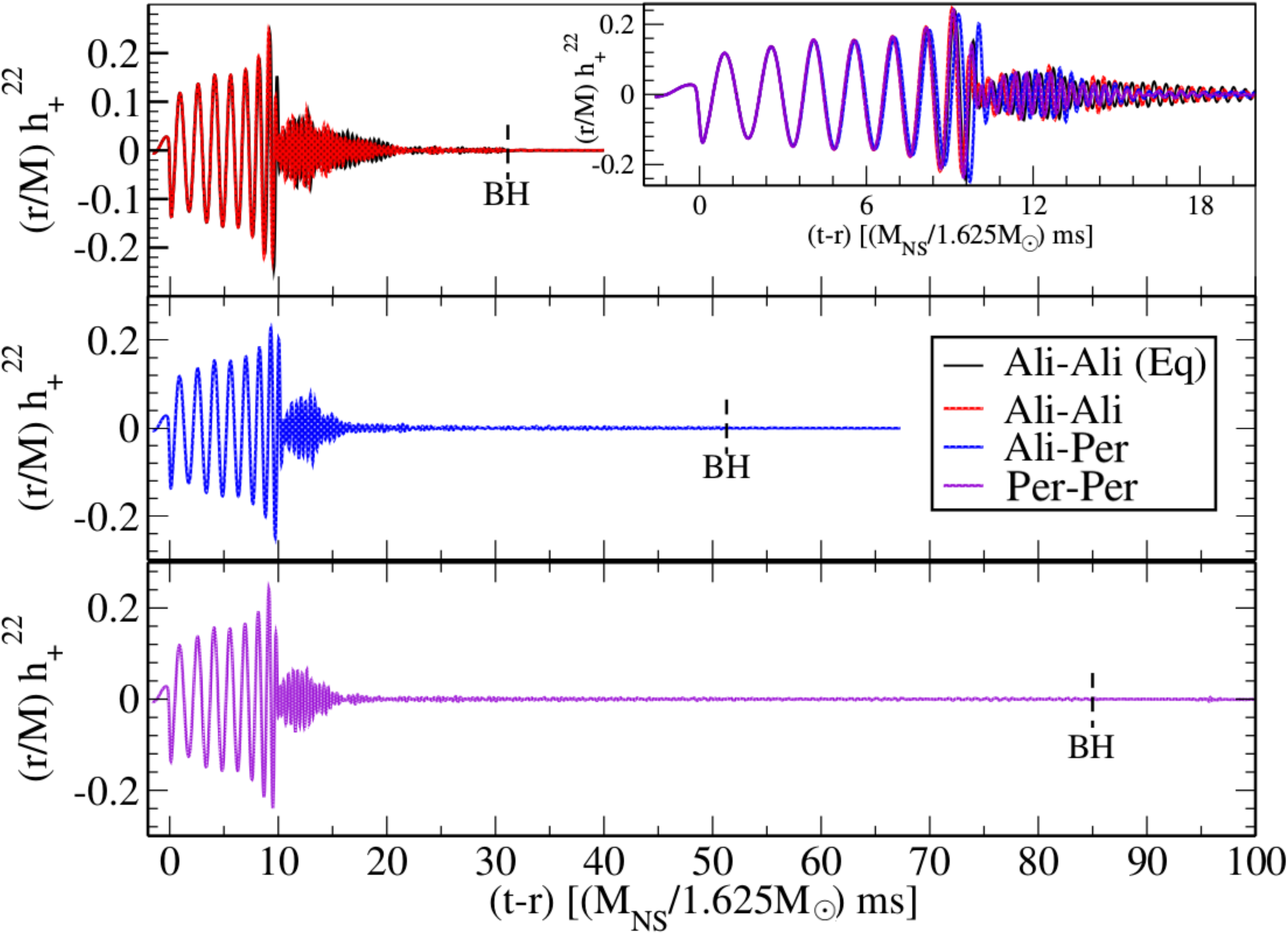}
    \caption{GW amplitude $h_+^{22}$ (dominant mode) as function of the retarded time, extracted
      at $r_{\rm ext}\approx 100M\sim 443(M_{\NS}/1.625M_\odot)\rm km$ for all cases listed in
      Table~\ref{table:summary_key}. The vertical dashed line denotes the coordinate time at
      which the BH horizon appears for the first time. The inset highlights the wavetrain
      during the inspiral, merger and HMNS ringdown.
    \label{fig:GW_allcase}} 
\end{figure}

\begin{figure}
  \centering
  \includegraphics[width=0.49\textwidth]{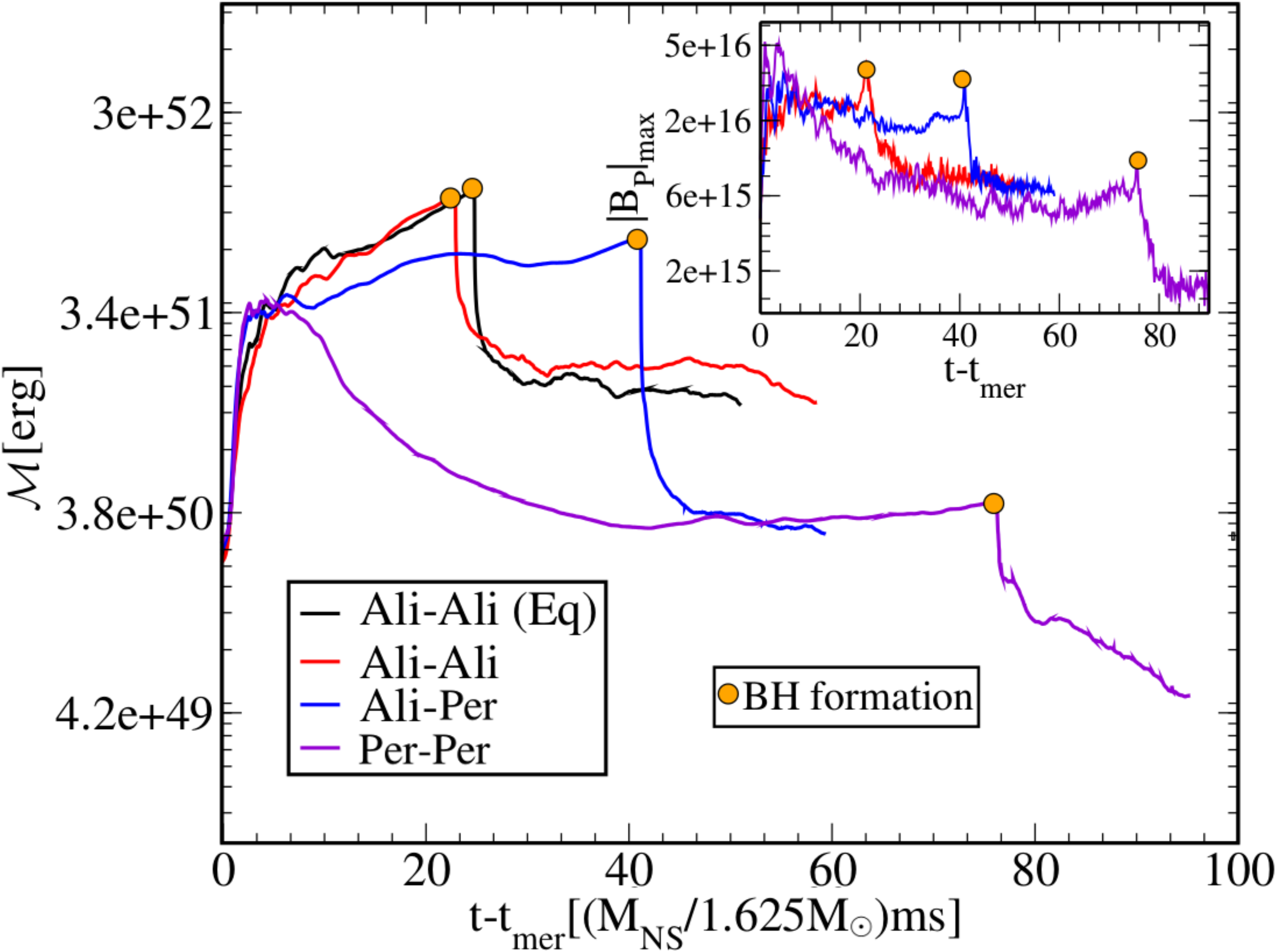}
  \caption{Evolution of the magnetic energy $\mathcal{M}$ for cases listed in
    Table~\ref{table:summary_key}. The coordinate time since merger is plotted.
    Dots mark the time at which the apparent horizon appears
    for the first time ($\Delta t_{\text{\tiny BH}}$).  The inset displays the evolution of the
    maximum value of the poloidal magnetic field component. Similar behavior is found
    in the toroidal component.
    \label{fig:EM_energy}}
\end{figure}

The basic dynamics of our NSNS configuration has been described in~\cite{Ruiz:2019ezy}.
As the GWs carry off energy and angular momentum, the orbital separation decreases. After
roughly $\sim 3.5$ orbits (or~$7$ GW cycles, see inset in Fig.~\ref{fig:GW_allcase})
the stars merge, forming a transient remnant with two massive central cores rotating about
each other. After $\gtrsim 300M\sim 4.4(M_{\NS}/1.625M_\odot)\rm ms$, these cores
collide and give birth to a magnetized and highly differentially rotating HMNS surrounded
by a Keplearian-like cloud of low density matter. As 
shown in the inset of Fig.~\ref{fig:EM_energy}, during the HMNS formation the strength of
the  poloidal magnetic field component peaks at $\gtrsim 10^{16} (1.625M_\odot/M_{\NS})
\rm G$, consistent with the values reported in the high-resolution studies in
\cite{Kiuchi:2015sga}. As pointed out in~\cite{Ruiz:2016rai}, the toroidal magnetic
field component is also amplified until it approximately equals the magnitude of the
poloidal one.

Effective turbulence induced by magnetic fields leads to the transport of angular momentum
from the rapidly rotating inner layers of the HMNS to the slowly rotating outer part~(see~Fig.
\ref{fig:Omega_AlivsPer}). It causes the central part of the HMNS to contract, forming a
nearly uniformly rotating, massive central core that contains roughly $\sim 88\%$ of the total
rest-mass of the system. As rigid rotation does not  provide enough centrifugal support
to the central core to prevent collapse (i.e. its mass exceeds $M_0\simeq 2.4(M_{\NS}/1.625
M_\odot)M_\odot$, the maximum mass allowed by a uniformly rotating $\Gamma=2$ star
\cite{Cook:1993qj,LBS2003ApJ}), it eventually collapses, forming a highly spinning BH
surrounded by a Keplerian accretion disk. As we show in the following section, only if the
NSNS has initially a large-scale strong poloidal component aligned to the 
orbital angular momentum of the system, does the BH + disk remnant eventually
launch a large-scale, magnetically-sustained outflow. The summary of the key results
is presented in~Table~\ref{table:summary_key}.
%
%%%%%%%%%%%%%%%%%%%%%
%%% GWs full + EQ %%%
%%%%%%%%%%%%%%%%%%%%%
%
%%%%%%%%%%%%%%%%%%%%%%%%%%
%%%  Magnetic Energy   %%%
%%%%%%%%%%%%%%%%%%%%%%%%%%
%
%%%%%%%%%%%%%%%%%%%%%%%%%%%%%%%
%%%  Equatorial vs 3D cases %%%
%%%%%%%%%%%%%%%%%%%%%%%%%%%%%%%
%
\subsection{Both aligned magnetic fields: equatorial symmetry vs. full 3D evolutions}
\label{sec:Alig_cases}
Fig.~\ref{fig:Eq_vs_full} shows a side-by-side comparison of the evolution of the
Ali-Ali configuration  in both equatorial symmetry (left column) and
full 3D (right column) at selected times. Colors depict the rest-mass density
$\rho_0$, normalized to its initial maximum  value $\rho^\text{max}_{0}= 
10^{14.78}(1.625\,M_\odot/M_{\NS})^2 \rm {g/cm}^3$, white lines shows the magnetic field
lines, while arrows indicate  plasma velocities.

As shown in the top panel of Fig.~\ref{fig:GW_allcase}, symmetries do not have a significant
effect on the dynamics of stars during the inspiral.
During this epoch, the binary companions orbit each other, dragging the frozen-in magnetic
field with them. Gravitational radiation extracts energy and angular momentum and drives
the system to an unstable orbit. The stars come into contact with another for the first
time at $t\gtrsim 580M\sim 8.5(M_{\NS}/1.625M_\odot)\rm ms$ (see Table \ref{table:summary_key}).
We observer that the  configuration in full 3D merges around $13M\sim 0.2(M_{\NS}/1.625M_\odot)
\rm ms$ earlier than in the equatorial case. Here, the merger time $t_{
  \text{\tiny mer}}$ is defined as the time of the peak amplitude of GWs~(see Fig.
\ref{fig:GW_allcase}). As shown in the second row of Fig.~\ref{fig:Eq_vs_full},
during  merger the stars become oblate with spiral arms that wrap around the nascent
remnant, forming a cloud of low-density matter. The central regions of the stars begin
to orbit around each other, and eventually, collide to form a magnetized HMNS. Due to
the KHI, along with the MRI, the magnetic energy
$\mathcal{M}$ is steeply enhanced~\cite{Kiuchi:2014hja,Kiuchi:2015sga}. Fig.
\ref{fig:EM_energy} shows that during the first $t-\tmer\simeq 200M\sim 3(M_{\NS}/1.625
M_\odot)\rm ms$, $\mathcal{M}$ is amplified by a factor of
$\sim 15$. Afterwards, and roughly during the next $t-\tbh \simeq 1700M\sim 25(M_{\NS}/
1.625M_\odot)\rm ms$ up to the BH formation, $\mathcal{M}$  grows by a factor of
$\lesssim 5$.  We notice that the magnetic field amplification saturates once the poloidal and
the toroidal components reach a magnitude of $\sim 10^{16}(1.625M_\odot/M_{\NS})\rm G$ 
(see inset of Fig.~\ref{fig:EM_energy}).

\begin{figure*}
  \centering
  \includegraphics[width=0.49\textwidth]{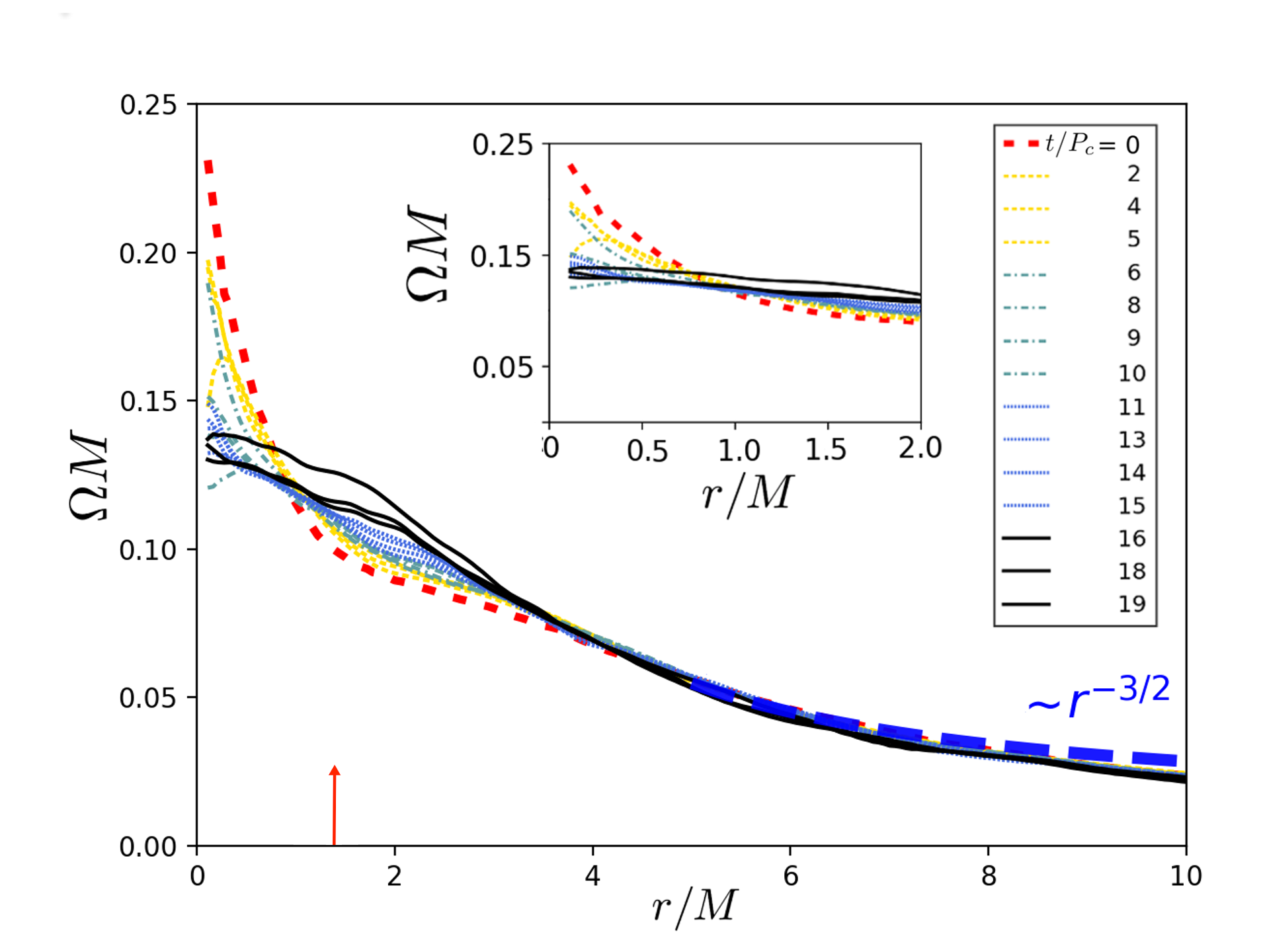}
  \includegraphics[width=0.49\textwidth]{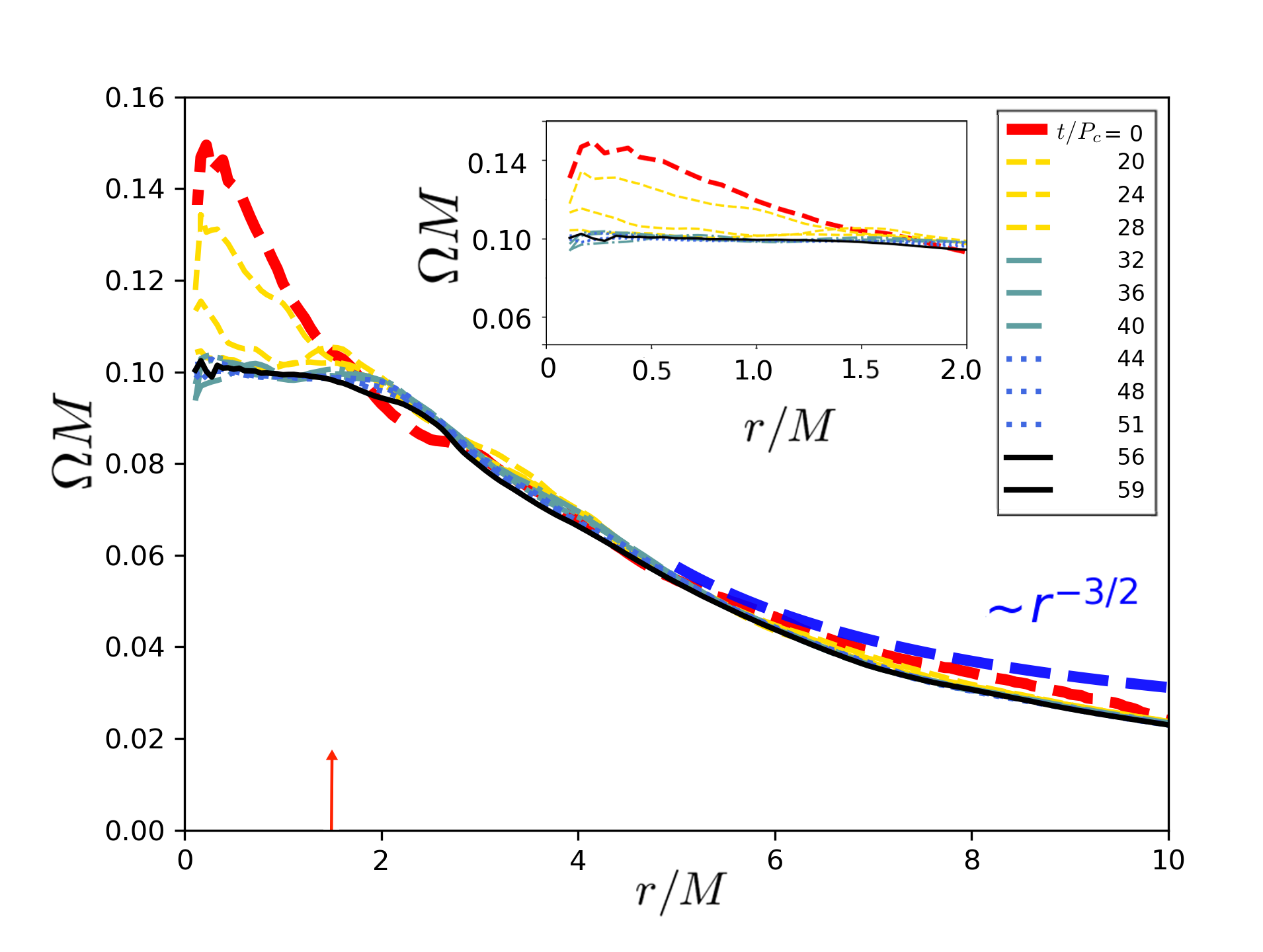}  
  \caption{
    Azimuthally-averaged rotation profile of the transient HMNS for the Ali-Ali case (left panel)
    and the Per-Per case (right panel) in the orbital plane at $\Delta t = t-t_{\text{\tiny HMNS}}$ (see
    Table~\ref{table:summary_key}). Here $t_{\text{\tiny HMNS}}$ is the HMNS formation time, and
    $P_c\simeq 50M\sim 0.7 (M_{\rm NS}/1.625M_\odot)\rm ms$ is the central HMNS period at $t=
    t_{\text{\tiny HMNS}}$. The red dashed line shows the initial differential rotation profile,
    while the continuous black curve display the final profile. The blue dashed curve exhibits a
    Keplerian angular velocity profile. The arrow marks the coordinate radius containing
    $\sim 50\%$ of the total rest-mass of the system. The inset highlights the rotation profile
    of the central core.~\label{fig:Omega_AlivsPer}
    { {Note that this rotation profile agrees with~\cite{Shibata:2002jb1} who adopts the
        same EOS, but
    in general it depends on the EOS (see e.g.~\cite{Shibata:2006nm1,Kastaun:2016yaf1}).}}
  }
\end{figure*}

Once the HMNS has settled down, we probe whether magnetic instabilities have been triggered.
We first compute the quality factor $Q_{\text{\tiny MRI}}$ and find that $\lambda_{\text{
    \tiny MRI}}$ of the fastest-growing MRI is resolved by $\gtrsim 10$ grid points and
it fits within the star. We conclude that MRI is resolved and operating in our system, as
expected from our previous simulations in~\cite{Ruiz:2019ezy} (see Fig.~9~in there).
Moreover, we find that for $t-\tmer\simeq 1360 M\sim 20(M_{\NS}/1.6M_\odot)\rm ms$,
the effective Shakura--Sunyaev viscosity  $\left<\alpha_{\text{\tiny SS}}
\right>_{P_c}$ parameter is $\sim 0.05$~(see Table~\ref{table:summary_key}). Here brackets
denote an average over one stellar rotation period~$P_c$ (see Fig.~\ref{fig:Omega_AlivsPer}).
Similar values
were reported in high-resolution simulations of strongly massive NS remnants
in~\cite{Kiuchi:2017zzg}. Thus, magnetic turbulence is likely to be fully developed
in the transient HMNS~\cite{Kiuchi:2017zzg}. However, magnetic turbulence can
be suppressed by numerical diffusion~\cite{Kiuchi:2017zzg,Hawley:2011ApJ,Hawley:2013lga} and,
therefore, the value of $\alpha_{\rm SS}$ in our simulations may be underestimated. As pointed
out in~\cite{Kiuchi:2017zzg},  higher resolutions are required to properly model magnetic
turbulence (see also~\cite{Hawley:2011ApJ,Hawley:2013lga} for a detailed discussion).

Differential rotation and turbulent magnetic viscosity
triggers magnetic braking~\cite{Sun:2018gcl} which causes the formation of a massive and
highly oblate central core. The core is immersed in a cloud of matter originating
from the expansion of the external layers of the HMNS (see the third row of
Fig.~\ref{fig:Eq_vs_full}). The flow in the central core drives the poloidal field
lines to a toroidal configuration (magnetic winding). The toroidal magnetic field is
then amplified until its magnitude equals the strength of the poloidal component. We
notice that by~$t-\tmer \simeq 950M\sim 14(M_{\NS}/1.625 M_\odot)\rm ms$,
the shape of the HMNSs in the two Ali-Ali cases (equatorial symmetry and full 3D case)
is basically the same, though the magnetic field, in the full 3D evolution, has been
wound into a larger-scale helical structure (see the third row of Fig.~\ref{fig:Eq_vs_full}).
%
%%%%%%%%%%%%%%%%%%%%%%%%%
%%% Omega Ali vs Per  %%%
%%%%%%%%%%%%%%%%%%%%%%%%%

By~$t-\tmer \simeq 1700M\sim 25(M_{\NS}/1.625M_\odot)\rm ms$~(see Table
\ref{table:summary_key}), the central core has approached rigid rotation (see left
panel in Fig.~\ref{fig:Omega_AlivsPer}), and collapses to a BH. Note that this timescale
is consistent with magnetic braking induced by magnetic winding~(the Alfv\'en timescale),
which~can be estimated~as~(see Eq.~2 in~\cite{Sun:2018gcl}):
\begin{eqnarray}
  \tau_{\text{\tiny A}}& \sim &\frac{R_{\text{\tiny HMNS}}}{v_{\text{\tiny A}}}\sim
  \\&&10{\rm ms}\,
  \left(\frac{\rho}{10^{14}\rm g/cm^3}\right)^{1/2}\,
  \left(\frac{|B|}{10^{15}G}\right)^{-1}\,
  \left(\frac{R_{\text{\tiny HMNS}}}{10^6\rm cm}\right)\,,
  \nonumber
\label{eq:t_a}
\end{eqnarray}
where $R_{\text{\tiny HMNS}}$ is the characteristic radius of the HMNS remnant and $v_{\text{\tiny A}}\sim
|B|/\sqrt{4\pi\rho}$ the Alfv\'en speed, with $|B|$ the strength of the magnetic field and $\rho$
the characteristic density of the remnant. Note that turbulent magnetic viscosity can also redistribute
angular momentum and damp the differential rotation on a viscous timescale
given by~(see Eq.~7 in~\cite{Sun:2018gcl})
\begin{eqnarray}
  \tau_{\rm vis} &\sim&{R_{\text{\tiny HMNS}}^{3/2}}\,{M^{-1/2}_{\text{\tiny HMNS}}
    \,\alpha_{\rm SS}^{-1}}\sim\\
  &&10{\rm ms}\,
  \left(\frac{\mathcal{C}}{0.3}\right)^{-3/2} \,
 \left(\frac{M_{\text{\tiny HMNS}}}{3.2M_\odot}\right)\,  
  \left(\frac{\alpha_{\rm ss}}{10^{-2}}\right)^{-1}
 \,,\nonumber
   \label{eq:tau_win}
\end{eqnarray}
where $M_{\text{\tiny HMNS}}$ is the characteristic mass of the HMNS and $\mathcal{C}=
M_{\text{\tiny HMNS}}/R_{{\text{\tiny HMNS}}}$ its compactness.

\begin{figure}
  \centering
  \includegraphics[width=0.50\textwidth]{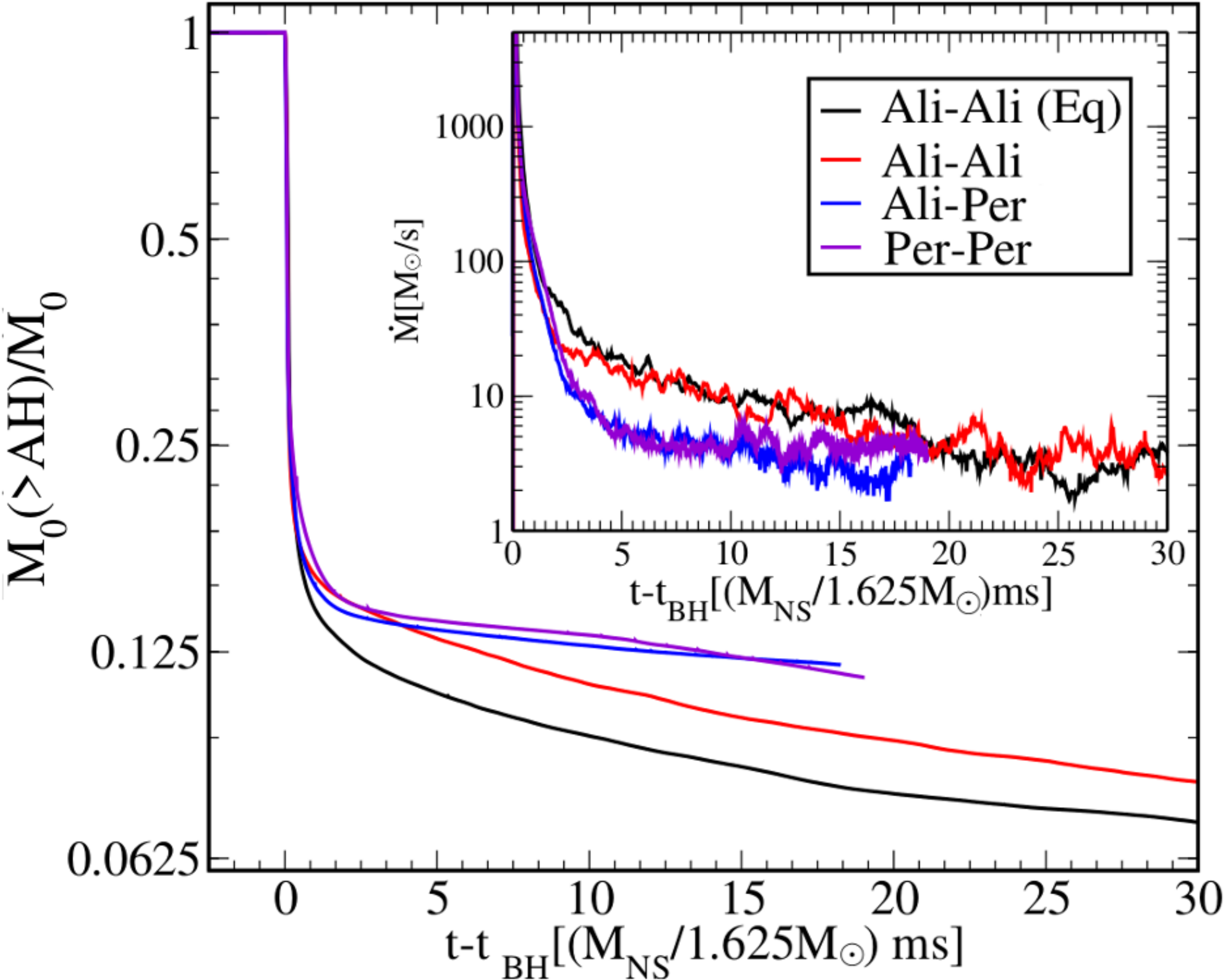}
  \caption{Evolution of the rest-mass fraction outside the apparent horizon $M_0(>{\rm AH})$
    normalized to its initial maximum value $M_0= 3.25M_{\odot}(k/k_{\text{\tiny L}})^{1/2}$
    for all cases listed Table~\ref{table:summary_key}. The inset shows the accretion
    rate history. The coordinate time since BH formation is plotted.
    \label{fig:M0_outside}}
\end{figure}

\begin{figure}
  \centering
  \includegraphics[width=0.41\textwidth]{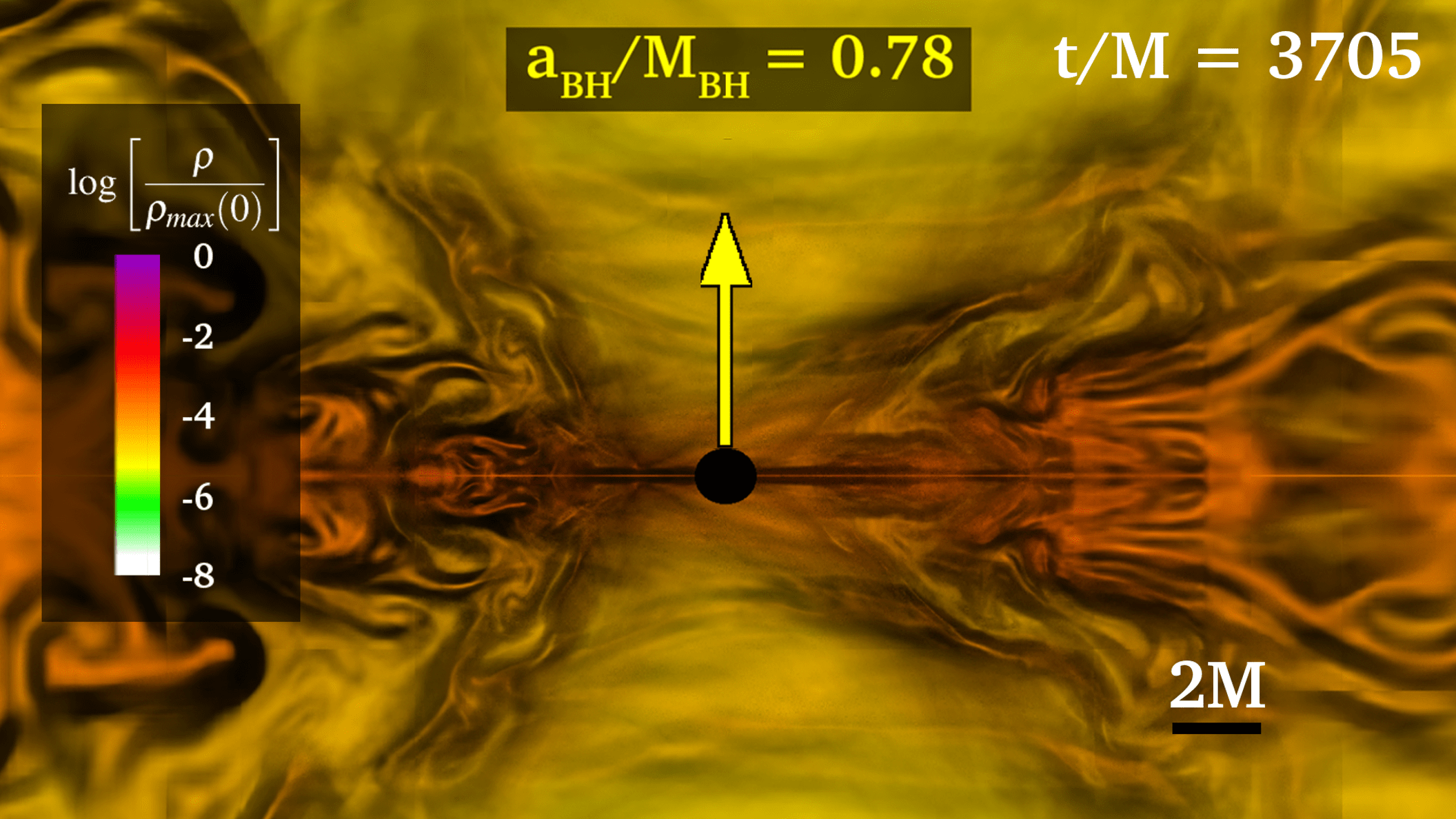}
  \includegraphics[width=0.41\textwidth]{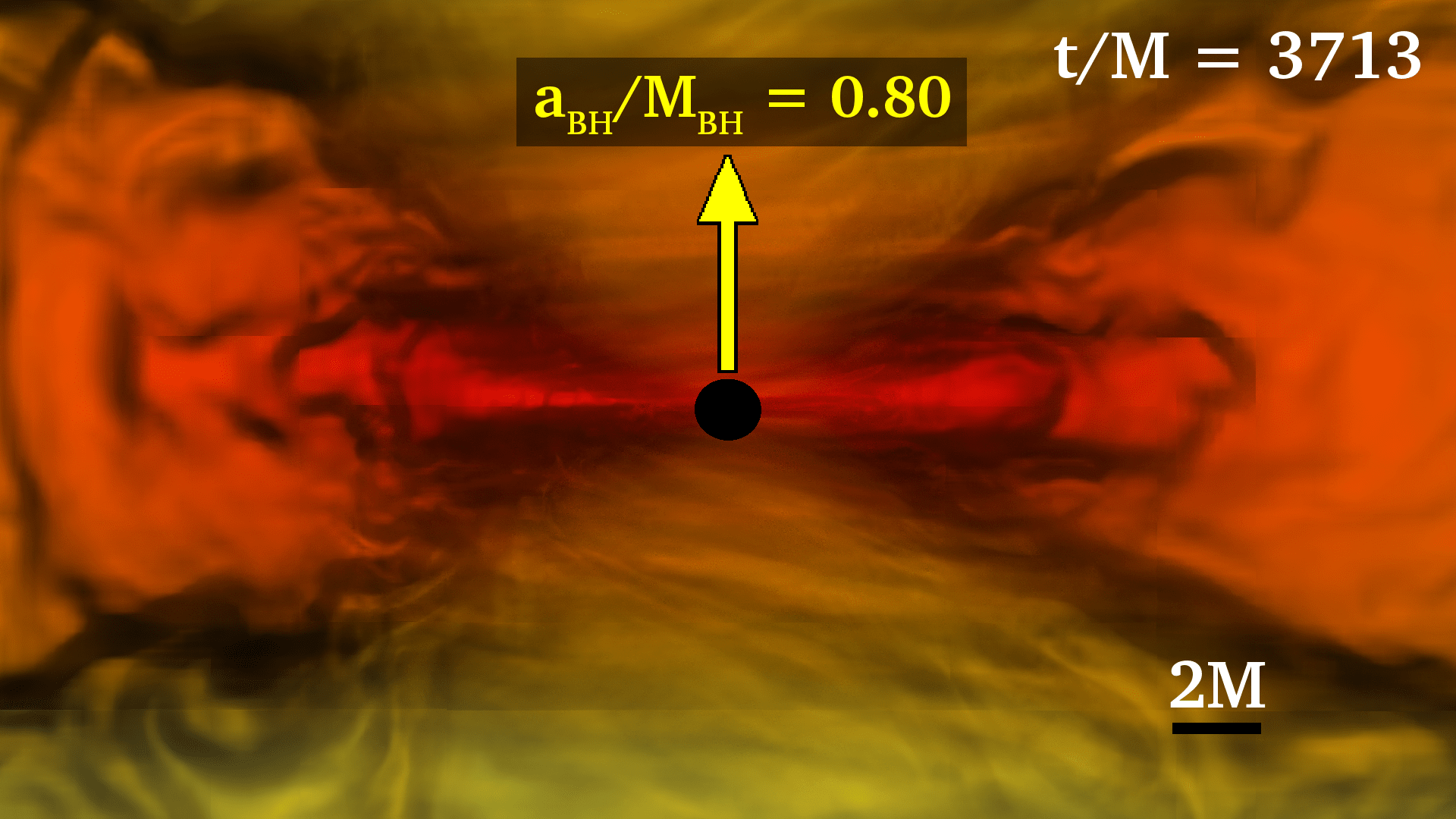}
  \includegraphics[width=0.41\textwidth]{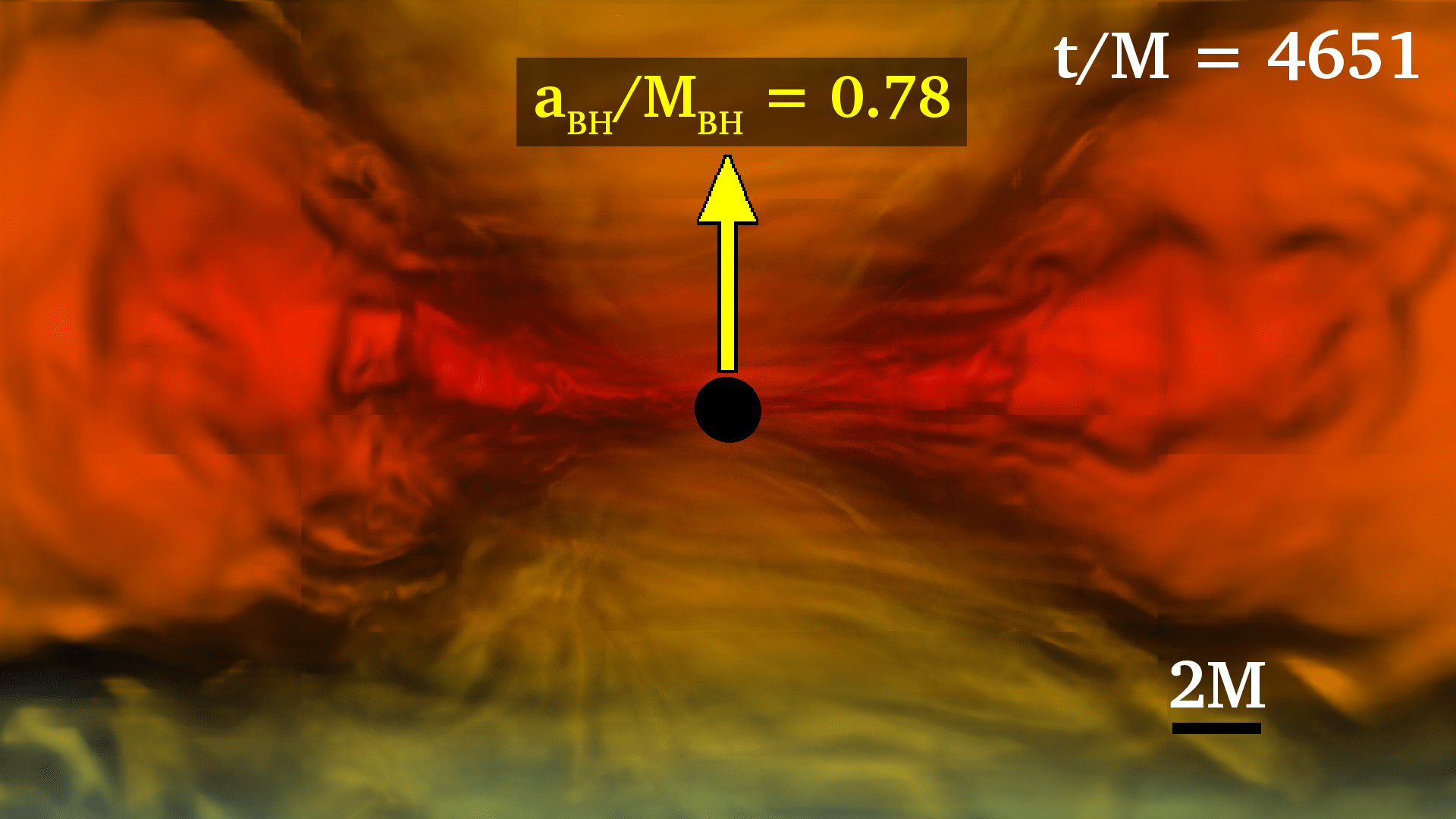}
  \includegraphics[width=0.41\textwidth]{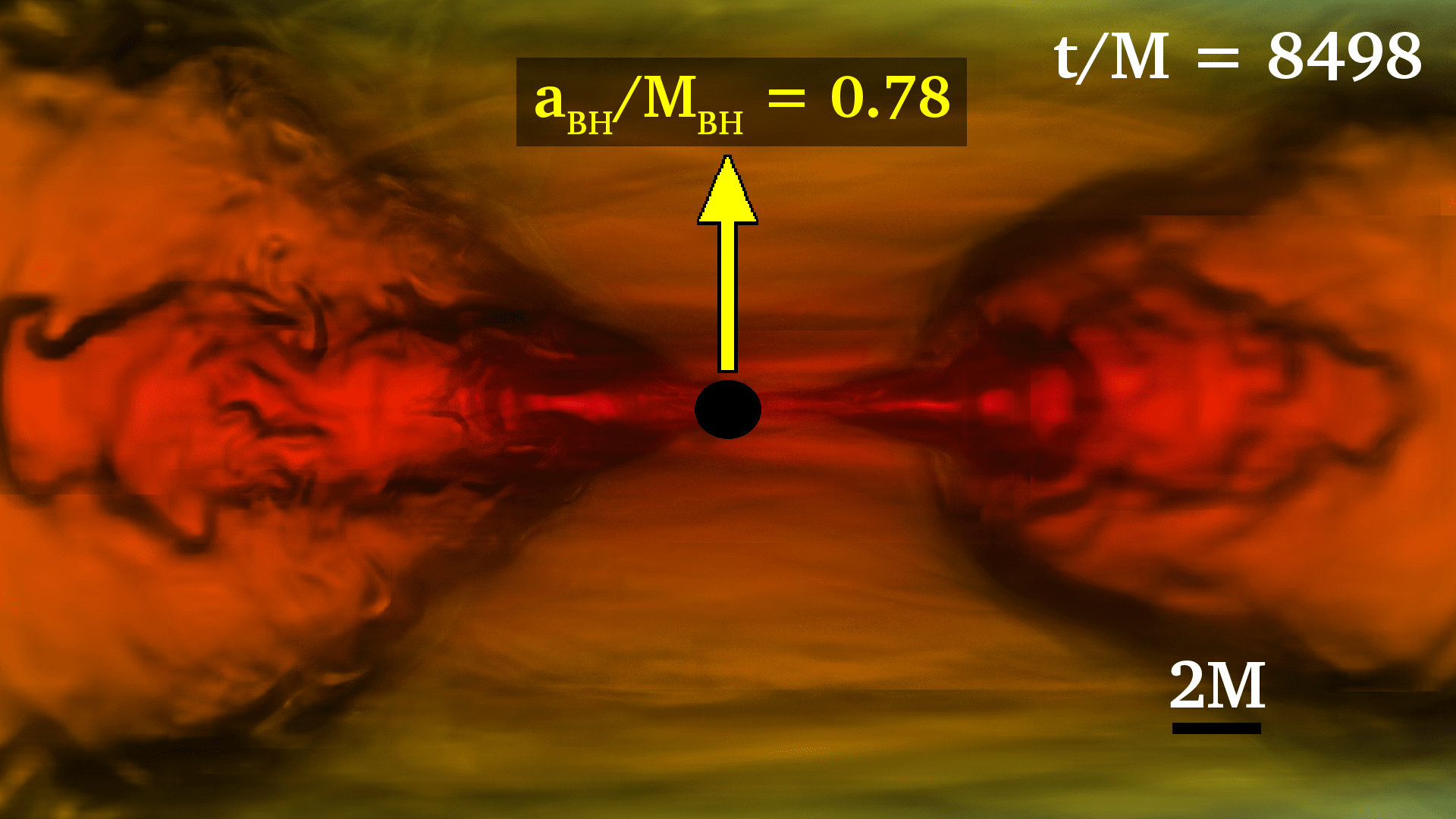}
  \caption{Meridional cut of the 3D density profile of the BH + disk remnant near
    the end of the simulations for all cases listed in~Table~\ref{table:summary_key}.
    From top to bottom the cases are Ali-Ali (Eq), Ali-Ali, Ali-Per
    and Per-Per, respectively. The BH horizon is shown as a black sphere, while
    the yellow arrow indicates the spin direction. Here $M=1.47\times 10^{-2}(M_{\NS}/
    1.625M_\odot)\rm ms$= $4.4288(M_{\NS}/1.625M_\odot)\rm km$.
    \label{fig:bh+disk_remnant}}
\end{figure}

In the two cases, the BH remnant has a mass of $M_{\text{\tiny BH}}\simeq 2.75M_\odot$
with spin parameter equal to either $a/M_{\text{\tiny BH}} \simeq 0.78$, in Ali-Ali (Eq),
or  $a/M_{\text{\tiny BH}}\simeq 0.80$, in the full 3D case. Note that a difference 
of $\sim 2.5\%$ is approximately at the same level as the accuracy of our numerical
simulations. As we state in the previous section, the violation of the conservation of
$J_{\rm ADM}$ is $\sim 4\%$, though the spins are determined by a local measurement
while $J_{\rm ADM}$ is determined by a global measurement.
Fig.~\ref{fig:M0_outside}
compares their rest-mass profiles following the accretion peaks (see also the fourth row
in Fig.~\ref{fig:Eq_vs_full}). We observe that although the
fractions of the total rest-mass outside the BH horizon are roughly the same~(see Table
\ref{table:summary_key}), the accretion disk in the full 3D case is a factor of~$\sim 2.5$
denser than in Ali-Ali (Eq). Fig.~\ref{fig:bh+disk_remnant} shows the
BH + disk remnant near the end of the simulations. Notice that in Ali-Ali, matter tends
to pile up around the orbital plane and closer to the BH, while in Ali-Ali (Eq), it is
concentrated further out in the bulk of the disk.

During the HMNS collapse, the inner core, which contains most of the magnetic energy
$\mathcal{M}$, is promptly swallowed by the BH, and hence $\mathcal{M}$ in the exterior
quickly decreases
by roughly one order of magnitude in only $\Delta t\simeq 68{\rm M} \sim 1(M_{\NS}/1.625
M_\odot)\rm ms$, and then it slightly decreases as the accretion proceeds reaching a
value of $10^{-3.7}M=10^{51.0}(M_{\NS}/1.625M_\odot)\rm erg$ (see~Fig.~\ref{fig:EM_energy}).
As shown in the inset of Fig.~\ref{fig:EM_energy},  following  BH formation,
the rms value of the poloidal magnetic field is $\simeq 10^{16} \rm G$, and remains
roughly constant. Similar behavior is observed in the toroidal component.

Following the HMNS collapse, the BH remnant is immersed in a heavy-loaded environment:
material ejected during the merger or radially blown out during the HMNS epoch begins
to rain down. However,  magnetic winding begins to build up magnetic pressure above
the BH poles until eventually it is large enough to balance the ram pressure of the
fall-back material. By~$t-\tbh \sim 1000M\sim15(M_{\NS}/1.625M_\odot)\rm ms$, the
magnetic pressure above the BH poles stops the inflow.
Simultaneously, field lines get wound into a helical structure. We observe that,
as displayed in third and fourth rows in~Fig.~\ref{fig:Eq_vs_full}, the field lines in
the Ali-Ali case are already tightly wound in regions above the BH that extend to heights~$\lesssim
10M\simeq 14\,r_{\text{\tiny BH}}$, { {where $r_{\text{\tiny BH}}$ is the
    coordinate radius of the apparent
horizon of the BH}},  while in the Ali-Ali (Eq) case the magnetic winding is still
underway. As the accretion proceeds, the atmosphere becomes thinner and hence
magnetically-dominated regions (force-free regions~where~$b^2/(2\rho_0)\gtrsim 1$) above
the BH progressively expand (see top panels in~Fig.~\ref{fig:b2rho_all3d}). Once $b^2/(2\rho_0)
\gtrsim 10$, the magnetic pressure above the BH poles is enough to overcome the ram pressure,
and a magnetically-sustained outflow is launched. By $t-\tmer\sim 1900M\sim 28(M_{\NS}/
1.625M_\odot) \rm ms$, unbound material with a Lorentz factor of
$\Gamma_L \gtrsim 1.26$ (see Table~\ref{table:summary_key}) expands beyond heights
$\gtrsim 100M \sim 430(M_{\NS}/1.625 M_\odot)~\rm km$ above the BH.  Therefore, we
conclude that at about $t-t_{\text{\tiny mer}}\sim3000M\sim 45(M_{\NS}/1.625 M_\odot)\rm ms$
following the gravitational radiation peak, an incipient jet is launched (see bottom panels in
Fig.~\ref{fig:Eq_vs_full}). { We note that, in the full 3D evolution there is an ``apparent'' delay
in the emergence  of the jet, which is launched only $\sim 60M\sim 0.9 (M_{\NS}/1.625M_\odot){\rm ms}$
earlier than in the equatorial case, though its funnel walls were formed $200M\sim 3 (M_{\NS}/1.625M_\odot){\rm ms}$
earlier}. This delay is likely due to its denser accretion
disk that requires longer evolution time for the emptying of the funnel. Nevertheless, field
lines are more tightly collimated  in the full 3D case (see top panels in Fig.~\ref{fig:b2rho_all3d}).
We estimate a funnel opening angle $\theta_{\rm jet}$ of~$\sim 20^\circ$ in Ali-Ali, and a
$\theta_{\rm jet}\sim 25^\circ$ in the equatorial case~(see also~\cite{Ruiz:2019ezy}).

%%%%%%%%%%%%%%%%%%%%%%%%%%%%%
%%%   Summary of results  %%%
%%%%%%%%%%%%%%%%%%%%%%%%%%%%%
%
\begin{table*}[]
  \begin{center}
    \caption{{\bf Summary of key results}. Here $t_{\text{\tiny mer}}$, $\Delta t_{\text{\tiny BH}}$,
      $t_{\text{\tiny evo}}$ are the coordinate time in units of~$(M_{\NS}/1.625M_\odot)\rm ms$ at which the
      binary merges, the apparent horizon appears (time measured after merger), and the full evolution time,
      respectively. $M_{\text{\tiny BH}}$ denotes the mass of the remnant BH in units of $M_{\odot}$ and
      $\tilde{a}\equiv a/M_{\text{\tiny BH}}$ its dimensionless spin parameter. $M_{\text{\tiny disk}}/M_0$
      is the accretion disk at $t_{\text{\tiny evo}}$, normalized to the initial total rest-mass of the system
      $M_0$, $\dot{M}$ is the rest-mass accretion rate once it has reached a quasi-stationary state in units of
      $(M_\odot/s)$, $\tau_{\text{\tiny disk}}\sim M_{\text{\tiny disk}}/\dot{M}$ is the disk lifetime in units
      of $(M_{\NS}/1.625M_\odot)\rm s$, $M_{\text{\tiny esc}}$ is the escaping rest-mass at $t_{\text{\tiny evo}}$.
      The fraction of the total energy and the fraction  of total angular momentum carried away by GWs are denoted
      by $\Delta \bar{E}_{\text{\tiny GW}}\equiv\Delta E_{\text{\tiny GW}}/M_{\text{\tiny ADM}}$ and $\Delta
      \bar{J}_{\text{\tiny GW}}\equiv\Delta J_{\text{\tiny GW}}/J_{\text{\tiny ADM}}$, respectively. $\left<
      \alpha_{\text{\tiny SS}}\right>_{P_c}$ denotes the Shakura--Sunyaev viscosity parameter averaged over one
      stellar rotation period once the HMNS has settled down, $B_{\text{\tiny
            rms}}$ is the rms value of the magnetic field above the BH poles in units of $(1.625M_\odot/M_{\NS})\rm G$. The
        Poynting luminosity (in units of $\rm erg/s$) driven by the incipient jet and its efficiency, time-averaged over the
        last $500M\sim 7.4 (M_{\NS}/1.625M_\odot)\rm ms$ of the evolution, are denoted by $L_{\text{\tiny EM}}$ and
        $\eta_{\text{\tiny EM}}\equiv L_{\text{\tiny EM}}/\dot{M}$. Finally, $\Gamma_{\text{\tiny L}}$ is the maximum fluid
        Lorentz factor at $t_{\text{\tiny evo}}$. If there is no corresponding value (absence of a jet), we write [N/A]
        for that case.
        \label{table:summary_key}}
        \begin{tabular}{ccccccccccccccccc}
        \hline\hline
        Model & $t_{\text{\tiny mer}}$ & $t_{\text{\tiny BH}}$ & $t_{\text{\tiny evo}}$ & $M_{\text{\tiny BH}}$ &  $\tilde{a}$   & $M_{\text{\tiny disk}}/{M_0}$ &
        $\dot{M}$  &  $\tau_{\text{\tiny disk}}$ & $M_{\text{\tiny esc}}/{M_0}$ & $\Delta \bar{E}_{\text{\tiny GW}}$ & $\Delta\bar{J}_{\text{\tiny GW}}$  &
        $\left<\alpha_{\text{\tiny SS}}\right>_{P_c}$&$B_{\text{\tiny rms}}$ & $L_{\text{\tiny EM}}$ & $\eta_{\text{\tiny EM}}$& $\Gamma_L$ \\
          \hline
          Ali-Ali (Eq)$^*$&11.2 & 25.0& 62  &$2.75$ & 0.78  & $7.82\%$     & 2.71    & 138.5 & $0.14\%$     & $0.76\%$  & $11.55\%$   & 0.04  &  $10^{15.8}$  & $10^{52.1}$  &   $0.3\%$  & 1.26 \\
          \hline
          \hline 
          Ali-Ali     &11.1 & 23.7& 62  &$2.75$ & 0.80  & $8.58\%$     & 2.86    & 143.9 & $0.16\%$     & $0.74\%$  & $11.33\%$  & 0.05  &  $10^{15.8}$  & $10^{52.3}$  &   $0.3\%$  & 1.29 \\
          Ali-Per     &11.7 & 41.9& 70  &$2.73$ & 0.78  & $11.43\%$    & 2.47    & 150.4 & $0.01\%$     & $0.70\%$  & $11.35\%$  & 0.02  &  $10^{15.7}$  & $10^{51.8}$  &   $0.2\%$  & 1.24 \\
          Per-Per     &11.2 & 76.5& 104 &$2.73$ & 0.78  & $11.37\%$    & 2.56    & 144.3 &$10^{-4}\%$  & $0.67\%$  & $11.08\%$   & 0.001 &  $10^{14.8}$  &  [N/A]       &   [N/A]     & [N/A] \\
          \hline
          \hline
    \end{tabular}
  \end{center}
  \begin{flushleft}
   $^{*}$ Case treated previously in~\cite{Ruiz:2019ezy}, and denoted as Msp0.36. 
  \end{flushleft}
\end{table*} 
%
%%%%%%%%%%%%%%%%%%%%
%%% M0 outside  %%%
%%%%%%%%%%%%%%%%%%%%
The incipient jet leads to an outgoing EM Poynting luminosity of $L_{\text{\tiny EM}}
\simeq 10^{52}\rm erg/s$ (see~Fig.~\ref{fig:Poynt_ejecta}) that is consistent with
typical sGRB luminosities~\cite{Ajello:2019zki}, and has an efficiency of~$\eta_{\text{
    \tiny EM}}\equiv L_{\text{\tiny EM}}/\dot{M}\sim 0.3\%$
(see Table~\ref{table:summary_key}). Similar values were reported in GRMHD
simulations of BH immersed in a magnetized accretion disk with similar spins (see e.g.~Eq.~3
in~\cite{McKinney:2005zw}).  Near the end of our simulations, the ratio $b^2/(2\,\rho_0)$,
which equals the maximum achievable Lorentz factor for Poynting-dominated jets
\cite{Vlahakis2003}, reaches values larger than $10^{2.5}$ above the BH poles (see~top
panel in Fig.~\ref{fig:b2rho_all3d}). Therefore, as it has been pointed out in~\cite{prs15},
the mildly relativistic magnetically-driven outflow found in our simulations may be
accelerated to Lorentz factors $\Gamma_L\gtrsim 10^2$, values required in sGRB
phenomenology.  Also, notice that the EM luminosity is also roughly consistent with
the theoretical expectation from the BZ mechanism~(see Eq.~5~in~\cite{Ruiz:2019ezy}).
as well as with the ``universal model'' common to all BH + disk systems formed through
the merger or collapse of compact objects~\cite{Shapiro:2017cny}.

As displayed in the inset of Fig.~\ref{fig:M0_outside}, by $t-\tbh\simeq 850 M\sim
12.5(M_{\NS}/1.625 M_\odot)\rm ms$, $\dot{M}$ begins to settle
down, and  gradually
decays afterwards. Once the magnetically-driven outflow is launched, we estimate that the
lifetime of the disk (fuel of the jet) in both case is $\tau_{\text{\tiny disk}}\sim
M_{\text{\tiny disk}}/\dot{M}$~$\simeq 140\rm ms$ (see~Table~\ref{table:summary_key}),
which is entirely consistent with the lifetime of sGRBs~\cite{Kann:2008zg}.
%
%%%%%%%%%%%%%%%%%%%%%%%%%%%%%
%%%  BH + disk eq vs full %%%
%%%%%%%%%%%%%%%%%%%%%%%%%%%%%
In addition to the EM outgoing luminosity, NSNS mergers can also give rise to detectable
kilonovae if the mass ejecta following the merger is larger than~$\sim 10^{-3}M_\odot$
(see~e.g.~\cite{Metzger:2011bv,Metzger:2016pju}). The inset in Fig.~\ref{fig:Poynt_ejecta}
shows the history of the rest-mass fraction of the escaping matter following the NSNS merger.
The  material ejected in Ali-Ali (Eq) is $M_{\text{\tiny ext}}\sim 10^{-2.34}M_\odot$, while
in Ali-Ali is $M_{\text{\tiny ext}}\sim 10^{-2.30}M_\odot$. So, Ali-Ali cases may lead
to kilonovae that can be potentially observed by current or planned telescopes
\cite{Metzger:2011bv}. 

The above results show that in a timescale of $\Delta t_{\text{\tiny evo}}\simeq 4200M\sim
62(M_{\NS}/1.625M_\odot){\rm ms}$, {\it there are no significant differences between 3D
  evolutions and those in which symmetry across the orbital plane is imposed}. In both
cases, the BH + disk remnant launches a magnetically-driven jet.
We do not find any evidence of additional magnetic instabilities that can be triggered
in full 3D evolutions (e.g.~poloidal field instability)~as have been previously suggested
in~\cite{Braithwaite:2005ps,Ciolfi:2012en,Ciolfi:2011xa}.
Notice that the resistive GRMHD studies reported in~\cite{Palenzuela:2013kra} that focus on
EM counterparts during the late inspiral and merger epoch of NSNSs, do not observe such 
instabilities either during an evolution time of~$\Delta t_{\text{\tiny evo}}\simeq 6\rm ms$,
which exceeds the instability growth (Alfv\'en time) of $3\rm ms$.
% 
%%%%%%%%%%%%%%%%%%%%%%%%%%%%%%%%
%%%  Aligned - Perpendicular %%%
%%%%%%%%%%%%%%%%%%%%%%%%%%%%%%%%
%
\subsection{One aligned and one perpendicular magnetic field}
\label{sec:Ali-Per_cases}
Fig.~\ref{fig:per_ali} summarizes the evolution of the rest-mass density along
with the magnetic field lines, and fluid velocities, of the  Ali-Per case. As in
the completely aligned cases, the magnetic field does not play a significant role
during the inspiral epoch (see inset in Fig~\ref{fig:GW_allcase}). The field lines
threading the bulk of each star are simply advected. However, we note that, as the
stars approach each other, the field lines connecting them are stretched and wound
and,  as the coordinate separation decreases, a strong toroidal magnetic field
component joining the bulk of the stars emerges.

As displayed in the second panel of Fig.~\ref{fig:per_ali}, { {the stars touch at
each other for the first time at $t\simeq
600M\sim 8.8(M_{\NS}/1.625M_\odot)\rm ms$, and merger roughly~$37M\sim 0.54(M_{\NS}/1.625 M_\odot)\rm
ms$ later than in the Ali-Ali case}}~(see Table~\ref{table:summary_key}), forming a double
central core that, after~$t-\tmer\simeq 360M\sim 5.3(M_{\NS}/1.625M_\odot)\rm ms$,
merges and forms a transient HMNS immersed in a low-density cloud of matter.
%
%%%%%%%%%%%%%%%%
%%% b2/2rho  %%%
%%%%%%%%%%%%%%%%%
\begin{figure*}
  \centering
  \includegraphics[width=0.49\textwidth]{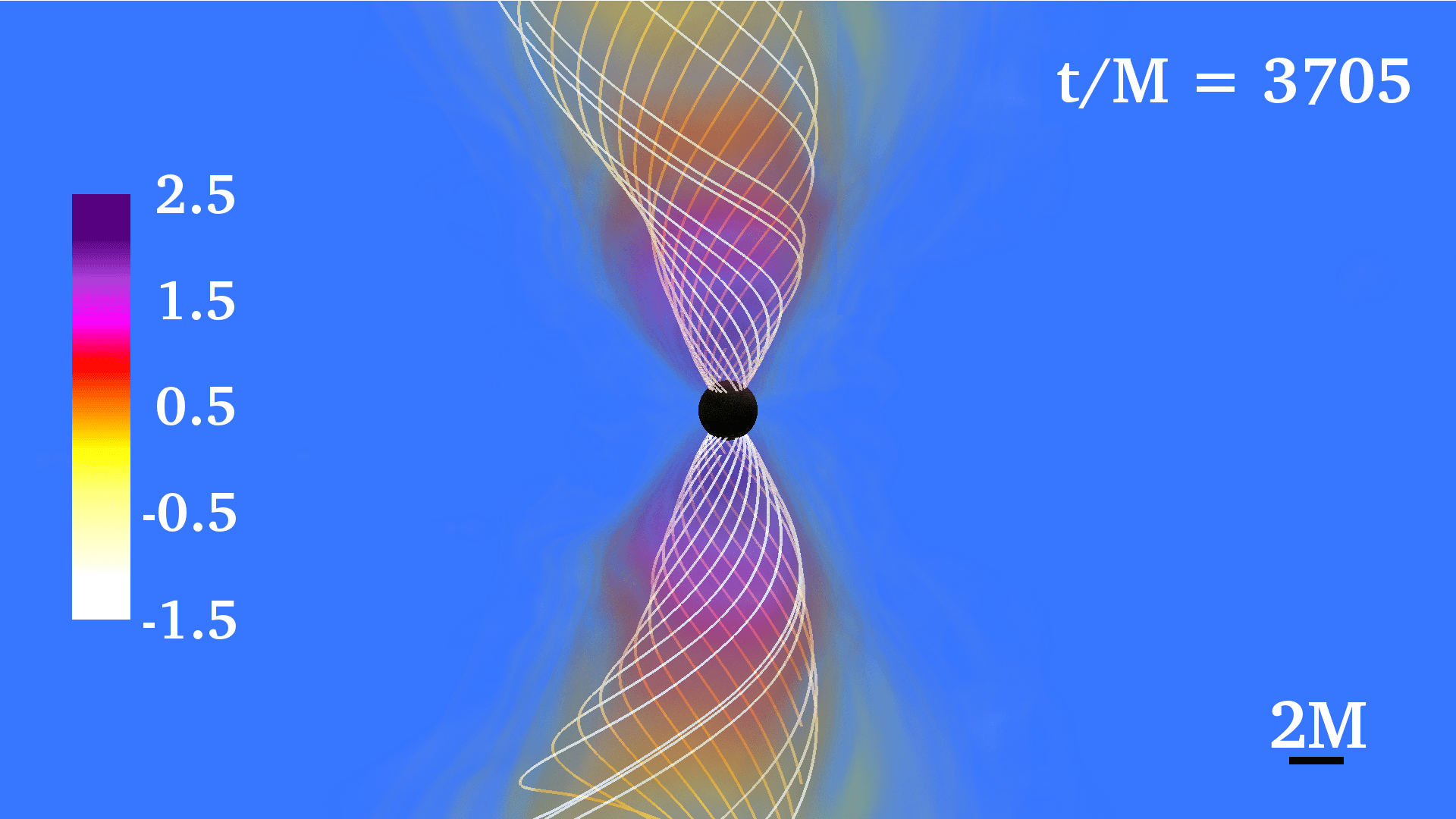}
  \includegraphics[width=0.49\textwidth]{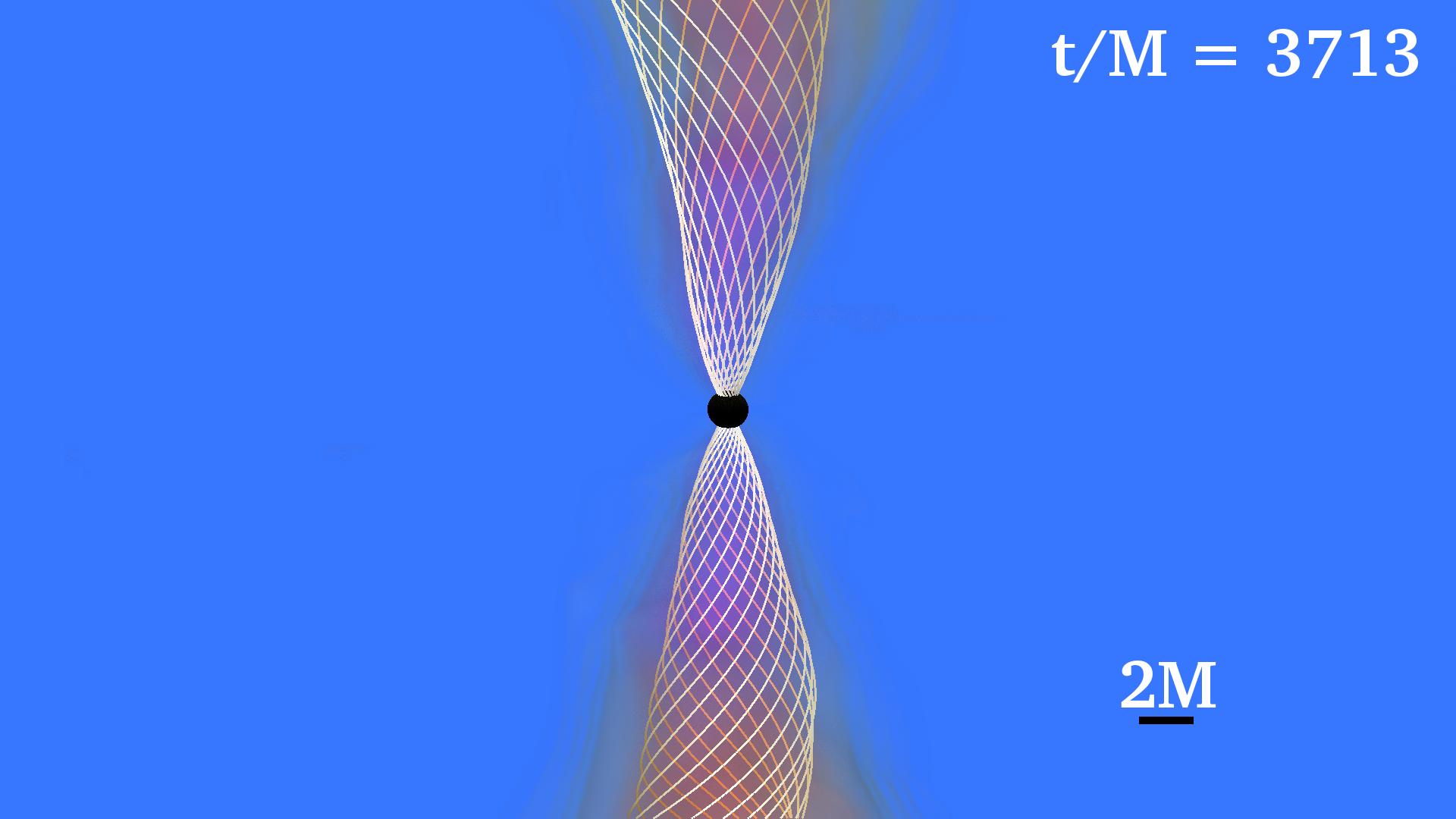}
  \includegraphics[width=0.49\textwidth]{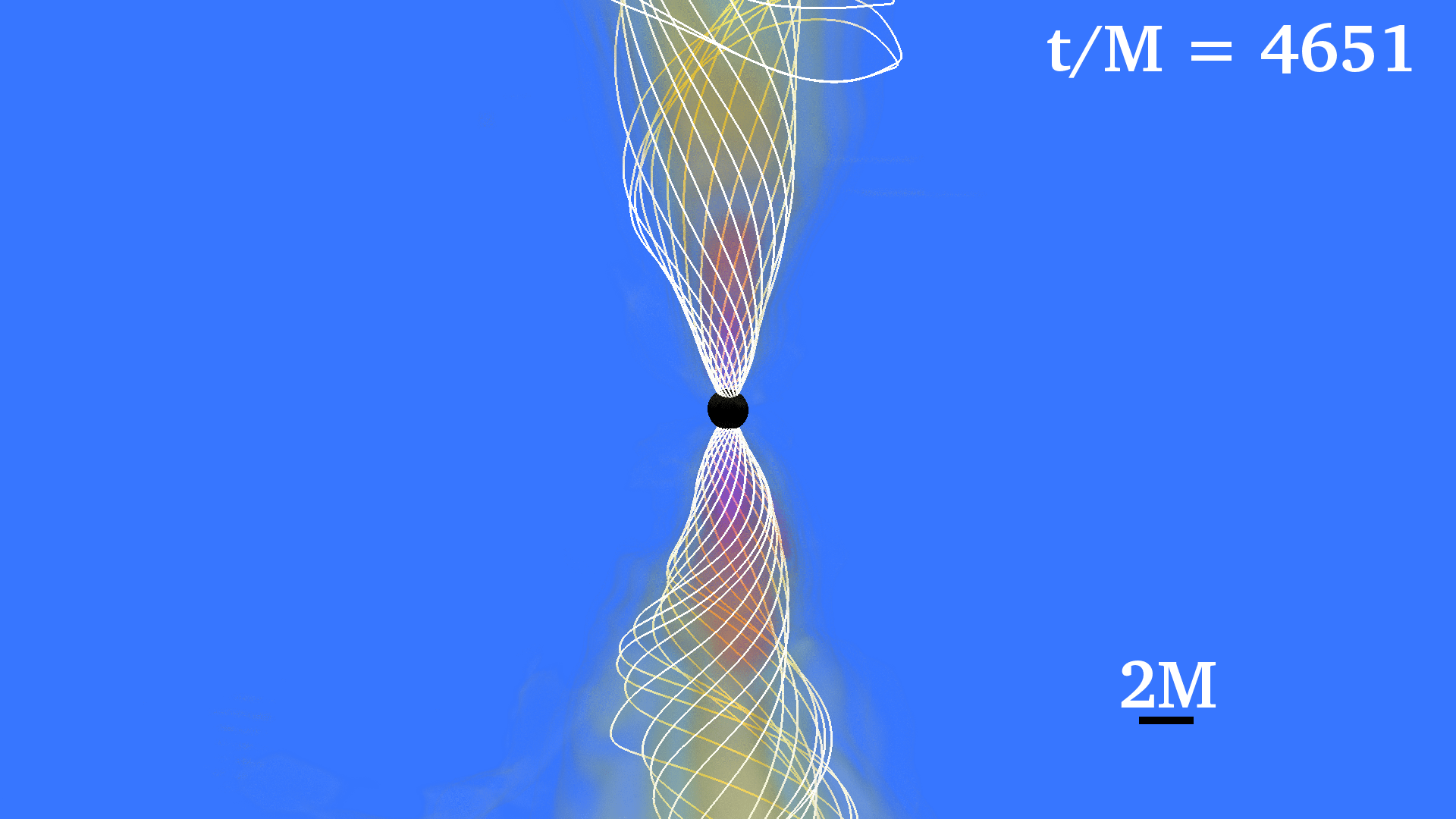}
  \includegraphics[width=0.49\textwidth]{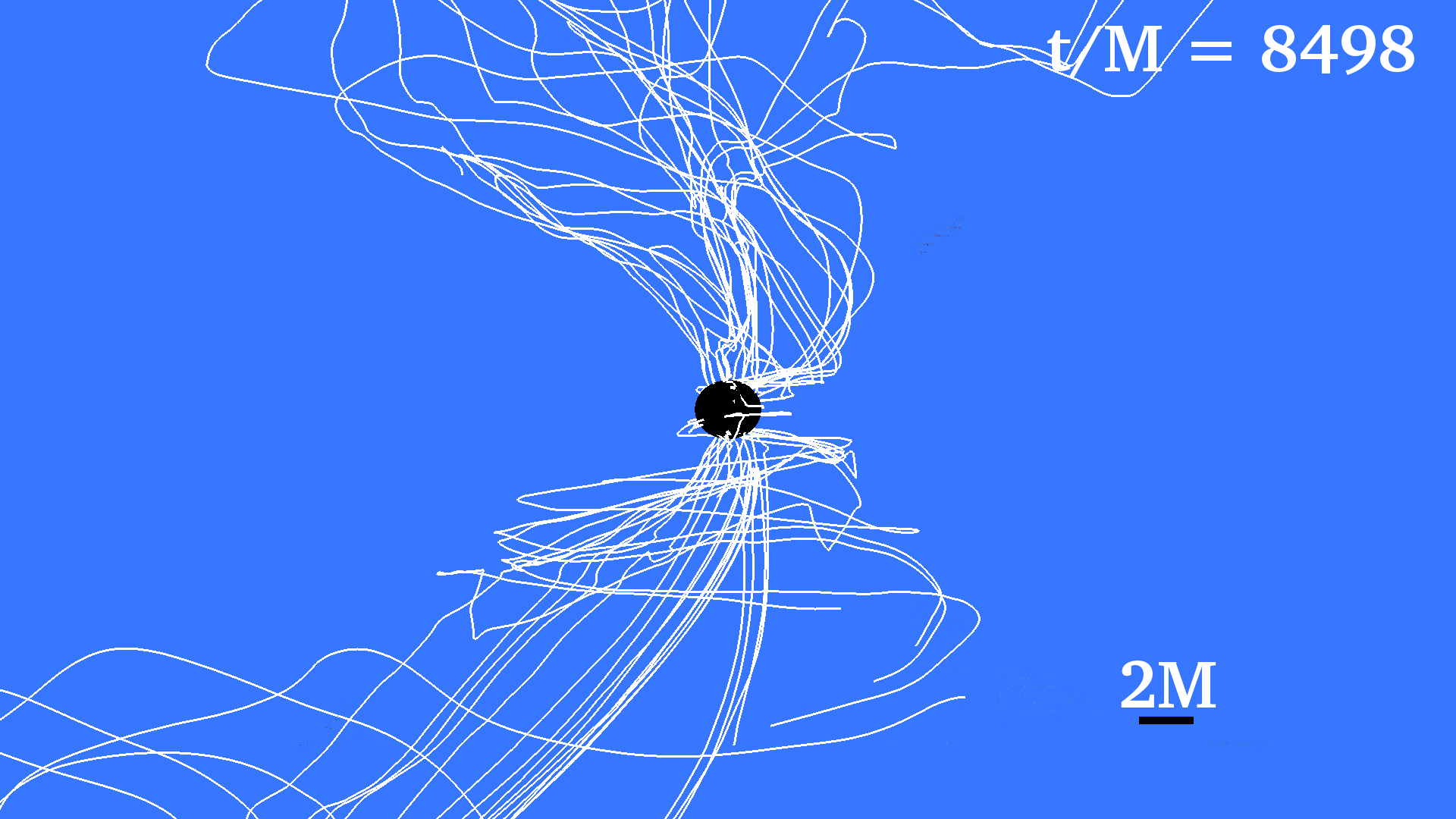}
  \caption{Volume rendering of the ratio~$b^2/2\rho_0$ (log scale) near the end of the
    simulations for the cases Ali-Ali (Eq) (left top panel), Ali-Ali (right top panel), Ali-Per
    (left bottom panel), and Per-Per (right bottom panel). Magnetic field lines, displayed as
    white lines, are plotted inside regions in which~$b^2/2\rho_0 \gtrsim
    10^{-2}$, our criterion for the funnel boundary. The BH horizon is shown as a
    black sphere. Here $M=4.4288(M_{\NS}/1.625M_\odot)\rm km.$
    \label{fig:b2rho_all3d}}
\end{figure*}

During the first~$t-\tmer \simeq 200M\sim 3(M_{\NS}/1.625M_\odot)\rm ms$,
magnetic instabilities amplify the magnetic energy $\mathcal{M}$
by a factor of $\sim 17$, which is slightly larger than in the Ali-Ali cases
(see Fig.~\ref{fig:EM_energy}), and, afterward by a factor of~$\lesssim 2$
during the next $t-\tbh\simeq 2260M\sim 33 (M_{\NS}/1.625M_\odot)\rm ms$ until the
catastrophic HMNS collapse. 
Once the HMNS settles down into a quasistationary state (see third panel in
Fig.~\ref{fig:per_ali}), we find that, as in Ali-Ali cases, the quality factor
$Q_{\text{\tiny MRI}}$ is $\gtrsim 10$ and $\lambda_{\text{\tiny MRI}}$ fits within
the star. However, we note that in this case $\left<\alpha_{\text{\tiny SS}}\right
>_{P_c}$ is $0.02$, a value slightly smaller than those in Ali-Ali cases~(see
Table~\ref{table:summary_key}), though it is still consistent with those typically
produced by turbulent magnetic viscosity~\cite{Kiuchi:2017zzg}.
Magnetic braking and magnetic turbulence bring the central region of
the HMNS into uniform rotation. The star collapses to a BH at $t-\tmer \simeq 2850M
\sim 42(M_{\NS}/1.625 M_\odot)\rm ms$. The BH has a mass of $M_{\text{\tiny
    BH}}\sim 2.73M_\odot$ and spin $a/M_\text{\tiny BH}=0.78$, roughly the
same values as those in the Ali-Ali cases~(see Table~\ref{table:summary_key}).

We note the lifetime of the HMNS remnant in Ali-Per is about
$1150M\sim 17(M_{\NS}/1.625 M_\odot)\rm ms$ longer than in Ali-Ali cases
(see~Table~\ref{table:summary_key}). This delay is likely due to a longer
viscous timescale~$t_{\rm vis}$: the smaller the
viscosity $\alpha_{\text{\tiny SS}}$ parameter the longer the time needed by
turbulent viscosity to damp differential rotation (see Eq.~\ref{eq:tau_win}).
Moreover, note that,
following~\cite{Sun:2018gcl}, the Shakura--Sunyaev parameter can be approximated
as~$\alpha_{\rm SS}\sim {B^{\hat{r}}B^{\hat{\phi}}}/{P}$, where $B^{\hat{r}}$
and $B^{\hat{\phi}}$ are the magnetic field components along the radial
and azimuthal directions. So, the induced-magnetic stresses that transport
angular momentum outward through the HMNS depend only on the magnitude of
the azimuthal and radial components of the magnetic field. As in 
Ali-Ali and Ali-Per the magnitudes of the poloidal and toroidal magnetic
components are similar (see inset~in~Fig.~\ref{fig:EM_energy}), but the magnetic
stresses are smaller in Ali-Per than Ali-Ali because the magnetic field component
along the orbital angular momentum prior to merger is smaller (see first row in Fig.
\ref{fig:Eq_vs_full} and top panel in Fig.~\ref{fig:per_ali}). Thus, 
it is expected then that in the Ali-Per case the HMNS collapses later,
as we found.

Following the collapse of the HMNS,
magnetic winding drives the field lines into a helical funnel (see fourth panel in
Fig.~\ref{fig:per_ali}), and starts to build magnetic pressure above the BH poles.
By $t-\tbh \simeq 1200M\sim 18 (M_{\NS}/1.625M_\odot)\rm ms$, a
magnetically-sustained outflow (see fifth panel in Fig.~\ref{fig:per_ali}), with a Lorentz
factor of $\Gamma_L \lesssim 1.24$ (see Table~\ref{table:summary_key}), is
launched. In this case, the time-delay between BH formation and jet
launching is about $\Delta t\simeq 400M\sim 6 (M_{\NS}/1.625M_\odot) \rm ms$ larger
than in the Ali-Ali cases, likely due to a denser environment: the accretion disk
in Ali-Per is denser than that in Ali-Ali (see Table \ref{table:summary_key}), and
so  it takes longer for the magnetic field to overcome the inertia of the ambient matter
(see second and the third panels of Fig.~\ref{fig:bh+disk_remnant}). As the
strength of the magnetic field does not increase following the BH formation (see
inset in Fig.~\ref{fig:EM_energy}), magnetic-dominated regions ($b^2/(2\rho_0)\gtrsim
1$) above the BH poles appear only after accretion empties the funnel. However,
the accretion rate is smaller in Ali-Per than Ali-Ali~(see Table~
\ref{table:summary_key}), and hence the BH + disk remnant in the first case
requires a longer time for the jet to emerge. This explains why the ratio $b^2/(2
\rho_0)$ in Ali-Per at the end of the simulation is smaller than that in Ali-Ali
(see right top and left bottom panels in Fig.~\ref{fig:b2rho_all3d}), though the
funnel opening angle is similar ($\theta_{\rm jet}\sim 20^\circ$).

\begin{figure}
  \centering
  \includegraphics[width=0.49\textwidth]{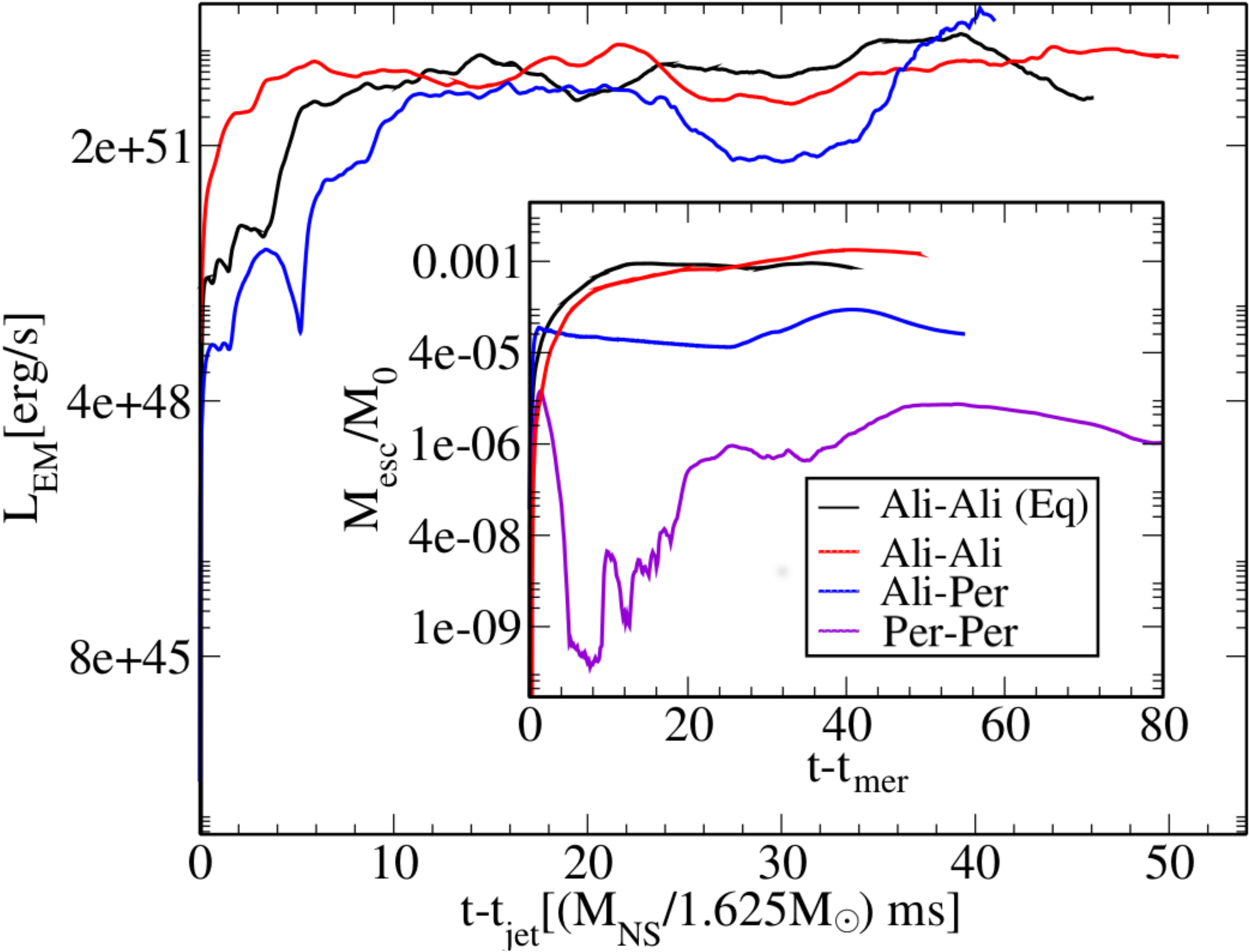}
  \caption{Outgoing EM (Poynting) luminosity driven by the incipient
    jet, and  computed on a sphere of coordinate radius~$r_{\text{\tiny
        ext}}=120M\sim 532(M_{\NS}/1.625M_\odot)\rm km$ for cases
    listed in Table~\ref{table:summary_key}. The inset focus on the
    rest-mass fraction of escaping matter following NSNS merger. 
    \label{fig:Poynt_ejecta}}
\end{figure}

The incipient jet leads to an outgoing EM Poynting luminosity of $L_{\text{
    \tiny EM}}\simeq 10^{51.8}\rm erg/s$ (see~Fig.~\ref{fig:Poynt_ejecta}) consistent
with typical sGRBs. Also, as displayed in the inset of Fig.~\ref{fig:M0_outside},
by $t-\tbh\simeq 510 M\sim 7.5(M_{\NS}/1.625M_\odot)\rm ms$,
$\dot{M}$ begins to settle down. We estimate the lifetime of the accretion disk to
be $\simeq 150(M_{\NS}/1.625 M_\odot)\rm ms$ (see Table \ref{table:summary_key}),
implying that the fuel of the jet (the disk) will be exhausted on a time scale
consistent with the duration of sGRBs~\cite{Ajello:2019zki}. It has an efficiency
of~$\eta_{\text{\tiny EM}}\simeq 0.2\%$, roughly the same as that in Ali-Ali (see Table
\ref{table:summary_key}).  As before, the incipient jet is consistent with
the BZ mechanism for launching jet and its Poynting luminosity.

We also find that the ejecta following the NSNS merger is $M_{\text{\tiny ext}}\sim
10^{-3.5}M_\odot$ (see inset in Fig.~\ref{fig:Poynt_ejecta}),  marginally below the
detection threshold. Thus, in contrast to the Ali-Ali cases, Ali-Per may not give
rise to a detectable kilonova by current or planned telescopes~\cite{Metzger:2011bv}.
The toroidal magnetic field component formed before the merger tends to trap the
outgoing material; fluid elements with a positive radial velocity can become more
easily unbound if they do not have to overcome large transverse magnetic stresses
arising from the toroidal stretching of field lines. So, magnetic field distributions
with a larger poloidal component along the orbital angular momentum of the system
lead to a larger ejecta.

The above results show that {\it a change in the initial aligned, poloidal field content
has a strong impact on the physical properties of the incipient jet, the magnetically-driven
outflow, and hence the likelihood of an observable counterpart kilonova}.

%
%%%%%%%%%%%%%%%%%%%
%%% Luminosity  %%%
%%%%%%%%%%%%%%%%%%%
%
%%%%%%%%%%%%%%%%%
%%% rho _max  %%%
%%%%%%%%%%%%%%%%%%
\begin{figure}
  \centering
  \includegraphics[width=0.49\textwidth]{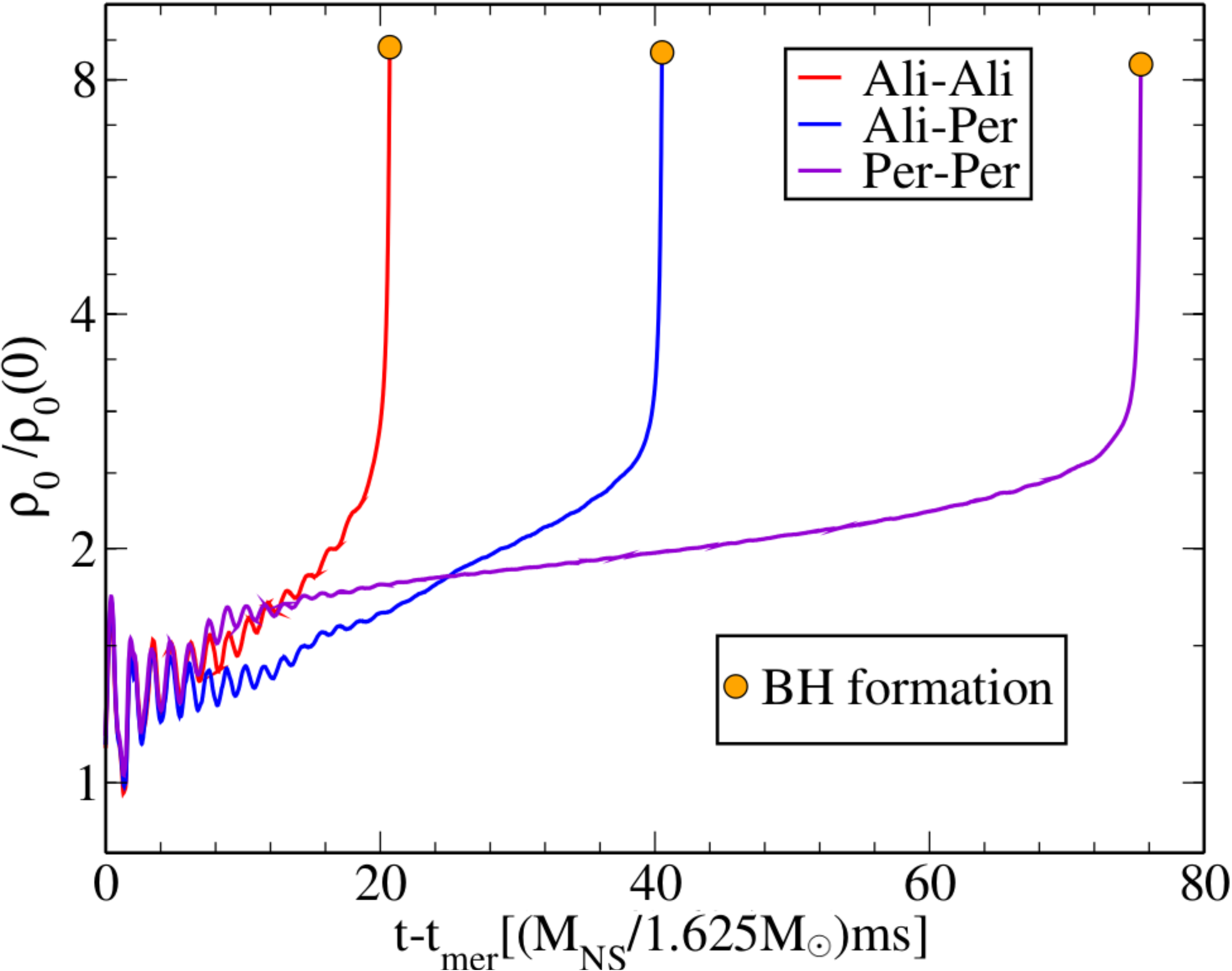}
  \caption{Evolution of the maximum value of the rest-mass density $\rho_0$
    (central density of the HMNS) normalized to its initial value
    $\rho^\text{max}_{0}\simeq 10^{14.78}(1.625\,
    M_\odot/M_{\NS})^2\text{g/cm}^3$ for cases in Table~\ref{table:summary_key}.
    Dots mark the time at which the apparent horizon appears for the first
    time ($\Delta \tbh$). The coordinate time since merger is plotted.
    \label{fig:rho_max}}
\end{figure}
%
%%%%%%%%%%%%%%%%%%%%%%%%%%%%%%%%%%%%%%
%%%  Perpendicular - Perpendicular %%%
%%%%%%%%%%%%%%%%%%%%%%%%%%%%%%%%%%%%%%
%
\subsection{Both perpendicular magnetic fields}
\label{sec:Per-Per_cases}
Fig.~\ref{fig:per-per} summarizes the evolution of the binary in which the dipole
magnetic moment in the two stars  is perpendicular to the direction of the total
orbital angular momentum of the system (see top left panel).  { {The stars merge
at $t\simeq 766M\sim 11.26(M_{\NS}/1.625M_\odot)\rm ms$, roughly %~$136M\sim 2.0(M_{\NS}/
%1.625 M_\odot)\rm ms$ later than
at the same time as in the Ali-Ali case}}~(see Table~\ref{table:summary_key}), 
forming a HMNS. We note that this transient is the more oblate remnant star of
the cases in Table~\ref{table:summary_key}, though the central angular
velocity is similar in all of them~(see Fig.~\ref{fig:Omega_AlivsPer}).

Following the NSNS merger, and during the next $t-t_{\text{\tiny mer}}\simeq 500M\sim 7(M_{
  \NS}/1.625M_\odot)\rm ms$, which approximately corresponds to one Alfv\'en time (see
section IV B in~\cite{Kiuchi:2017zzg}), the magnetic energy  is exponentially amplified
from $10^{50.3}\rm erg$ to $10^{51.6}\rm erg$ (similar amplification factor was observed
in Ali-Per). During this period, the poloidal magnetic field peaks at about $10^{16.7}
(1.625M_\odot/M_{\NS}) \rm G$, the largest magnetic field strength of the all full 3D cases
(see inset
of Fig.~\ref{fig:EM_energy}). Following this amplification period, and in contrast to
the other cases, the instability saturates more quickly and dies way. The poloidal magnetic
field component then decays and relaxes to $\sim 10^{15.7} (1.625M_\odot/M_{\NS})\rm G$,
its original strength just before the NSNS merger. Similar behavior is observed in the toroidal
component. 
Magnetic winding ceases to enhance the toroidal component once the angular velocity becomes
nearly constant  in the central core of the HMNS (see right panel of Fig.~\ref{fig:Omega_AlivsPer}).
Similar behavior was found in axisymmetric simulations of highly spinning NSs reported
in~\cite{dlsss06b}, where the stellar radius is resolved by $\sim 90$ grid points, a
resolution factor of $\sim 1.36$ higher than in our case.

Once the HMNS settles down, we find that $\lambda_{\text{\tiny MRI}}$ is only resolved by 
$\lesssim 6$ grid points, and hence it is at most marginally resolved. By~$t-t_{\text
  {\tiny mer}}\simeq 850M\sim 12(M_{\NS}/1.625M_\odot)\rm ms$, the viscosity
parameter $\left<\alpha_{\text{\tiny SS}}\right>_{P_c}$ is $\sim 0.001$~(see Table
\ref{table:summary_key} for a comparison with the other cases). Although we observe
evidence of magnetic  turbulence, it may not be fully developed.

Fig.~\ref{fig:rho_max} shows the evolution of the maximum value of the rest-mass density
$\rho_0$, normalized to its initial value $\rho^{\text{max}}_{0}(0)\simeq 10^{14.78}(1.625\,
M_\odot/M_{\NS})^2\text{g/cm}^3$ for cases in Table~\ref{table:summary_key}. Note that
this quantity coincides with the central value of the rest-mass density of the HMNS
after $t-\tmer \sim 200M\sim 3(M_{\NS}/1.625M_\odot)\rm ms$.
During the first $t-t_{\text{\tiny mer}}\simeq 850M\sim 12(M_{\NS}/1.625M_\odot)\rm ms$,
the dynamics of $\rho_0$ in cases Ali-Ali and Per-Per is fairly similar. Consistent with magnetic
braking~(see~section IV B in~\cite{Kiuchi:2017zzg}), angular momentum is  transferred
from  the inner to the outer layers of the HMNS. It forms a massive central core with
a central rest-mass density $\rho_0\simeq 1.9\rho_0(0)$  surrounded by a cloud of
matter (see second panel in Fig.~\ref{fig:per-per}). Afterwards, and in contrast to
$\rho_0$ in Ali-Ali that speedily increases, the central density of Ali-Per slowly
increases for the next $\Delta t \simeq4570M\sim 64(M_{\NS}/1.625M_\odot)\rm ms$ until
$\rho_0 \simeq 3\rho_0(0)$, where the catastrophic collapse is triggered. This is a time
span roughly consistent with the viscous timescale $t_{\text{\tiny vis}}$ (see Sec.
\ref{sec:Ali-Per_cases}). During this period, the central core shrinks $0.5M\simeq
2.2(M_{\rm NS}/1.625M_\odot)\rm km$, while the cloud of matter expands by
$\simeq 3M\sim 13(M_{\rm NS}/1.625M_\odot)\rm km$ (see forth panel in
Fig.~\ref{fig:per-per}). Moreover, after $\Delta t\sim 56P_c$ (see right panel
of Fig.~\ref{fig:Omega_AlivsPer}),  magnetic viscosity has damped the differential
rotation in the central core . Here $P_c\simeq 50M\sim 0.7(M_{\rm NS}/1
.625M_\odot)\rm ms$ is the central period of the transient HMNS.

As magnetic turbulence may be suppressed by numerical diffusion, we also probe the effect
of the resolution on $\alpha_{\text{\tiny SS}}$.  We rerun the Per-Per case with a
resolution factor of $1.25$ higher than before, the highest factor we can afford with
the finite computational resources at our disposal. We find that the values of the
viscosity  parameter within the HMNS turn out to be roughly insensitive to this change
in the resolution.

By~$t-\tmer\simeq 5465M\sim 76.5 (M_{\rm NS}/1.625M_\odot)\rm ms$, the collapse
is triggered (see Fig.~\ref{fig:rho_max}). { {In the high resolution case, we found
that the collapse is triggered roughly $\Delta t\sim 648M \sim 9.5(M_{\NS}/1.625M_\odot)
\rm ms$ later than before. The sensitivity of the collapse time for HMNSs to the magnetic
field is physical and has been observed previously~(see e.g.~\cite{grb11}).
Its dependence on resolution,  even  in  purely  hydrodynamic  simulations, has been noted
as well (see e.g.~\cite{PEFS2015,PEFS2016}).}}
The HMNS collapses to a BH with mass
$M_{\text{\tiny BH}}\sim 2.73M_\odot$ and  spin parameter $a/
M_\text{\tiny BH}=0.78$. Similar values were found in all the previous cases~(see
Table~\ref{table:summary_key}). 

Following collapse, the magnetic energy decreases
even further (see the inset in Fig.~\ref{fig:EM_energy}). After $t-t_{\text{\tiny BH}}
\simeq 1965M\sim 27.5(M_{\rm NS}/1.625M_\odot)\rm ms$, near to the end of the simulation
$\mathcal{M}\simeq 10^{-5.1}M=10^{49.6}(M_{\NS}/1.625M_\odot)\rm erg$. We also note that,
following the accretion peak, the rms
value of the magnetic field in regions directly above the pole is $\simeq 10^{14.8}(1.625M_\odot/
M_{\NS})\rm G$ (see~Table~\ref{table:summary_key}) and remains roughly constant until the
end of the simulation.

As displayed in Fig.~\ref{fig:M0_outside}, following the accretion peak, the  BH remnant is immersed
in an accretion disk with a rest-mass $\sim 11.37\%$ of the total initial rest-mass of the system,
a value slightly smaller than the one in the Ali-Per case (see~Table~\ref{table:summary_key}). The inset of Fig.
\ref{fig:M0_outside} shows that  by $t-t_{\text{\tiny BH}}\simeq 360 M\sim 5(M_{\NS}/1.625
M_\odot)\rm ms$ the accretion rate  begins to settle down, and then gradually decays thereafter.
We estimate that the disk will be accreted in $\tau_{\text{\tiny disk}}\sim 144\rm ms$.
Notice that this timescale is nearly the same as that in the Ali-Ali case (see~Table~\ref{table:summary_key}),
which may indicate that the accretion in Per-Per is also driven by magnetic stresses in the
bulk of the disk.

Fig.~\ref{fig:bh+disk_remnant} displays the final configuration of the BH + disk remnant.
We note that even after $t-\tbh \simeq 1965M\sim 27.5(M_{\rm NS}/1.625M_\odot)\rm ms$ , there
is still a dense cloud of fall-back material raining down into the BH
remnant.  By this time, we do not observe magnetically dominant force-free regions with $b^2/(2\rho_0)
\gtrsim 1$.
As is shown in the bottom left panel of Fig.~\ref{fig:b2rho_all3d}, near to the end of the
simulation, we find that this ratio is $\simeq 10^{-3.7}$ above the BH poles.  We do not find
any evidences of an outflow or a large-scale magnetic field collimation (see panel fifth and
sixth in Fig.~\ref{fig:per-per}). This result is consistent with the GRMHD simulations of NSNS
mergers undergoing prompt collapse in~\cite{Ruiz:2017inq}, which suggest that there is a threshold
value of the magnetic energy ($\mathcal{M}/M\simeq 10^{-3}$) below which the BH + disk remnant
does not launch a magnetically-supported jet. Finally, the material ejected following the merger
is $M_{\text{\tiny esc}}\sim 10^{-6}M_\odot$ (see inset in Fig.~\ref{fig:Poynt_ejecta}), and
hence it is unlikely to produce any detectable EM counterpart.

These results suggest that {\it the lack of a large-scale, aligned poloidal field component
  in the initial system may suppress detectable EM counterparts, such as a magnetically-supported
  jets or ejecta that can give rise to GRBs or detectable kilonovae}.
%
%%%%%%%%%%%%%%%%%%%%%%%%%%%%%%%%%%%%
%%%  Gravitational wave analysis %%%
%%%%%%%%%%%%%%%%%%%%%%%%%%%%%%%%%%%%
%
\subsection{Distinguishability of the gravitational waves}
\label{sec:GW_ftt}
\begin{figure}
  \centering
  \includegraphics[width=0.49\textwidth]{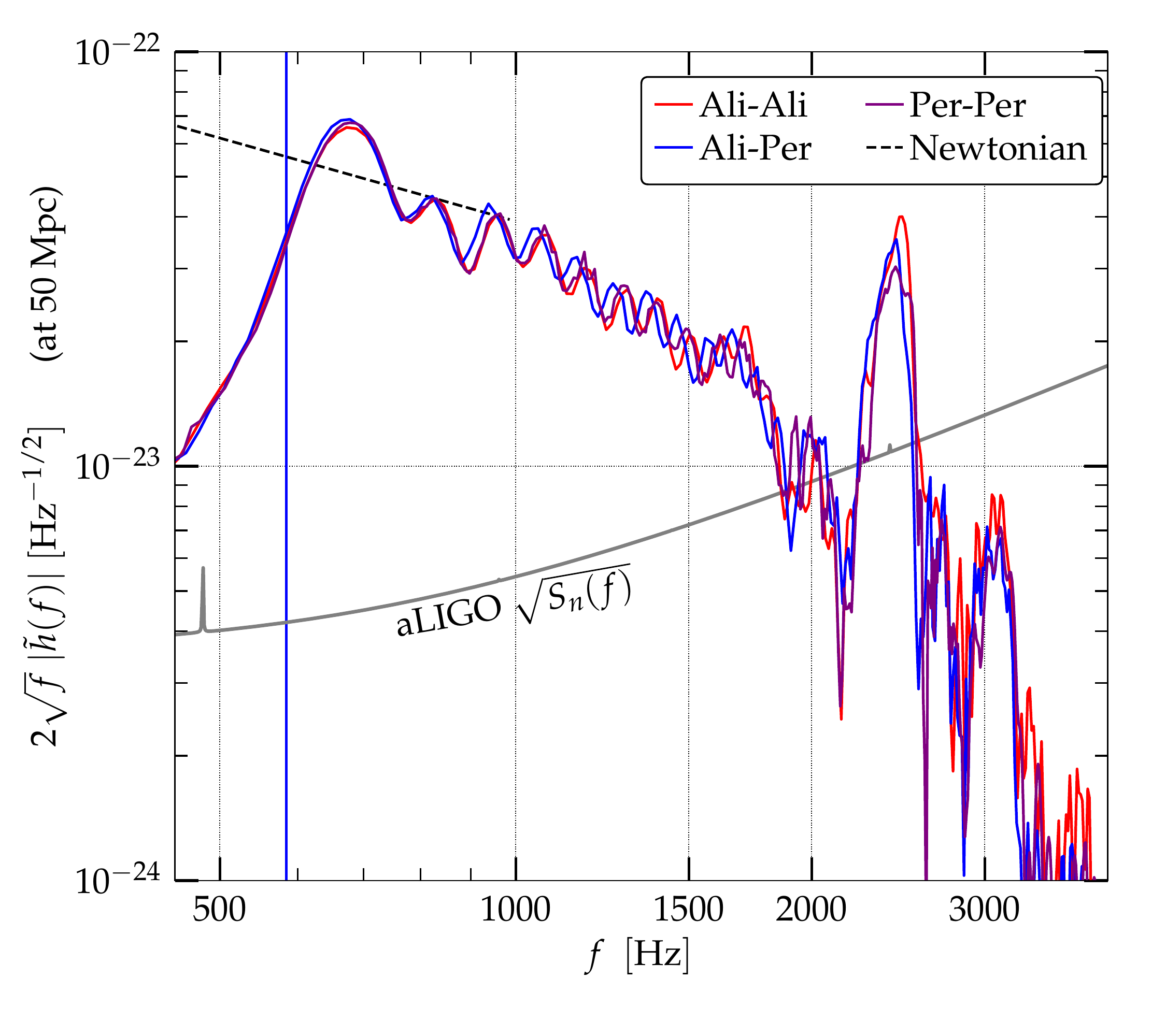}
  \caption{Gravitational-wave power spectrum of the dominant mode $(l,m)=(2,2)$
    at a source distance of $50\rm Mpc$ for all the full 3D cases listed in Table~\ref{table:summary_key},
    along with the aLIGO noise curve.  This curve corresponds to the {\tt ZERO$\_$DET$\_$HIGH$\_$P}
    configuration~\cite{Shoemaker09L}. Dashed lines display the Newtonian prediction~\cite{Cutler:1994ys}. 
     The vertical blue line marks the initial GW frequency.
    \label{fig:fft}}
\end{figure}

By $t-\tmer\sim 850M\sim 12.5(M_{\NS}/1.625M_\odot)\rm ms$  we notice that, in
all cases listed in Table~\ref{table:summary_key}, the HMNS remnant has reached a quasiaxisymmetric
configuration and, as shown in Fig.~\ref{fig:GW_allcase},  no longer emits significant gravitational
radiation. This time corresponds to a frequency of~$f_{\text{\tiny axym}}\simeq 2800(M_{\NS}/
1.625M_\odot)^{-1}\rm Hz$, near to the edge of the aLIGO high frequency band~\cite{Shoemaker09L}.
So, for frequencies~$f\leq f_{\text{\tiny axym}}$, we probe if our different seed magnetic field
configurations are distinguishable by their gravitational waveforms. Fig.~\ref{fig:fft} shows the
gravitational-wave power spectrum of the dominant mode $(l,m)=(2,2)$ at a source distance of
$50\rm Mpc$ for all the full 3D cases listed in Table~\ref{table:summary_key}, along with the aLIGO
noise curve of the {\tt ZERO$\_$DET$\_$HIGH$\_$P} configuration~\cite{Shoemaker09L}. Dashed lines
displays the Newtonian prediction~\cite{Cutler:1994ys}, while the vertical blue line marks the
dominant GW frequency of NSNS configurations at the start of the simulations. Note that the spectral
features of the waveforms are roughly the same. 
%
%%%%%%%%%%%%
%%% fft  %%%
%%%%%%%%%%%%
We also compute the match function $\mathcal{M}_{\text{\tiny{GW}}}$~\cite{Allen2012},
\begin{equation}
\mathcal{M}_{\text{\tiny{GW}}} =  \underset{(\phi_c,t_c)}{{\rm max}}
\frac{({h}_1|{h}_2(\phi_c,t_c))}
     {\sqrt{({h}_1|{h}_1)({h}_2|{h}_2)}}\,,
\label{eq:match}
\end{equation}
between two given waveforms.
The maximization is taken over a large set of phase shifts $\phi_c$ and time
shifts $t_c$. Here $({h}_1|{h}_2)$ denotes the standard noise-weighted inner
product (see Appendix C~ in \cite{Allen2012})
\begin{equation}
(h_1|h_2 )= 4\,{\rm Re}\int_0^\infty\frac{\tilde{h}_1(f)\,\tilde{h}^*_2(f)}
  {S_h(f)}\,df\,,
\end{equation}
where $h=h_+-i\,h_\times$, $\tilde{h}$ is the Fourier transform of the
strain amplitude $\sqrt{\tilde h_+(f)^2 +\tilde h_{\times}(f)^2}$
of the dominant mode $(l,m)=(2,2)$, and $S_h(f)$ is the power spectral
density of the aLIGO noise. Using the aLIGO configuration
{\tt ZERO$\_$DET$\_$HIGH$\_$P}, we find that $\mathcal{M}_{\text{\tiny{GW}}}=0.9974$
between the waveforms of Ali-Ali and Ali-Per, $\mathcal{M}_{\text{\tiny{GW}}}=0.9993$
between the waveforms of Ali-Ali and Per-Per, and $\mathcal{M}_{\text{\tiny{GW}}}=0.9977$
between the waveforms of Ali-Per and Per-Per. These results imply that the GWs
of Ali-Ali and Per-Per will be potentially distinguishable for a signal-to-noise ratio $> 25$,
while in the other cases, they can be  potentially  distinguished with a signal-to-noise ratio $> 15$
\cite{Chatziioannou17,Harry_2018}. Note that the first BHBH gravitational wave event
was observed with a signal-to-noise ratio of $\sim 24$~\cite{LIGO_first_direct_GW}, while
GW170817 was detected
with signal-to-noise ratio of $\sim 32.5$~\cite{TheLIGOScientific:2017qsa}.
{ {For the numbers quoted above we used the full  waveform at the same
resolution to compute the match numbers. On the other hand, if one focuses in the 
postmerger epoch alone (i.e. $f>1850(M_{\NS}/1.625M_\odot)^{-1}\rm Hz$ and a distance
of 50 Mpc) the match numbers  reduce significantly due to the small signal power above the
aLIGO sensitivity level,  and distinguishability is lost. However, at smaller distances
(e.g. 10 Mpc) distinguishability may again be possible.
Our results
suggest then that, although in principle, {\it aLIGO may be able to distinguish waveforms arising
  from different magnetic field configurations,
that would practically be very difficult. Future detectors will have better
chances to detect magnetic field orientiation effects.}}}
  
%
%%%%%%%%%%%%%%%%%%%
%%% Conclusions %%%
%%%%%%%%%%%%%%%%%%%
\section{Summary and Conclusions}
\label{sec:conclusion}
The merger of a binary neutron stars is likely to be the progenitor of the coincident
gravitational wave event GW170817 with EM counterparts across the spectrum. It is the
likely progenitor of an sGRB~(event GRB170817A~\cite{Monitor:2017mdv}). Such a sGRB counterpart
was originally proposed in~\cite{Pac86ApJ,EiLiPiSc,NaPaPi}, and recently demonstrated by
self-consistent GRMHD simulations of binary neutron star mergers whose remnant undergoes
delayed collapse
in~\cite{Ruiz:2018wah,Ruiz:2017inq}. These multimessenger signals have been used to impose
some constraints on the physical properties of a neutron star (see~e.g.~\cite{Margalit:2017dij,
  Shibata:2017xdx,Ruiz:2017due,Rezzolla:2017aly,Most:2018hfd,
  TheLIGOScientific:2017qsa, Abbott:2018exr,Radice:2017lry,Bauswein:2017vtn} and references
therein), such as the maximum mass of a spherical star, tidal deformability, equation of state,
radius of a neutron star, etc. 

To solidify the role of binary neutron star mergers as multimessenger sources, we
studied in this paper the impact of different orientations of seed magnetic field configurations.
We focused on the emergence of a magnetically-driven jet and the ejecta that may
give rise to a kilonova detectable by current telescopes, such as the Large Synoptic
Survey Telescope~\cite{Shibata:2017xdx,Rosswog}. We considered spinning binary
neutron stars initially on a quasicircular orbit undergoing merger and delayed collapse
to a BH. The binaries consisted of two identical $\Gamma=2$ polytropes with spin $\chi_{
  \text{\tiny{NS}}}= 0.36$ aligned along the direction of the total orbital angular momentum
of the system~$L$. Each star is initially threaded by an interior and exterior magnetic
field resembling that of pulsars, and whose dipole moment $\mu$ is either aligned or perpendicular
to $L$. For comparison purposes, we also considered the binary evolution in which $\mu$ in both stars
is aligned along $L$ but where we imposed symmetry across the orbital plane (equatorial symmetry),
to calibrate what is done in numerous simulations.

We found that following merger, in all cases listed in Table~\ref{table:summary_key},
magnetic braking in the bulk of the HMNS remnant induces the formation of a nearly uniformly
rotating central core immersed in a low-density Keplerian cloud of matter that eventually
collapses to a BH. Depending on the initial poloidal field content along $L$, the HMNS collapses
in a timescale ranging between $t-\tmer\sim 24(M_{\NS}/
1.625M_\odot)\rm ms$, when $\mu$ in both stars is aligned along $L$, to $t-\tmer\sim 76(M_{\NS}/
1.625M_\odot)\rm ms$, when it is perpendicular in both of them. Nevertheless, the mass
[$M_{\text{\tiny BH}}\simeq 2.75M_\odot$] and the spin parameter [$a/M_{\text{\tiny BH}}
  \simeq 0.78$] of the BH remnant, as well as the rest-mass of the accretion disk
[$M_{\text{\tiny disk}}/M_0\gtrsim 9\%$], are roughly independent of the initial magnetic field
orientation.

We noticed that the final magnetic energy $\mathcal{M}$ is also highly affected by the content of
the large-scale, aligned poloidal magnetic field  prior to merger. As shown in Fig.~\ref{fig:EM_energy},
the larger the component, the larger the final magnetic energy. Consistent with the GRMHD simulations of
binary neutron star mergers undergoing prompt collapse in~\cite{Ruiz:2017inq}, which suggest that
there is a threshold value of the magnetic energy  below which
the BH + disk remnant does not launch an incipient jet, we found that only in the cases in which the magnetic
energy becomes larger than~$\gtrsim 10^{-3}M$, where $M=10^{54.7}(M_{\NS}/1.625M_\odot)\rm erg$ is the ADM mass
of the system, does a magnetically-supported jet emerge. The lifetime [$\Delta t\gtrsim 140(M_{\NS}/
  1.625M_\odot)\rm ms$] and Poynting luminosities~[$L_{\text{\tiny EM}}\simeq 10^{52}$erg/s]
of the jet  are consistent with typical short gamma ray bursts, as well as with the
Blandford--Znajek mechanism for launching jets. Moreover, as shown in~Fig.\ref{fig:Eq_vs_full},
symmetries do not play  a significant role  in the binary evolution: the final outcome in the
equatorial case is roughly the same as that in the full 3D case. We also noticed that the magnetic
field configuration does have a strong affect on the material ejected following the merger. We found that,
only in cases where $\mu$ in both stars is aligned  with $L$, the computed ejecta is  
$\gtrsim 10^{-3}M_\odot$, the value required to give rise to a detectable kilonova~\cite{Shibata:2017xdx,Rosswog}.
In the case where $\mu$ in one star is aligned with $L$ and in the other star perpendicular to it, 
the ejecta is marginally below the detectability threshold value, and hence the kilonova may not be
detected by current or planned telescopes. In the case where $\mu$ in both stars is perpendicular to $L$,
the ejecta is negligible.

Our  preliminary results indicate that the binary neutron star merger models without magnetic
fields used to explain the early part of the radioactively powered kilonova signal (blue luminosity)
linked to GW170817 may overestimate the amount of escaping matter and, therefore, its corresponding
luminosity.  Higher resolution studies involving more general magnetic configurations may be
required to obtain solid estimates.

We also probed whether different seed magnetic field orientations could be distinguishable by
aLIGO. For that we computed the match  function~$\mathcal{M}_{\text{\tiny{GW}}}$
(see Eq.~\ref{eq:match}).  We find that $\mathcal{M}_{\text{\tiny{GW}}}=0.9974$
between the waveforms of Ali-Ali and Ali-Per, $\mathcal{M}_{\text{\tiny{GW}}}=0.9993$
between the waveforms of Ali-Ali and Per-Per, and $\mathcal{M}_{\text{\tiny{GW}}}=0.9977$
between the waveforms of Ali-Per and Per-Per. These results imply that, the GWs
of Ali-Ali and Per-Per will be distinguishable for a signal-to-noise ratio $> 25$,
while in the other cases, they can be distinguished with a signal-to-noise ratio $> 15$.
Hence current detectors  may, in principle, be able to distinguish different
magnetic configurations.
%
%%%%%%%%%%%%%%%%%%%%%%
%   Acknowledgments
%%%%%%%%%%%%%%%%%%%%%%

\acknowledgements
We thank the Illinois Relativity Group REU team, G. Liu, K. Nelli, M. N.T
Nguyen, and S. Qunell for assistance with some of the visualizations.
This work was supported by NSF grant PHY-1662211 and NASA grant  80NSSC17K0070 to the
University of Illinois at Urbana-Champaign. This work made use of the
Extreme Science and Engineering Discovery Environment (XSEDE), which is supported by National
Science Foundation grant number TG-MCA99S008. This research is part of the Blue Waters
sustained-petascale computing project, which is supported by the National Science Foundation
(awards OCI-0725070 and ACI-1238993) and the State of Illinois. Blue Waters
is a joint effort of the University of Illinois at Urbana-Champaign and its National Center
for Supercomputing Applications.  Re-sources supporting this work were also provided by the
NASA High-End Computing (HEC) Program through the NASA Advanced  Supercomputing  (NAS)
Division at Ames Research Center.
%
%%%%%%%%%%%%%%%%%%%%%%%%%%%%%%%%%%%%%%%%
\bibliographystyle{apsrev4-1}        %%%
\bibliography{references}            %%%
%%%%%%%%%%%%%%%%%%%%%%%%%%%%%%%%%%%%%%%%
\end{document}